%% file: uranieStart.tex
\documentclass[preprint,3p,twocolumn]{elsarticle}

\usepackage{amssymb}
\usepackage{subfig}
\usepackage{mathptmx}
\usepackage{hyperref}
\usepackage{UranieShortcut}
\usepackage{amsmath}
\usepackage{array}

\newcolumntype{C}[1]{>{\centering\arraybackslash}m{#1}}
\newcolumntype{L}[1]{>{\raggedright\arraybackslash}m{#1}}
\newcolumntype{N}{@{}m{0pt}@{}}

\graphicspath{{figure/}{macros/PDFfiles/}}

\begin{document}

\begin{frontmatter}

\title{The \uranie{} platform: an \openS{} software for optimisation, meta-modelling and uncertainty analysis.}

\author[cea]{J-B.~Blanchard\corref{jbb}}
\ead{jean-baptiste.blanchard@cea.fr}
\author[cea]{G.~Damblin}
\ead{guillaume.damblin@cea.fr}
\author[cea]{J-M.~Martinez}
\ead{jean-marc.martinez@cea.fr}
\author[cea]{G.~Arnaud}
\ead{gilles.arnaud@cea.fr}
\author[cea]{F.~Gaudier}
\ead{fabrice.gaudier@cea.fr}

\address[cea]{Den-Service de thermo-hydraulique et de m\'ecanique des fluides (STMF), CEA, Universit\'e Paris-Saclay, F-91191, Gif-sur-Yvette, France}
\cortext[jbb]{Corresponding author}

\begin{keyword}
  uncertainty quantification \sep propagation \sep optimisation \sep EGO \sep sensitivity analysis \sep \surmod \sep kriging \sep neural
  network \sep \doe \sep \openS \sep \proglang{C++} \sep \proglang{Python}
\end{keyword}


\begin{abstract}
  The high-performance computing resources and the constant improvement of both numerical simulation accuracy and the
  experimental measurements with which they are confronted, bring a new compulsory step to strengthen the credence given to
  the simulation results: uncertainty quantification. This can have different meanings, according to the requested goals
  (rank uncertainty sources, reduce them, estimate precisely a critical threshold or an optimal working point) and it could
  request mathematical methods with greater or lesser complexity.  This paper introduces the \uranie{} platform, an \openS{}
  framework which is currently developed at the Alternative Energies and Atomic Energy Commission (CEA), in the nuclear
  energy division, in order to deal with uncertainty propagation, \surmod{s}, optimisation issues, code calibration\ldots
  This platform benefits from both its dependencies, but also from personal developments, to offer an efficient data handling
  model, a \cpp{} and \python{} interpreter, advanced graphical tools, several parallelisation solutions\ldots These methods
  are very generic and can then be applied to many kinds of code (as \uranie{} considers them as black boxes) so to many
  fields of physics as well. In this paper, the example of thermal exchange between a plate-sheet and a fluid is introduced
  to show how \uranie{} can be used to perform a large range of analysis. The code used to produce the figures of this paper
  can be found in \url{https://sourceforge.net/projects/uranie/} along with the sources of the platform. This paper has been
  submitted to Computer Physics Communication.
\end{abstract}
\end{frontmatter}

\section{Introduction} 
\label{intro}

\input{Introduction}

\section{Surrogate model generation} 
\label{model}
\input{Modelling}

\section{Uncertainty propagation}
\label{uncert}

\input{UncertPropag}

\section{Sensitivity analysis} 
\label{sensi}
\input{SensiAnalysis}

\section{Optimisation} 
\label{optim}

\input{Optimization}

\section{Perspectives}
\label{final}
\input{Broadening}

\bibliographystyle{elsarticle-num}
\bibliography{uranieStart}
\end{document}

%% file: Introduction.tex
Uncertainty quantification is the science of quantitative characterisation and reduction of uncertainties in both
computational and real world applications. This procedure usually requests a great number of code runs to get reliable
results, which has been a real drawback for a long time. In the past few years many interesting developments have been
brought to try to overcome this, these improvements coming both from the methodological and computing side. Among the
interesting features oftenly used to perform uncertainty quantification, one can state, for instance, sensitivity analysis to
get a rough ranking of uncertainty sources and \surmod{} generation to emulate the code and perform a complete analysis on it
(uncertainty propagation, optimisation, calibration)\ldots Knowing this and with the increasing number of resources available
to assess complex computations (fluid evolution with a fine mesh, for instance), physicists should know whether or not it
might be useful to increase the mesh resolution. It could instead be more relevant to reduce a specific uncertainty source, or
add new locations to be included in a learning database for building a \surmod.

The \uranie{} platform has been developed in order to gather the methodological developments coming both from the academic
and the industrial world and provide them to the broadest audience possible. This is done, keeping in mind few important
aspects such as:
\begin{itemize}
\item Open-source: the platform can be used by anyone, and every motivated person can investigate the code and propose
  improvements or corrections.
\item Accessibility: the platform is developed on Linux but a windows-porting is performed. Even though it is written in
  \cpp, it can also be used through \python.
\item Modularity: the platform is organised in modules so that one should only load requested modules and that analysis can
  be organised as a compilation of fundamental bricks. The modules are introduced in \Fig{modorga} and discussed later on.
\item Genericity: the platform can work on an explicit function but it can also handle a code considering it as a black-box
  (as long as communications can be done through file, for instance). This assures that \uranie's methods are non-intrusive
  and that it can be applied to all science fields.
\end{itemize}

In this section, the \uranie{} platform is introduced, from its internal organisation to its dependencies.  The physical
\usecase{} of this paper, a simplified mono-dimensional thermal exchange model, is later discussed along with a classical
strategic plan to tackle uncertainty analysis.

\subsection[The uranie platform]{The \uranie{} platform}
\label{intro_Uranie}

\uranie{} (the version under discussion here being \uraniev) is an \openS{} software dedicated to perform studies on
uncertainty, sensitivity analysis, \surmod{} generation and calibration, optimisation issues\ldots Developed by the nuclear
energy division of the CEA\footnote{Alternative Energies and Atomic Energy Commission, Saclay, France. The nuclear energy
division is usually referred as Den.} and written in \cpp{}, it is based on the \ROOT~\cite{brun:1997pa} platform (discussed
in \Sect{extdep}).
  
\begin{figure}[!h]
  \begin{center}
    \includegraphics[width=1.05\linewidth,keepaspectratio=true]{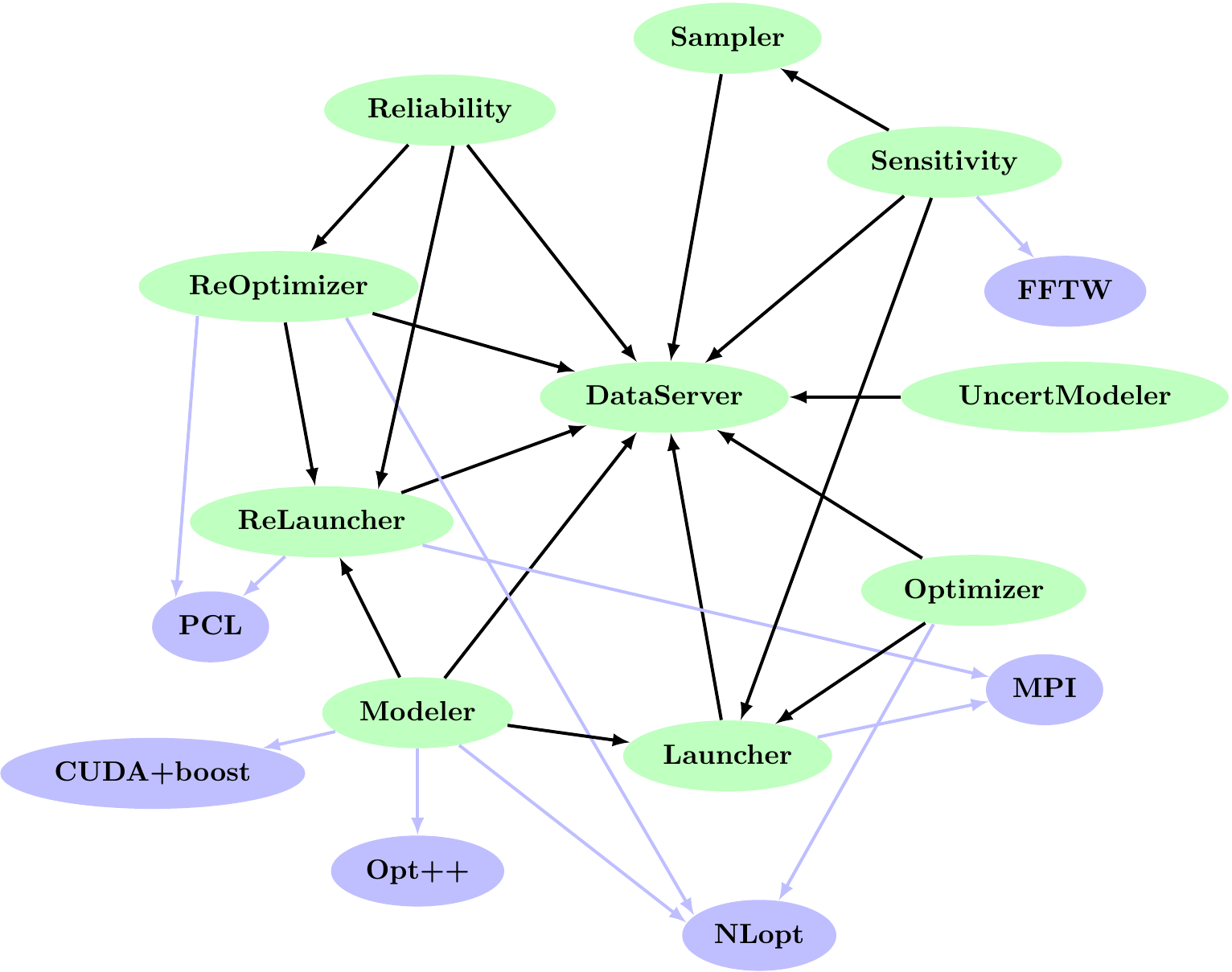}
  \end{center}
  \caption{Organisation of the \uranie-modules (green boxes) in terms of inter-dependencies. The external dependencies are
    shown as light-purple boxes.\label{fig_modorga}}
\end{figure}

The platform consists in a set of so-called technical libraries, usually referred as modules (represented as the green boxes
in \Fig{modorga}), each performing a specific task. Some of them are considered low-level, in the sense that they are the
foundation bricks upon which rely the rest of the modules, which can be considered more methodologically-oriented (dedicated
to a specific kind of analysis).

In the rest of this section, the main modules, used throughout this paper, will be briefly described (in terms of role)
starting with the \dataserver{} one, which is the spine of the \uranie{} project, as shown in \Fig{modorga}.

\subsubsection{DataServer module}
The \dataserver{} library is the core of the \uranie{} platform. It contains all the necessary information about the
variables of a problem (such as the names, units, probability laws, data files, and so on\ldots), the data itself (if
information have been brought or generated) and it allows to perform very basic statistical operations (computing averages,
standard deviations, quantiles\ldots).

\subsubsection{Sampler module}
The \sampler{} library allows to create a large variety of \doe{} depending on the problem to deal with (uncertainty
propagation, \surmod{} construction, \ldots). Some of these methods are mainly present to be embedded by more complicated
methods (such as designs developed in the Fourier-conjugate space, discussed later on in \Sect{FastIntro}).

\subsubsection{Modeler module}
The \modeler{} library allows the construction of one or more \surmod{s}. The idea is to provide a simpler, and hence faster,
function to emulate the specified output $Y$ of a complex model (which is generally costly in terms of resources) for a given
set of input factors $X_i$ (for $i=1,\ldots,n_X$, $n_X$ being the number of input factors). In this paper, the following
\surmod{s} will be introduced: chaos polynomial expansion, artificial neural network, gaussian process, also known as
kriging.

\subsubsection{Optimizer and Reoptimizer modules}
The \optimizer{} and \reoptimizer{} libraries are dedicated to optimisation and model calibration. Model calibration
basically consists in setting up the \dof{} of a model such that simulations are as close as possible to an experimental
database. The optimisation is a complex procedure and several techniques are available to perform single-criterion or
multi-criteria analysis, with and without constraint, using local or global approaches.

\subsubsection{Sensitivity module}
The \sensitivity{} library allows to perform sensitivity analysis (SA) of one or several output response $Y$ of a code, with
respect to the chosen input factors $X_i$ (for $i=1,\ldots,n_X$, $n_X$ being the number of input factors). A glimpse of the
very basic concepts of sensitivity analysis is introduced along with the method used throughout this paper: a screening one
(the Morris method) and two different estimation of the \sobol{} coefficients.
 
\subsection[Uranie installation and external dependencies]{\uranie{} installation and external dependencies}
\label{extdep}%

Even though the \uranie{} platform is developed on Linux operating systems, a Windows-version has also been made. In order to
check and guarantee the best portability possible, the platform is tested daily on seven different Linux distributions and
Windows 7. Getting the source of the \uranie{} platform can be done at the Sourceforge web page:
\href{https://sourceforge.net/projects/uranie/}{https://sourceforge.net/projects/uranie/}.

Once the sources have been retrieved, it is highly-advised to follow the instruction listed in the \emph{README} file to
perform the installation. On top of the code itself, this installation brings \uranie{} documentation, among which:
\begin{itemize}
\item a methodological manual (both \proglang{html} and \proglang{pdf} format, \cite{metho}). It gives a shallow introduction
  to the main methods and algorithms, from a mathematical point of view, and provides references for the interested reader,
  to get a deeper insight on these problematics.
\item a user manual (both \proglang{html} and \proglang{pdf} format, \cite{userM}). It gives explanations on the
  implementations of methods along with a large number of examples.
\item a developer manual. This is a description of methods, from the computing point of view, obtained thanks to the
  \pkg{Doxygen} platform~\cite{van2008doxygen}.
\end{itemize}

In the case of the Windows version, an installation can be done from the previously-introduced archive, but a dedicated free
standing archive is specifically-produced by the \uranie{} support team and is provided on request\footnote{mailto:
  \href{mailto:support-uranie@cea.fr}{support-uranie@cea.fr}.}.

In any case, \uranie{} has few dependencies to external packages. They are sorted in two categories: the compulsory and
optional ones.  The latter are shown as light purple boxes in \Fig{modorga} and will only prevent, if not there, some methods
from being used. \uranie, on the other hand, can simply not work without the compulsory ones.  Both types are listed and
briefly discussed below.

\subsubsection{Compulsory dependencies}
\label{compuls_dep}%

\begin{itemize}
\item \ROOT: Discussed thoroughly below, the version used here is \ROOTv.
\item \pkg{Cmake}: Free and \openS{} software for managing the build process of compiled software, the version used
  here is v3.7.1~\cite{martin2007open}.
\item \pkg{CPPUnit}: Unit testing framework for \cpp{} programming, the version used here is
  v1.13.1~\cite{feathers2002cppunit}.
\end{itemize}

The \ROOT{} system is an \openS{} object oriented framework for large scale data analysis. It started as a private project in
1995 at CERN\footnote{European Organisation for Nuclear Research, Geneva, Switzerland.} and grew to be the officially
supported LHC analysis toolkit. \ROOT{} is written in \cpp{}, and contains, among others, an efficient hierarchical
object-oriented database, a \cpp{} interpreter, advanced statistical analysis (multi-dimensional histogramming, fitting,
minimisation, cluster finding algorithms) and visualisation tools. The user interacts with \ROOT{} via a graphical user
interface, the command line or batch scripts. The command and scripting language is \cpp{} (using the interpreter) and large
scripts can be compiled and dynamically linked in. The object-oriented database design has been optimised for parallel access
(reading as well as writing) by multiple processes.

The \ROOT{} system is developed in \cpp{} (but can be called with other languages such as \python{} or \proglang{Ruby}
though) and is well maintained and documented. \uranie{} is built as a layer on \ROOT{} and, as a result, it benefits from
numerous features of \ROOT{}, among which:
\begin{itemize}   
\item the \cpp{} interpreter (CINT); 
\item the \python{} interface: it provides an automatic transcription of \uranie-classes into \python  
\item an access to SQL databases; 
\item many advanced data visualisation features;
\item and much more\ldots
\end{itemize}

\subsubsection{Optional dependencies}
\label{option_dep}%

\noindent
\begin{itemize}
\item \pkg{OPT++}: Libraries that include non linear optimisation algorithms written in \cpp{}, the version used here is
  v2.4~\cite{meza2007objectoriented}. As this package is not maintained anymore, a patched (and recommended) version is
  included in the \uranie{} archive.
\item \pkg{FFTW}: Library that computes the discrete Fourier transform (DFT) in one or more dimensions, of arbitrary
  input size, the version used here is v3.3.4~\cite{FFTW05}.
\item \NLopt: Library for nonlinear optimisation, the version used here is v2.2.4~\cite{nlopt}.  
\item \pkg{PCL}: (Portable Coroutine Library) Implements the low level functionality for coroutines, the version used here
  is v2.2.4.
\item \pkg{MPI}: (Message Passing Interface) Standardised and portable message-passing system needed to run parallel
  computing, the version used here is v1.6.5~\cite{gabriel04:_open_mpi}.
\item \pkg{CUDA}: (Compute Unified Device Architecture) Parallel computing platform and programming model invented by
  NVIDIA to harness the power of the graphics processing unit (GPU), the version used here is
  v8.0~\cite{nvidia2011nvidia}. If requested, it should be used with the \pkg{boost} library, with a version greater than
  v1.47.
\end{itemize}

\subsection{The uncertainty general methodology}
\label{intro_metho}

\begin{figure*}
  \begin{center}
    \includegraphics[width=0.75\textwidth]{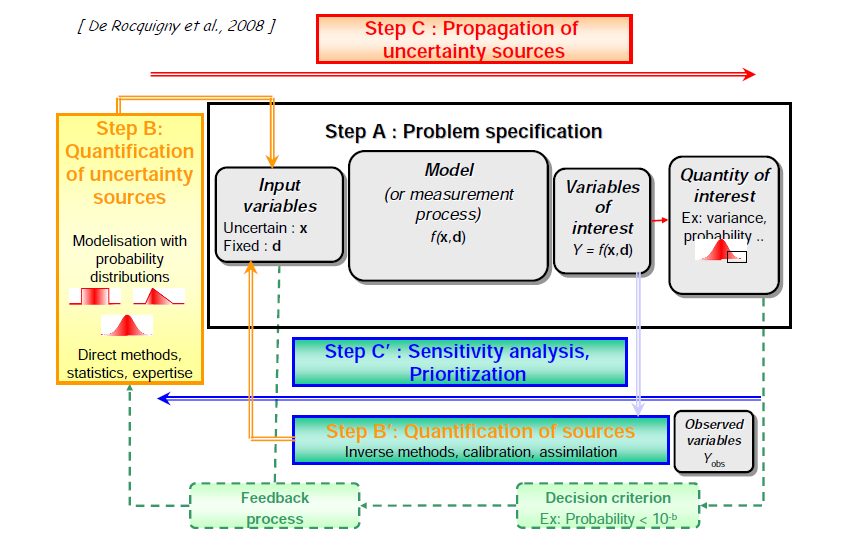}
  \end{center}
  \caption{Sketch that represents in few boxes the different steps that can compose an uncertainty propagation or
    quantification analysis~\cite{de2008uncertainty}.\label{fig_UncertAna}}
\end{figure*}

Many issues related to uncertainty treatment of computer code simulations share the same framework. It can be sketched in a
few key steps, gathered for illustration purpose in \Fig{UncertAna}~\cite{de2008uncertainty} and described below.
\begin{description}
\item[The problem specification (A).] This step is the starting point of a great deal of study as it is when the number of
  input variable is defined, along with the variable of interest and the corresponding quantity of interest (a quantile, a
  mean, a standard deviation\ldots). All these are linked through a model that can be a function, a code or even a \surmod{}
  (which can use instead of the code). One can write the general equation that links the model ($f$), the input variables,
  both uncertain ($\mathbf{x} = (x_{1}, \hdots, x_{n_{X}})$) and fixed ($\mathbf{d} = (d_{1}, \hdots, d_{n_{D}})$) and the
  variable of interest $y$, as
  \begin{equation}
    y = f(\mathbf{x},\mathbf{d})
  \end{equation}
\item[The quantification of uncertainty source (B).] In this step, the statistical laws followed by the different input
  variables are chosen along with their characteristics (mean, standard deviation\ldots). The possible correlations between
  inputs can also be defined here.
\item[The propagation of uncertainty sources (C).] Given the choice made in steps A and B, the uncertainties on the input
  variables are propagated to get an estimation of the resulting uncertainty on the output under study. This can be
  performed, for instance, with analytic computation, using Monte-Carlo approach through a \doe\ldots
\item[The inverse quantification of sources (B').] Given the definition of the problem in step A and a provided set of
  experiments, one can measure the mean value and/or the uncertainty of the input variables, in order, for instance, to spot
  which experiment should be run to constrain the largest one, or to calibrate the model.
\item[The sensitivity analysis (C').] Given the choice made in steps A and B, this analysis can be used to rank the input
  variables with respect to the impact of their uncertainty on the uncertainty of the variable of interest. Some methods even
  provide a quantitative illustration of this impact, for instance as a percentage of the output standard deviation.
\end{description}

This is a very broad description of the kind of analysis usually performed when discussing uncertainty quantification. All
these steps, can indeed be combined, or replaced, once or in an iterative way, to get a more refined analysis.

\subsection{The thermal exchange model}
\label{TherExchMod}
In this part, the physical equations of the \usecase{} used throughout this paper are laid out in a simple way, discussing
first the physical equations. This model will be more precisely detailed and also refined as required by the studies
performed in the following sections.

\subsubsection{Introduction}

\begin{figure}[h!]
  \begin{center}
    \includegraphics[width=0.95\linewidth]{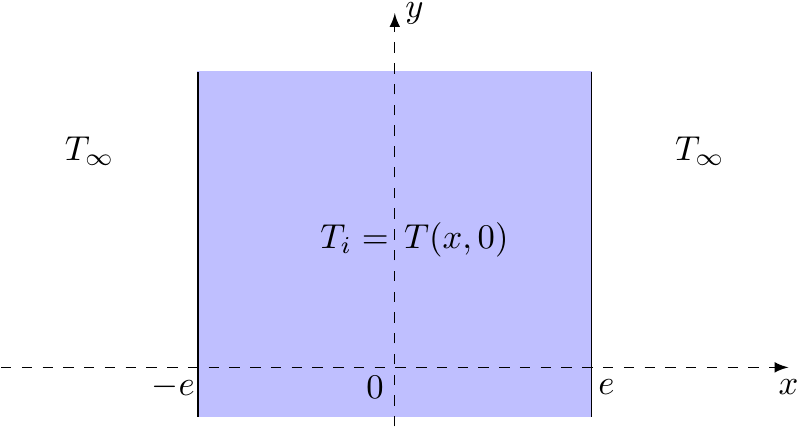}    
  \end{center}
  \caption{Simplified sketch of the thermal exchange problem \label{fig_ExpeSketch}}  
\end{figure}

The experimental setup is depicted in \Fig{ExpeSketch} and is composed of a planar sheet whose width is 2$e$ (along the
$x$-direction) while its length is considered infinite (represented without boundaries along the $y$-direction). At $t=0$
this sheet, whose initial temperature is set to $T_i$, is exposed to a warmer fluid (whose temperature is written as
$T_\infty$). The aim of this problem is to represent the temperature profiles, depending on the time and the position within
the sheet, using different materials for the sheet, and to investigate the impact of various uncertainty sources these
temperature profiles.

Studying the evolution of the temperature within the sheet in fact consists in solving the heat equation which can be
written as follows, if we consider the mono-dimensional problem as depicted in \Fig{ExpeSketch}:
\begin{equation}
  \label{eq_heat}
  \frac{\partial T}{\partial t} = \alpha \frac{\partial^2 T}{\partial x^2}
\end{equation}
In this equation $\alpha$ [m$^2$.s$^{-1}$] is the thermal diffusivity which is defined by 
\begin{equation}
  \label{eq_TherDiff}
  \alpha = \frac{\lambda}{\rho C_{\rho}}
\end{equation}
where $\lambda$ is the thermal conductivity [W.m$^{-1}$.K$^{-1}$], $C_{\rho}$ is the massive thermal capacity
[J.kg$^{-1}$.K$^{-1}$] and $\rho$ is the volumic mass [kg.m$^{-3}$]. There are three conditions used to resolve the heat
equation, the first one being the initial temperature
\begin{equation}
  \label{eq_initTemp}
  T(x,t=0) = T_i
\end{equation}
the second one relies on the flow being null at the centre of the sheet
\begin{equation}
  \label{eq_centerFlow}
  \frac{\partial T}{\partial x}\bigg|_{x=0} = 0 
\end{equation}
while the last one relies on the thermal flow equilibrium at the surface of the sheet
\begin{equation}
  \label{eq_surfaceFlow}
  -\lambda \frac{\partial T}{\partial x}\bigg|_{x=e} = h(T(x=e,t)-T_\infty)
\end{equation}

Usually, the thermal coupling between a fluid and a solid structure is characterised by the thermal exchange coefficient h
[W.m$^{-2}$.K$^{-1}$]. This coefficient allows to free oneself from a complete description of the fluid, when one is only
interested in the thermal evolution of the structure (and \textit{vice-versa}). Its value depends on the dimension of the
complete system, on the physical properties of both the fluid and the structure, on the liquid flow, on the temperature
difference\ldots The thermal exchange coefficient is characterised by the Nusselt number ($N_u$), from the fluid point of
view, and by the Biot number ($B_i$), from the structure point of view. In the rest of this paper, the latter will be
discussed and used thanks to the relation
\begin{equation}
  \label{eq_BiDef}
  B_i = \frac{he}{\lambda}
\end{equation}

\subsubsection{Analytic model}
\label{AnalyticModel}
In the specific case where the thermal exchange coefficient, $h$ and the fluid temperature $T_\infty$ can be considered
constant, \Eqn{heat} has an analytic solution for all initial conditions (all the more so for the one stated
in \Eqn{initTemp}), when it respects the flow conditions defined in \Eqns{centerFlow}{surfaceFlow}. The resulting analytic
form is usually express in terms of thermal gauge $\theta$, which is defined as
\begin{equation}
  \label{eq_thermalGauge}
  \theta(x,t)=\frac{T(x,t)-T_i}{T_\infty-T_i}  
\end{equation}
The complete form is the following infinite serie
\begin{equation}
  \label{eq_serieGauge}
  \theta(x_{ds},t_{ds})=2\sum^{\infty}_{n=1} \beta_n \cos(\omega_n x_{ds})\exp(-\frac{1}{4}\omega^2_n t_{ds})
\end{equation}
where the original parameters have been changed to dimensionless ones
\begin{eqnarray}
  x_{ds} & = & x / e \\
  t_{ds} & = & \frac{ t }{t_D} = t\times \frac{4\alpha}{ e^2 } = t\times \frac{4 \lambda}{e^2 \rho C_\rho}
\end{eqnarray}
Given this, the elements in the serie (\Eqn{serieGauge}) can be written
\begin{equation}
  \label{eq_BetanDef}
  \beta_n=\frac{\gamma_n\sin(\omega_n)}{\omega_n(\gamma_n+B_i)}
\end{equation}
where 
\begin{equation}
  \label{eq_GammanDef}
  \gamma_n=\omega_n^2+B_i^2
\end{equation}
and $\omega_n$ are solutions of the following equation
\begin{equation}
  \label{eq_OmeganDef}
  \omega_n\tan(\omega_n)=B_i
\end{equation}

This model has been implemented in \uranie{} and tested with two kinds of material to get an idea of the temperature profile
in the structure.

\subsubsection{Looking at PTFE and iron}

In this part, two very different kinds of plate-sheets are compared: a composite one, made out of PTFE (whose best known
brand name is Teflon) and an iron one. The main properties (of interest for our problem) of the sheets are gathered in
\Tab{metalCara} side-by-side for both PTFE and iron. The last column shows the relative uncertainty found in the literature
(or chosen in the case of the thickness) for the iron case. They will be applied as well on the PTFE.  The last three lines
are the properties that are computed from the first four ones and once the thermal exchange coefficient has been set to a
constant value (here 100), as stated in \Sect{AnalyticModel}.

\begin{table*}
  \begin{center}
    \begin{tabular}{L{7.2cm}C{1.7cm}C{1.7cm}C{0.01cm}C{2.8cm} N}
      \cline{2-5}
      \multicolumn{1}{l}{} & PTFE & Iron && Uncertainty (\%) &\\[4pt]
      \hline
      Thickness [m]: \color{blue}{e} & 10$\times$10$^{-3}$ & 20$\times$10$^{-3}$ & &0.5 &\\[4pt]
      Thermal conductivity [W.m$^{-1}$.K$^{-1}$]: \color{blue}{$\lambda$} & 0.25 & 79.5 && 0.6 &\\[4pt]
      Massive thermal capacity [J.kg$^{-1}$.K$^{-1}$]: \color{blue}{$C_{\rho}$} & 1300 & 444 & &1.2&\\[4pt]
      Volumic mass [kg.m$^{-3}$]: \color{blue}{$\rho$} & 2200 & 7874 &&0.2&\\[4pt]
      \hline
      \multicolumn{5}{l}{}\\[-10.2pt]
      Thermal diffusivity [m$^2$.s$^{-1}$]: \color{blue}{$\alpha$} & 8.7$\times$10$^{-8}$ & 2.27$\times$10$^{-5}$ & \multicolumn{2}{c}{} &\\[4pt]
      Diffusion thermal time [s]: \color{blue}{$t_{D}$} & 287 & 4.4 & \multicolumn{2}{c}{}&\\[4pt]
      Biot number (for $h=100$), [$\varnothing$]: \color{blue}{$B_i$} & 4 & 0.025 & \multicolumn{2}{c}{}&\\[4pt]
      \cline{1-3}
    \end{tabular}      
  \end{center}  
  \caption{Summary of both PTFE and iron characteristics. The last column shows the relative uncertainty found in the
    literature (or chosen in the case of the width) for the iron case. They will be applied as well on the
    PTFE. \label{tab_metalCara}}
\end{table*}

\begin{figure*}
  \begin{center}
    \subfloat[$\theta(x_{ds},t_{ds})$ v.s. $x_{ds}$ for PTFE]{
      \includegraphics[page=3,width=.5\textwidth]{{PTFECasePlotTest_xr_0_1_s_0.1_tr_0_10_s_1}.pdf}
      \label{fig_subPTFEGaugexds}
    }
    \subfloat[$\theta(x_{ds},t_{ds})$ v.s. $x_{ds}$ for iron]{
      \includegraphics[page=3,width=.5\textwidth]{{IronCasePlotTest_xr_0_1_s_0.1_tr_0_10_s_1}.pdf}
      \label{fig_subironGaugexds}
    }

    \subfloat[$\theta(x_{ds},t_{ds})$ v.s. $t_{ds}$ for PTFE]{
      \includegraphics[page=2,width=.5\textwidth]{{PTFECasePlotTest_xr_0_1_s_0.1_tr_0_10_s_1}.pdf}
      \label{fig_subPTFEGaugetds}
    }
    \subfloat[$\theta(x_{ds},t_{ds})$ v.s. $t_{ds}$ for iron]{
      \includegraphics[page=2,width=.5\textwidth]{{IronCasePlotTest_xr_0_1_s_0.1_tr_0_10_s_1}.pdf}
      \label{fig_subironGaugetds}
    }
    
    \subfloat[$\theta(x_{ds},t_{ds})$ for PTFE]{
      \includegraphics[page=1,width=.5\textwidth]{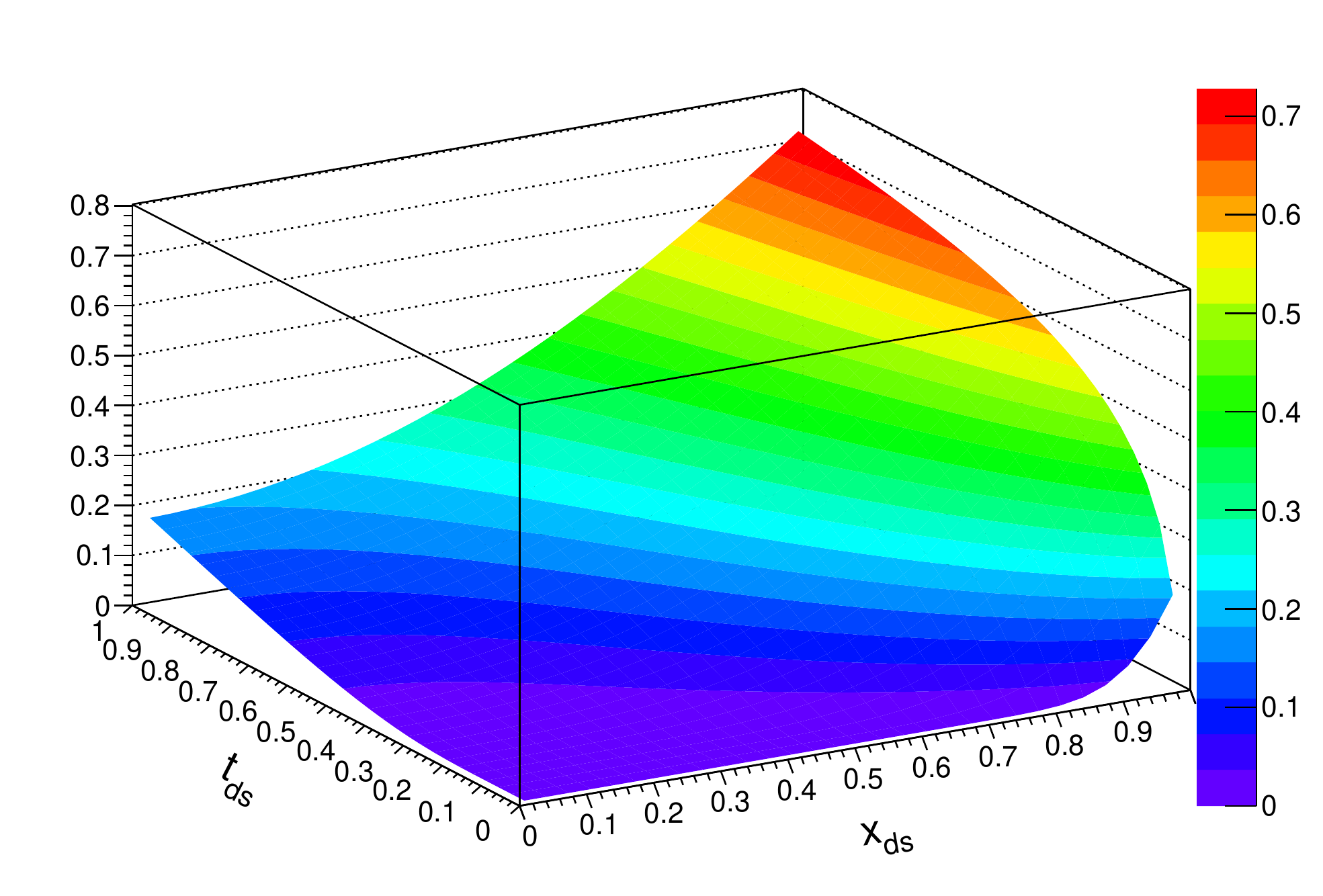}
      \label{fig_subPTFEGauge}
    }
    \subfloat[$\theta(x_{ds},t_{ds})$ for iron]{
      \includegraphics[page=1,width=.5\textwidth]{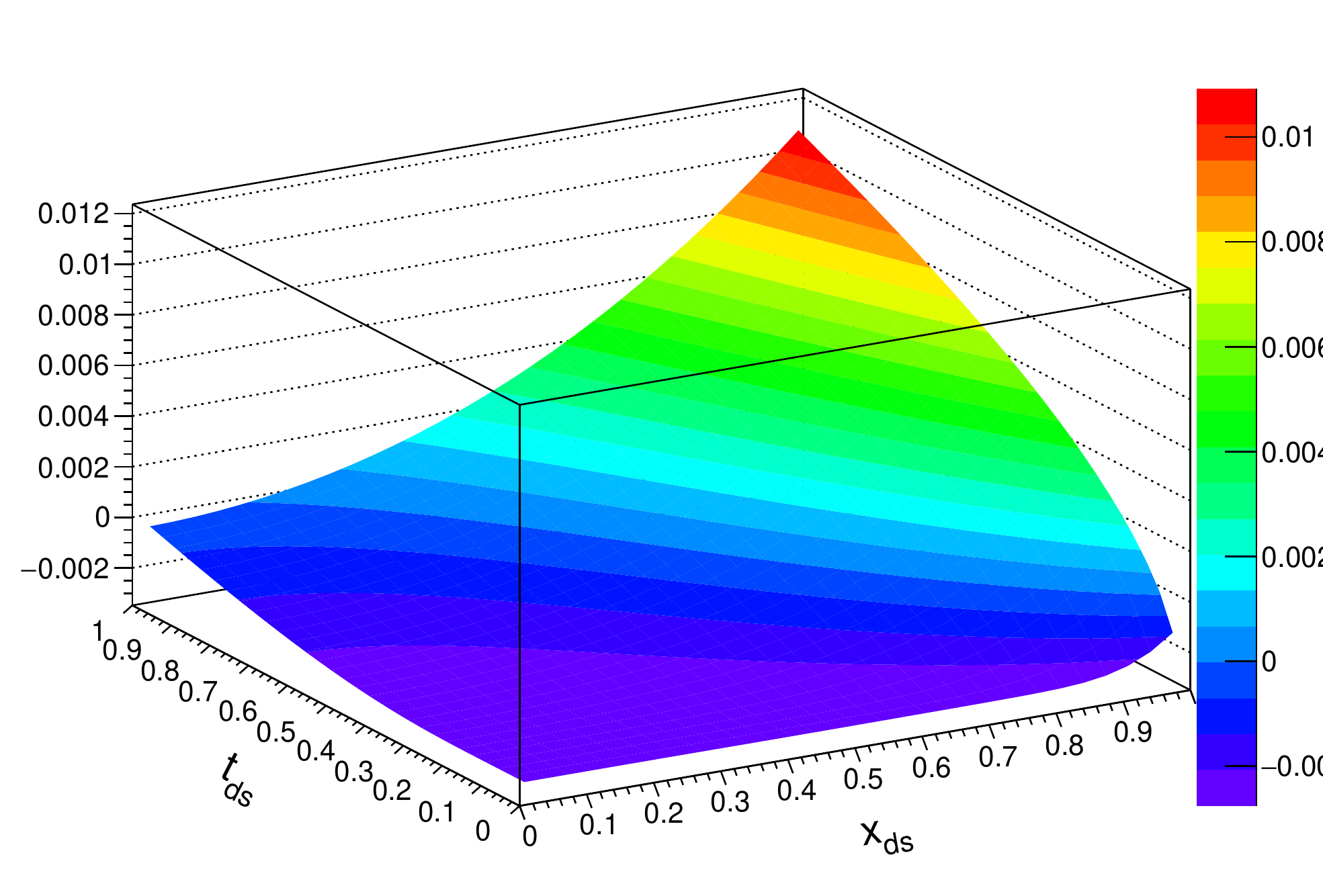}
      \label{fig_subironGauge}
    }

  \end{center}
  \caption{Evolution of the thermal gauge as a function of either the position for different time steps (a,b), the time for
    different positions (c,d) or depending on both parameters (e,f), for PTFE (left) and iron (right).\label{fig_allGauge}}
\end{figure*}

Given these properties, several plots have been produced to characterise the evolution of the temperature profiles in the
sheet matter and are gathered in \Fig{allGauge}. Looking at these plots, a major difference can be drawn between the two
sheets: in the PTFE case, the gauge is very different between two positions at a same time and this difference varies also
through time (see \Figs{subPTFEGaugexds}{subPTFEGaugetds}). For the iron, on the other hand, the differences through time and
space are very small. This is even more important when considering that the range over which the gauge is displayed is
significantly reduced. The iron thermal gauge is actually far from reaching the value 1, even after 10 diffusion thermal
time, whereas this is the case for PTFE.\par

These differences could have been foreseen, looking at the properties gathered in \Tab{metalCara}: the previously discussed
observations are the expected ones once one considers the value of the Biot number. For a material with a Biot number greater
than 1, as the PTFE sheet, the thermal conduction is small within the sheet, leading to temperature gradient within the
structure. On the other hand, when the material has a Biot number significantly smaller than 1, as the iron plate, the
temperature is expected to be quite similar at the surface and in the centre of the sheet.\par

\subsection{Paper layout}

This paper will describe several typical analysis that can be run using the \uranie{} platform. It is not meant to be fully
exhaustive concerning the methodology behind the introduced techniques, but also concerning the methods and options
implemented in \uranie. In many cases, though, a sub-part called \emph{to go further} will introduce briefly the important,
yet not discussed, options and solutions that can be offered to the user. Given what has been seen in \Fig{allGauge}, the
\usecase{} used throughout the rest of this paper will be the PTFE case.

The first introduced concept will be the generation of \surmod{} (see \Sect{model}). Three different techniques will be
applied on an pre-produced \doe, that will be called the training database, describing the lowest level of complexity of the
\usecase{} (only considering the dimensionless variables $x_{ds}$ and $t_{ds}$). A more global picture will be used to take
into account the uncertainties introduced in \Tab{metalCara} and see how to transpose them into uncertainty on the thermal
gauge in \Sect{uncert}. The impact of every uncertainty source will be ranked but also numerically estimated thanks to
various sensitivity analysis in \Sect{sensi}. Finally a calibration of some of the model parameters is performed using
different techniques in \Sect{optim}, also questioning the fact that the thermal exchange coefficient $h$ can be considered
constant. Finally, some important left-over concepts are discussed along with the actual perspectives in \Sect{final}.

%% file: Modelling.tex

In this part, different \surmod{s} will be introduced to reproduce the behaviour of a given code or function. The aim of this
step is to obtain a simplified model able to mimic, within a reasonable acceptance margin, the output of both a training and
a test database, along with an important improvement in terms of time and memory consumption.

The full analytic model, detailed in \Eqn{serieGauge}, plays the role of the complex model that should be approximated. To do
so, a training database will be produced, composed of n$_S$ locations (n$_S$ being set to 40 here), written as
\begin{equation}
  \label{eq_trainDB}
  \mathcal{L} =  \lbrace (\mathbf{x}^{j}, y^{j}), j=1\ldots n_{S} \rbrace
\end{equation}
where $\mathbf{x}^{j} = (x_{ds}^{j}, t_{ds}^{j})$ and $y^{j} = \theta(x_{ds}^{j},t_{ds}^{j}, B_{i}=4)$. The Biot number is
set to 4 as only the PTFE case will be considered (see \Tab{metalCara}).
    
Three different techniques will be applied: the polynomial chaos expansion, the artificial neural network and the kriging
approximation. Each and every method will have a brief introduction before being applied to our \usecase. The interested
readers are invited to go through the references for a more meticulous description. An example of making practical use of the
the kriging surrogate is provided in \Sect{optim}. The starting point will always be the loading of the training database in
an \code{TDataServer} object which is the spine of \uranie.

\begin{lstcolored}[firstnumber=1,]
//Create the TDataServer object
TDataServer *tds = new TDataServer("tdsObs", "The training DB");
//Load the data in the dataserver
tds->fileDataRead("training_database.dat");
\end{lstcolored}

\subsection{Polynomial chaos expansion}

\subsubsection{Introduction}
The concept of polynomial chaos development relies on the homogeneous chaos theory introduced by Wiener~\cite{Wiener38} and
further developed by Cameron and Martin~\cite{Cameron47}. Using polynomial chaos (later referred to as PC) in numerical
simulation has been brought back to the light by Ghanem and Spanos~\cite{Ghanem91}. The basic idea is that any
square-integrable function can be written as
\begin{equation}
  \label{eq_PCdecomp}
  f(\mathbf{x})=\sum_{\alpha}f_{\alpha}\Psi_{\alpha}(\mathbf{x})
\end{equation}

where $\lbrace f_{\alpha}\rbrace$ are the PC coefficients, $\lbrace \Psi_{\alpha}\rbrace$ is the orthogonal
polynomial-basis. The index over which the sum is done, $\alpha$, corresponds to a multi-index whose dimension is equal to
the dimension of the input vector $\mathbf{x}$ (\ie{} here $n_X$) and whose L1 norm ($|\alpha|_{1} = \sum_{i=1}^{n_X}
\alpha_i$) is the degree of the resulting polynomial.

From this development, it becomes clear that a threshold must be chosen on the order of the polynomials used, as the number
of coefficient will growing quickly, following this rule $N_{\rm coeff} = \frac{(n_{X}+p)!}{n_{X}!\,p!}$, where $p$ is the
cut-off chosen on the polynomial degree.


\subsubsection[Implementation in Uranie and application to the \usecase]{Implementation in \uranie{}  and application to the \usecase}
\label{PCimple}
In \uranie, the implementation of the polynomial chaos expansion method is done through the \NISP{} library
(\cite{baudininria-00494680}), \NISP{} standing for \emph{Non-Intrusive Spectral Projection}.  Originally written to deal
with normal laws, for which the natural orthogonal basis is Hermite polynomials, this decomposition can be applied to few
other distributions, using other polynomial orthogonal basis, such as Legendre (for uniform and log-uniform laws), Laguerre
(for exponential law), Jacobi (for beta law)\ldots

The PC coefficients are estimated through a regression method, simply based on a least-squares approximation: given the
training database $\mathcal{L}$, the vector of output $\mathbf{y}(n_{S})$ is computed with the code. The regression are
estimated, given that one calls the correspondence matrix $H(n_{S},p)$ and the coefficient-vector $\boldsymbol{\beta}$, by a
minimisation of $|| \mathbf{y} - H \boldsymbol{\beta} ||^{2}$, where
\begin{equation}
  \arraycolsep=1.2pt \mathbf{y} = \left (\begin{array}{cccc} y^{1} & y^{2} & \hdots & y^{n_{S}} \end{array} \right ),
\end{equation}
\begin{equation}
  \arraycolsep=1.2pt H=
  \left ( \begin{array}{ccc} \Psi_{1}(\mathbf{x}^{1}) & \hdots & \Psi_{p}(\mathbf{x}^{1}) \\ \Psi_{1}(\mathbf{x}^{2}) &
    \hdots & \Psi_{p}(\mathbf{x}^{2}) \vrule height 0.4cm width 0pt\\ \vdots & \ddots & \vdots \\ \Psi_{1}(\mathbf{x}^{n_S})
    & \hdots & \Psi_{p}(\mathbf{x}^{n_S})\\ \end{array} \right),
\end{equation}
\begin{equation}
  \boldsymbol{\beta} = \left
  ( \begin{array}{c} \beta_{1} \\ \beta_{2} \vrule height 0.4cm width 0pt\\ \vdots \\ \beta_{p} \end{array} \right )
\end{equation}
This leads to write the general form of the solution as $\boldsymbol{\beta} = (H^{T}H)^{-1}H^{T}\mathbf{y}$ which means that
the estimation of the points using the \surmod{} are given through $\mathbf{\hat y} = H \boldsymbol{\beta} =
H(H^{T}H)^{-1}H^{T}\mathbf{y} = P\mathbf{y}$, where $P = H(H^{T}H)^{-1}H^{T}$. Here, the P matrix links directly the output
variable and its estimation through the \surmod: this formula is useful as it can be used to compute the \Loo{} uncertainty.

\begin{figure}[h!]
  \begin{center}
    \includegraphics[page=1,width=.8\linewidth,keepaspectratio=true]{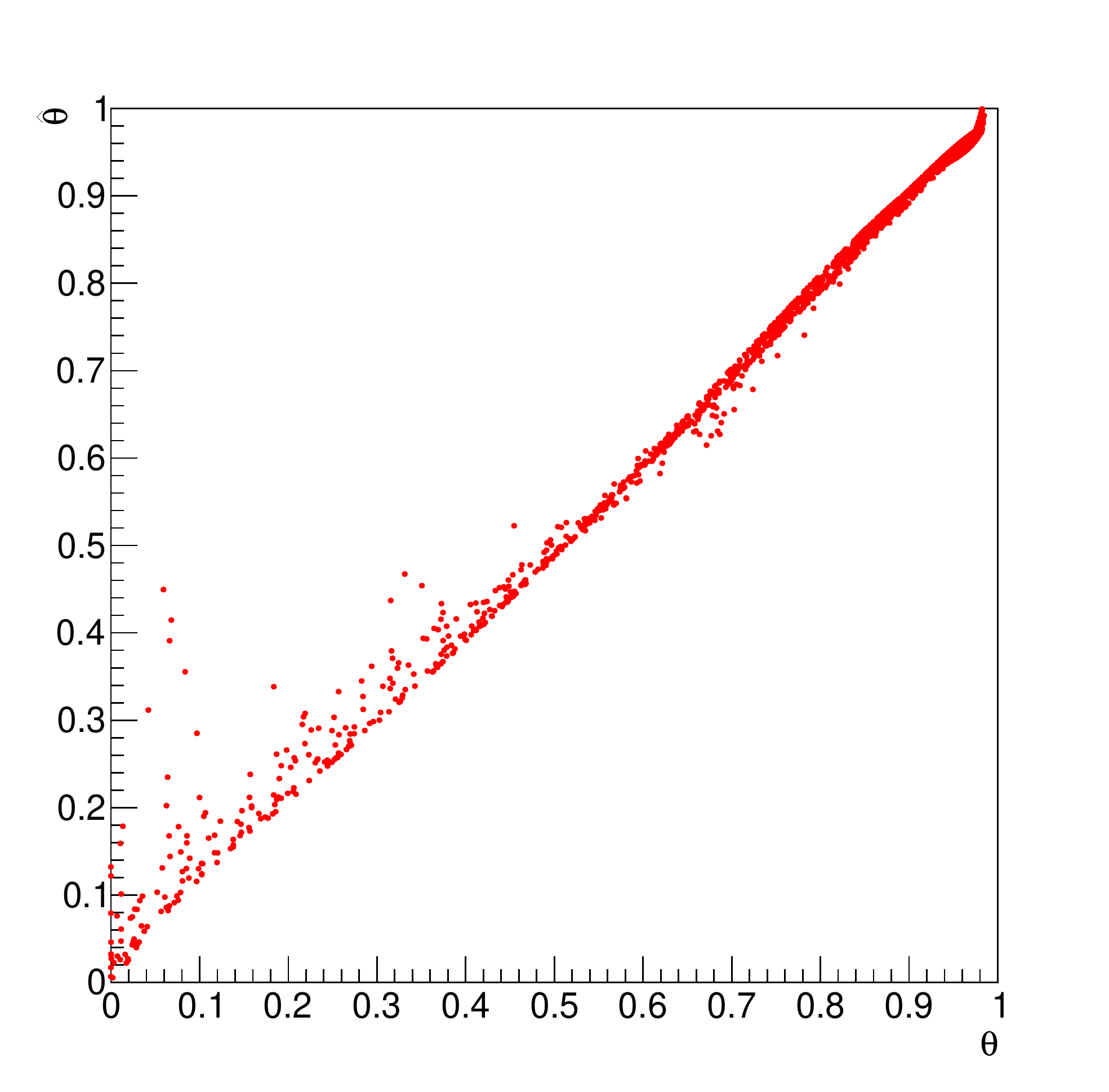}
  \end{center}
  \caption{Distribution of the thermal gauge values estimated by the \surmod{} ($\hat\theta$) as a function of the ones
    computed by the complete model ($\theta$) in a test database, not used for the training. \label{fig_PCresults}}
\end{figure}

\Fig{PCresults} represents the distribution of the thermal gauge values (as defined in \Eqn{thermalGauge}) estimated by the
\surmod{} ($\hat\theta$) as a function of the ones computed by the complete model ($\theta$) in a validation database containing
2000 locations, not used for the training. A nice agreement is found on the overall range.

In practice, the main steps used to get the PC expansion are gathered in the following block:
\begin{lstcolored}[firstnumber=1,]
//Create the TNisp object
TNisp *nisp = new TNisp(tds);
//Give sample size (tsize) and method (LHS)
nisp->setSample("Lhs",tsize);
//Create the PC object, given dataserver and nisp object
TPolynomialChaos * pc = new TPolynomialChaos(tds,nisp);   
//Set the degree of the PC
pc->setDegree(degree); 
//Estimate the coefficient 
pc->computeChaosExpansion("Regression");
\end{lstcolored}

\subsubsection{To go further }
\label{PCfurther}
There are several points not discussed in this section but which can be of interest for users:
\begin{itemize}
\item Based on regression method explained in \Sect{PCimple}, \uranie{} also provides a method to estimate the best degree
  possible, relying on the \Loo{} estimation, limiting the range of tested degree, given the learning database size ($n_S$).
\item PC coefficients can be estimated using an integration methods (instead of the regression) which relies on specific
  \doe{} (usually sparse-grids) that are oftenly smaller, in terms of computations, than the regularly-tensorised approaches
 ~\cite{baudininria-00494680}.
\item When the PC development is done on the natural polynomial basis of the stochastic laws (listed in \Sect{PCimple}), the
  PC coefficients can be combined and transformed into \sobolCs{} (discussed in \Sect{sensi}) providing both a \surmod{} and
  a sensitivity analysis.
\end{itemize}

\subsection{Artificial neural networks}
The Artificial Neural Networks (ANN) in Uranie are Multi Layer Perceptron (MLP) with only one hidden layer (containing
$n_{H}$ neurons) limited to one output variable ($n_Y=1$).

\subsubsection{Introduction}

\begin{figure}[h!]
  \begin{center}
    \includegraphics[width=0.65\linewidth,keepaspectratio=true]{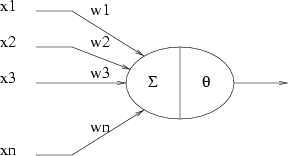}
  \end{center}
  \caption{Schematic description of a formal neuron~\cite{McCulloch1943}.}
  \label{fig_neuronsketch}
\end{figure}

The concept of formal neuron has been proposed after observing the way biological neurons are intrinsically connected
(\cite{McCulloch1943}). This model is a simplification of the various range of functions dealt by a biological neuron, the
formal one (displayed in \Fig{neuronsketch}) being requested to satisfy only the two following purposes:

\begin{itemize}
\item summing the weighted input values, leading to an output value, called neuron's activity, $a = \sum^{n_X}_{i=i} \omega_i
  x_i$, where $\omega_{1}, \ldots, \omega_{n_{X}}$ are the synaptic weights of the neuron.
\item emitting a signal (whether the output level goes beyond a chosen threshold or not) $s = f( a + \theta)$ where $f$ and
  $\theta$ are respectively the transfer function and the bias of the neuron.
\end{itemize}

\begin{figure}[h!]
  \begin{center}
    \subfloat[tanh]{
      \includegraphics[width=0.75\linewidth,keepaspectratio=true]{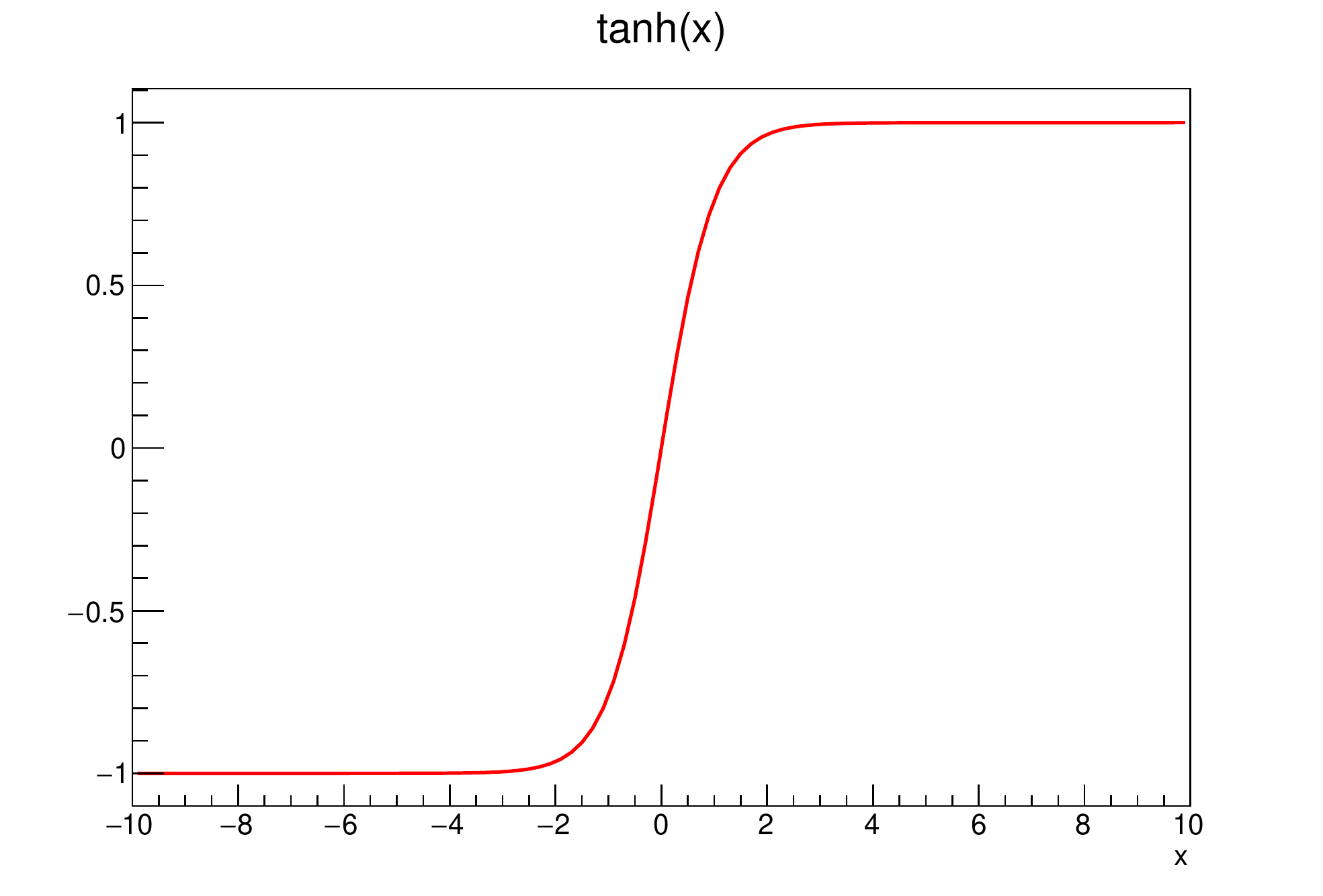}
      \label{fig_functTransfert_tanh}
    }
    
    \subfloat[logistic]{
      \includegraphics[width=0.75\linewidth,keepaspectratio=true]{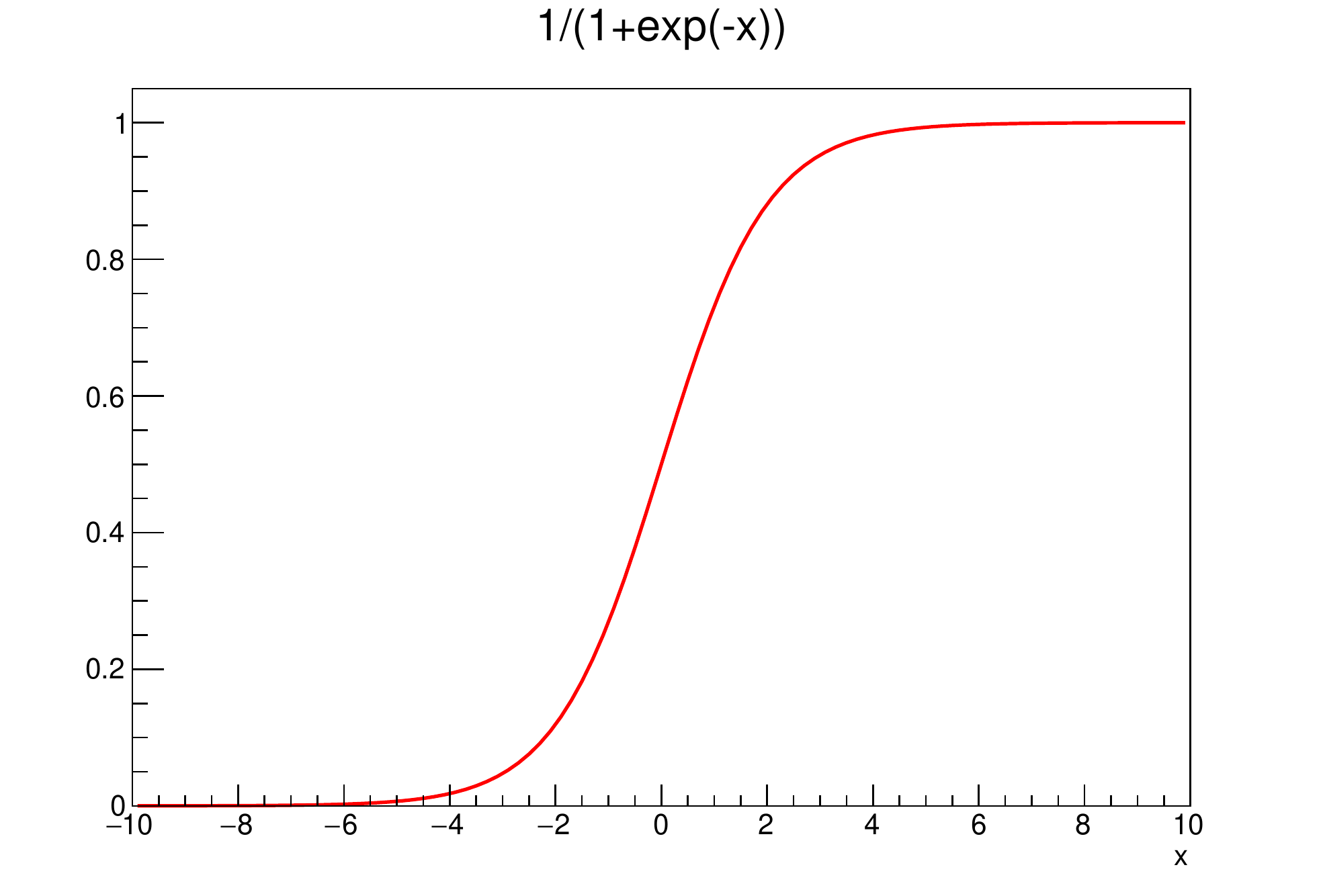}
      \label{fig_functTransfert_sig}
    }
  \end{center}
  \caption{Example of transfer functions: the hyperbolic tangent (left) and the logistical one (right).\label{fig_functTransfert}}%
\end{figure}

One can introduce a shadow input defined as $x_0 = 1$ (or -1), which is a way to consider the bias as another synaptic weight
$\omega_0 = \theta$. The resulting emitted signal is written as \[s = f( \sum^{n_X}_{i=0} \omega_ix_i)\] There are a large
variety of transfer functions possible, and an usual starting point is the sigmoid family, defined with three real
parameters, c, r and k, as $f_{c,k,r} (x) = c\frac{e^{kx}-1}{e^{kx}+1} + r$. Setting these parameters to peculiar values
leads to known functions such as the hyperbolic tangent and the logistic function, shown in \Fig{functTransfert} and defined
as \[f_{1,2,0} (x) = \frac{e^{2x}-1}{e^{2x}+1} = \frac{e^x-e^{-x}}{e^x+e^{-x}} = \tanh(x) \] and \[ f_{1/2,1,1/2}
(x) = \frac{1}{2}\frac{e^{x}-1}{e^{x}+1} + \frac{1}{2} = \frac{1}{1+e^{-x}}\]

The first artificial neural network conception has been proposed and called the perceptron~\cite{rosenblatt196}. The
architecture of a neural network is the description of the organisation of the formal neurons and the way they are connected
together. There are two main topologies:

\begin{itemize}
\item complete: all the neurons are connected to the others.
\item by layer: neurons on a layer are connected to all those on the previous and following layer.
\end{itemize}

\subsubsection[Implementation in Uranie and application to the \usecase]{Implementation in \uranie{}  and application to the \usecase}

\begin{figure*}
  \begin{center}
    \includegraphics[width=0.65\linewidth,keepaspectratio=true]{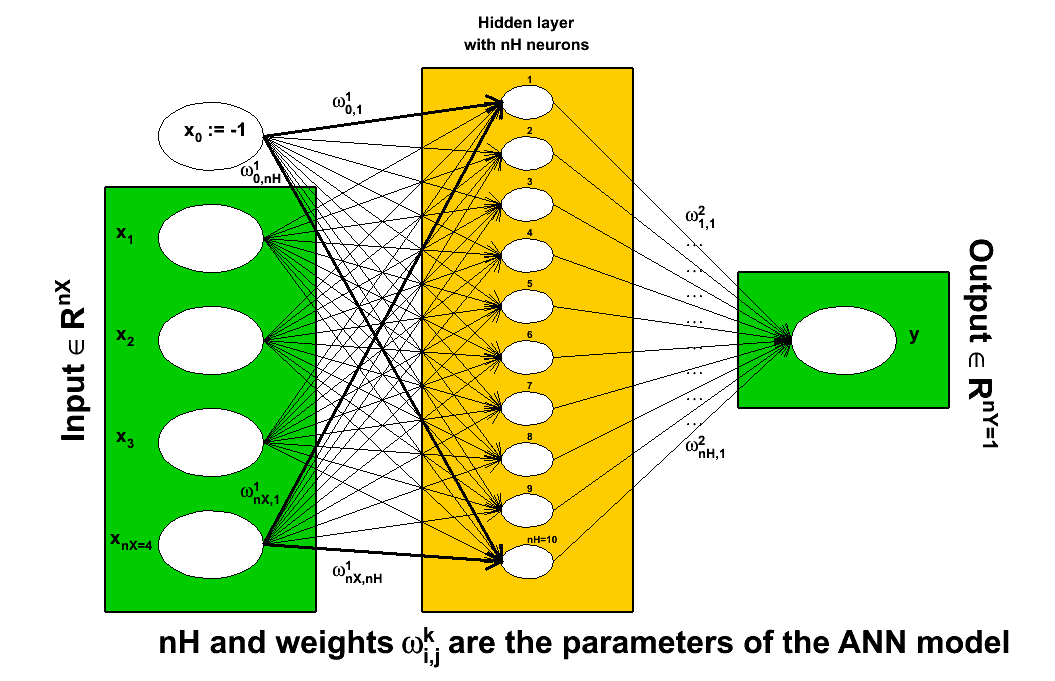}
  \end{center}
  \caption{Schematic description of the working flow of an ANN as used in \uranie. \label{fig_ANNdef}}%
\end{figure*}

The general organisation of \uranie's ANN is detailed in three steps in the following part and displayed in \Fig{ANNdef}.
The first layer, where the vector of entries is stored, is called the input layer. The last one, on the other hand, is called
the output layer while in between lies the single hidden layer, composed of $n_H$ hidden neurons.

The first step is the definition of the problem: what are the input variables under study, how many neurons will be created
in the hidden layer, what is the chosen activation function.
 
The second step is the training of the ANN. Using the full database $\mathcal{L}$, two mechanisms are run simultaneously:

\begin{itemize}
\item the learning itself.  By varying all the synaptic weights contained in the parameter $\Xi$, the aim is to produce the
  output set $\hat{y} = f_{\Xi} (\mathbf{x})$, that would be as close as possible to the output stored in $\mathcal{L}$ then
  keep the best configuration (denoted as $\Xi^*$). The difference between the real output and the estimated one is
  measured through a loss function which could be, in the case of regression, a quadratic loss function such
  as \[L(y,\hat{y}) = \frac{1}{2}|| y - \hat{y} ||^2 \] From there, one can define the risk function $R(\Xi)$ used to
  transform the optimal parameters search into a minimisation problem.  The empirical risk function can indeed be written
  as \[R(\Xi) = \frac{1}{n_S} \sum^{n_S}_{i=1} L(y_i,f_{\Xi}(\mathbf{x_i}))\]
\item the regularisation. Since the ANN is trained only on the $\mathcal{L}$ ensemble, the \surmod{} could be trained too
  specifically for this sub-part of the input space which might not be representative of the overall input space. To avoid
  this, the learning database is split into two sub-parts: one for the training (see previous bullet), and one to prevent the
  over-fitting to happen. For every newly tested parameter set $\Xi$, the generalised error (computed as the average error
  over the set of points not used in the training procedure) is determined. While it is expected that the risk function is
  becoming smaller when the number of optimisation steps is getting higher, the generalised error is also becoming smaller at
  first, but then it should stabilise and even get worse. This flattening or worsening is used to stop the optimisation.
\end{itemize}

This procedure is stochastic: the splitting of the $\mathcal{L}$ ensemble is done using a random generator, so does the
initialisation of the synaptic weights for all the formal neurons. It is important then to export the constructed neural
network as running twice the same methods will not give the same performances.

\begin{figure}[h!]
  \begin{center}
    \includegraphics[page=1,width=.8\linewidth,keepaspectratio=true]{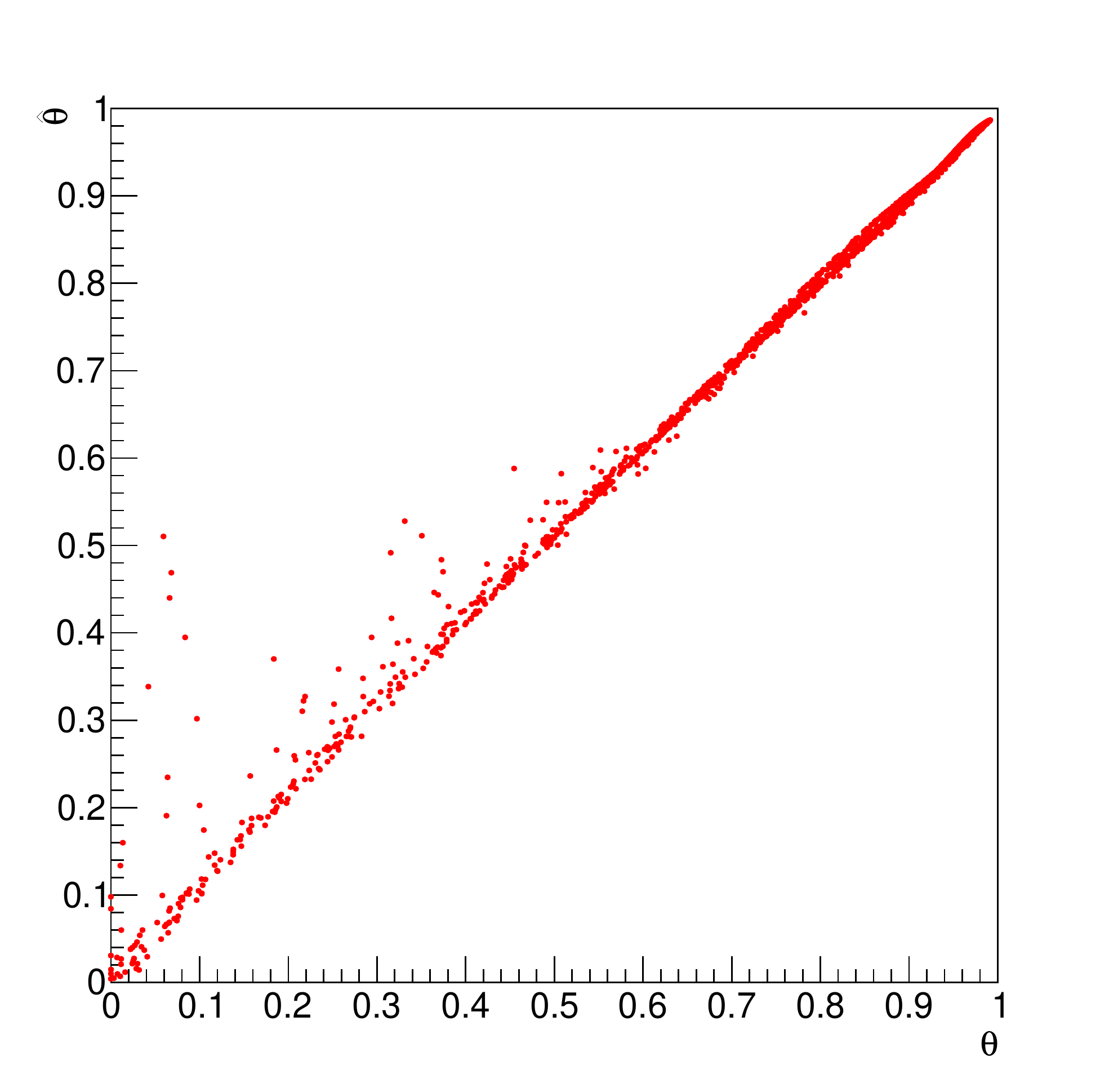}
  \end{center}
  \caption{Distribution of the thermal gauge values estimated by the \surmod{} ($\hat\theta$) as a function of the ones
    computed by the complete model ($\theta$) in a test database, not used for the training. \label{fig_ANNresults}}
\end{figure}

\Fig{ANNresults} represents the distribution of the thermal gauge values (as defined in \Eqn{thermalGauge}) estimated by
the \surmod{} ($\hat\theta$) as a function of the ones computed by the complete model ($\theta$) in a test database
containing 2000 points, not used for the training. A nice agreement is found on the overall range.

In practice, the main steps used to get the neural network trained are gathered in the following block:
\begin{lstcolored}[firstnumber=1,]
//Create ANN object, providing dataserver and its architecture (the inputs, the number of neurons and the output)
TANNModeler* tann = new TANNModeler(tds, Form("xad:tad,\%d,theta", nH));
//Define the activation function
tann->setNormalization(TANNModeler::kMinusOneOne);
//Train the ANN (5 learning/regularisation splitting and 10 weight initialisation)
tann->train(5, 10,);
\end{lstcolored}

\subsubsection{To go further }

There are several points not discussed in this section but which can be of interest for users:
\begin{itemize}
\item The learning step can be run in parallel on graphics processing units (GPU) which can boost it considerably.
\item Even though this \surmod{} is not yet implemented for several outputs, one can create an ANN which embeds the results of
  other neural networks.
\end{itemize}
Several investigations are ongoing to improve this technique, among which:
\begin{itemize}
\item Implement a multi-output approach
\item Implement a multi-hidden layers approach
\item Use Hamiltonian Markov Chain for the synaptic weight~\cite{neal2011mcmc}
\end{itemize}

\subsection{Kriging}
\label{KrigingIntro}
First developed for geostatistic needs, the kriging method, named after D. Krige and also called Gaussian Process method
(denoted GP hereafter) is another way to construct a \surmod. It recently became popular thanks to a series of interesting
features:

\begin{itemize}
\item it provides a prediction along with its uncertainty, which can then be used to plan simulations and therefore improve
  predictions of the \surmod
\item it relies on relatively simple mathematical principle
\item some of its hyper-parameters can be estimated in a Bayesian fashion to take into account \emph{a priori} knowledge.
\end{itemize}
  
Kriging is a family of interpolation methods developed for the mining industry~\cite{Matheron70}. It uses information about
the "spatial" correlation between observations to make predictions along with a confidence interval at new locations.  In
order to produce the prediction model, the main task is to produce a spatial correlation model. This is done by choosing a
correlation function and search for its optimal set of parameters, based on a specific criterion.

\subsubsection{Introduction}
\label{kriging_intro}

The modelisation relies on the assumption that the deterministic output $y(\mathbf{x})$ can be written as a realisation of a
gaussian process $Y(\mathbf{x})$ that can be decomposed as $Y(\mathbf{x})=m(\mathbf{x}) + Z(\mathbf{x})$ where
$m(\mathbf{x})$ is the deterministic part, called hereafter deterministic trend, that describes the expectation of the
process and $Z(\mathbf{x})$ is the stochastic part that allows the interpolation. This method can also take into account the
uncertainty coming from the measurements. In this case, the previously-written $Y(\mathbf{x})$ is referred to as $Y_{\rm
  Real}(\mathbf{x})$ and the gaussian process is then decomposed into $Y_{\rm Obs}(\mathbf{x})=m(\mathbf{x}) + Z(\mathbf{x})
+ \epsilon(\mathbf{x})$, where $\epsilon(\mathbf{x})$ is the uncertainty introduced by the measurement.
 
To construct the model from the training database $\mathcal{L}$, a parametric correlation function can be chosen along with a
deterministic trend (to bring more information on the behaviour of the output expectation). These steps define the list of
hyper-parameters to be estimated ($\Xi$) by the training procedure. The best estimated hyper-parameters ($\Xi^*$) constitute
then the kriging model that can then be used to predict the value of new points.

To end this introduction, it might be useful to show a very-general correlation function: the Matern function, called
hereafter $K_\nu$. It uses the Gamma function $\Gamma$ and the modified Bessel function of order $\nu$. This $\nu$ parameter
describes the regularity (or smoothness) of the trajectory (the larger, the smoother) which should be greater than 0.5. In
one dimension, with $\delta x$ the distance, this function can be written as

\begin{equation}
  c(\delta x) = \frac{1}{\Gamma(\nu)2^{\nu-1}}\left(2\sqrt{\nu}\frac{\delta x}{l}\right)^\nu K_\nu \left(2\sqrt{\nu}\frac{\delta x}{l}\right).
  \label{eq_matern}
\end{equation}

In this function, $l$ is the correlation length parameter, which has to be positive. The larger $l$ is, the more $Y$ is
correlated between two fixed locations $x_1$ and $x_2$ and hence, the more the trajectories of $Y$ vary slowly with respect
to $x$.

\subsubsection[Implementation in Uranie and application to the \usecase]{Implementation in \uranie{}  and application to the \usecase}

The kriging approximation in \uranie{} is provided through the \gpLib{} library~\cite{gpLib}. Based on the gaussian process
properties of the kriging~\cite{bacho2013}, this library can estimate the hyper-parameters of the chosen correlation
function in several possible ways, then build the prediction model. More details about these steps are provided hereafter and
in the \gpLib{} tutorial~\cite{KrigUranie}.

The first step is to construct the model from a training database $\mathcal{L}$, by choosing a parametric correlation
function, amongst the list below, for which $l$ is the vector of correlation lengths and $\nu$ is the vector of regularity
parameters:
\begin{itemize}
\item Gauss: defined with one parameter per dimension, as $c(\delta \mathbf{x}) = \exp\left[-\sum_{k=1}^{n_X}
  \left(\frac{\delta x_k}{l_k}\right)^2\right]$.
\item Isogauss: defined with one parameter only, as $c(\delta \mathbf{x}) = \exp\left[-\frac{ |\delta \mathbf{x}|^2 }{l^2}
  \right]$.
\item Exponential: defined with two parameters per dimension, as $c(\delta \mathbf{x}) = \exp\left[-\sum_{k=1}^{n_X}
  \left(\frac{|\delta x_k|}{l_k}\right)^{p_k}\right]$, where $p$ are the power parameters. If $p=2$, the function is
  equivalent to the Gaussian correlation function.
\item MaternI: the most general form, defined with two parameters per dimension, as
  $c(\delta \mathbf{x}) = \prod_{k=1}^{n_X} \frac{1}{\Gamma(\nu_k)2^{\nu_k-1}} \left(2\sqrt{\nu_k}\frac{\delta x_k}{l_k}\right)^{\nu_k} K_{\nu_k}
  \left(2\sqrt{\nu_k}\frac{\delta x_k}{l_k}\right).$
\item MaternII: defined as maternI, with only one smoothness (leading to $n_X+1$ parameters).
\item MaternIII: the distance $\boldsymbol{\delta} = \sqrt{\sum_{k=1}^{n_X}\left(\frac{\delta x_k}{l_k}\right)^2}$ is put in
  \Eqn{matern} instead of $\delta x$ (leading to $n_X+1$ parameters).
\item Matern3/2: equivalent to maternIII, when $\nu = 3/2$.
\item Matern5/2: equivalent to maternIII, when $\nu = 5/2$.
\item Matern7/2: equivalent to maternIII, when $\nu = 7/2$.
\end{itemize}

The next step is to find the optimal hyper-parameters ($\Xi^*$) of the correlation function and the deterministic trend (if
one is prescribed), which can be done in \uranie{} by choosing:
\begin{itemize}
\item an optimisation criterion (in the example: the \emph{log-likelihood} function);
\item the size of the \doe{} used to define the best starting point for the optimisation;
\item an optimisation algorithm configured with a maximum number of runs;
\end{itemize}

Once the ``best'' starting point is found, the chosen optimisation algorithm is used to seek for an optimal
solution. Depending on various conditions, convergence can be difficult to achieve. Once done, the kriging \surmod{} can be
applied to the testing database to get predicted output values and their corresponding uncertainties.

It is, however, possible, even before using a testing database, to check the specified covariance function at hand, using
the \Loo{} technique (Loo). This method consists in the prediction of a value for $y_i$ using the rest of the known values in
the training site, \emph{i.e.}  $y_1,\,\ldots,\, y_{i-1},\, y_{i+1},\, \ldots,\, y_{n_S}$ for $ i=1,\ldots, n_S$. From there,
it is possible to use the \Loo{} prediction vector $(y^{Loo}_i)_{i=1,\, \ldots,\, n_S}$ and the expectation $\bar{y}$ to
calculate two criteria: the \emph{Mean Square Error} (MSE) and the quality criteria $Q^2_{Loo}$ defined as \[MSE_{Loo}
= \frac{1}{n_S} \sum_{i=1}^{n_S} (y_i-y^{Loo}_i)^2 \] and \[ Q^2_{Loo} = 1- \sum_{i=1}^{n_S} \frac{(y_i-y^{Loo}_i)^2}{(y_i
- \bar{y})^2}.\] The first criterion should be close to 0 while, if the covariance function is correctly specified, the
second one should be close to 1. Another possible test to check whether the model seems reasonnable consists in using the
predictive variance vector $(\sigma^2_{y^{Loo}_{i}})_{i=1,\, \ldots,\, n_S}$ to look at the distribution of the ratio
$(y_i-y^{Loo}_i)^2/\sigma^2_{y^{Loo}_{i}}$ for every point in the training site. A good modelling should result in a standard
normal distribution.

\begin{figure}[h!]
  \begin{center}
    \subfloat[$\theta(x_{ds}=0.5,t_{ds},B_i=4)$ and the kriging model]{
      \includegraphics[width=.8\linewidth]{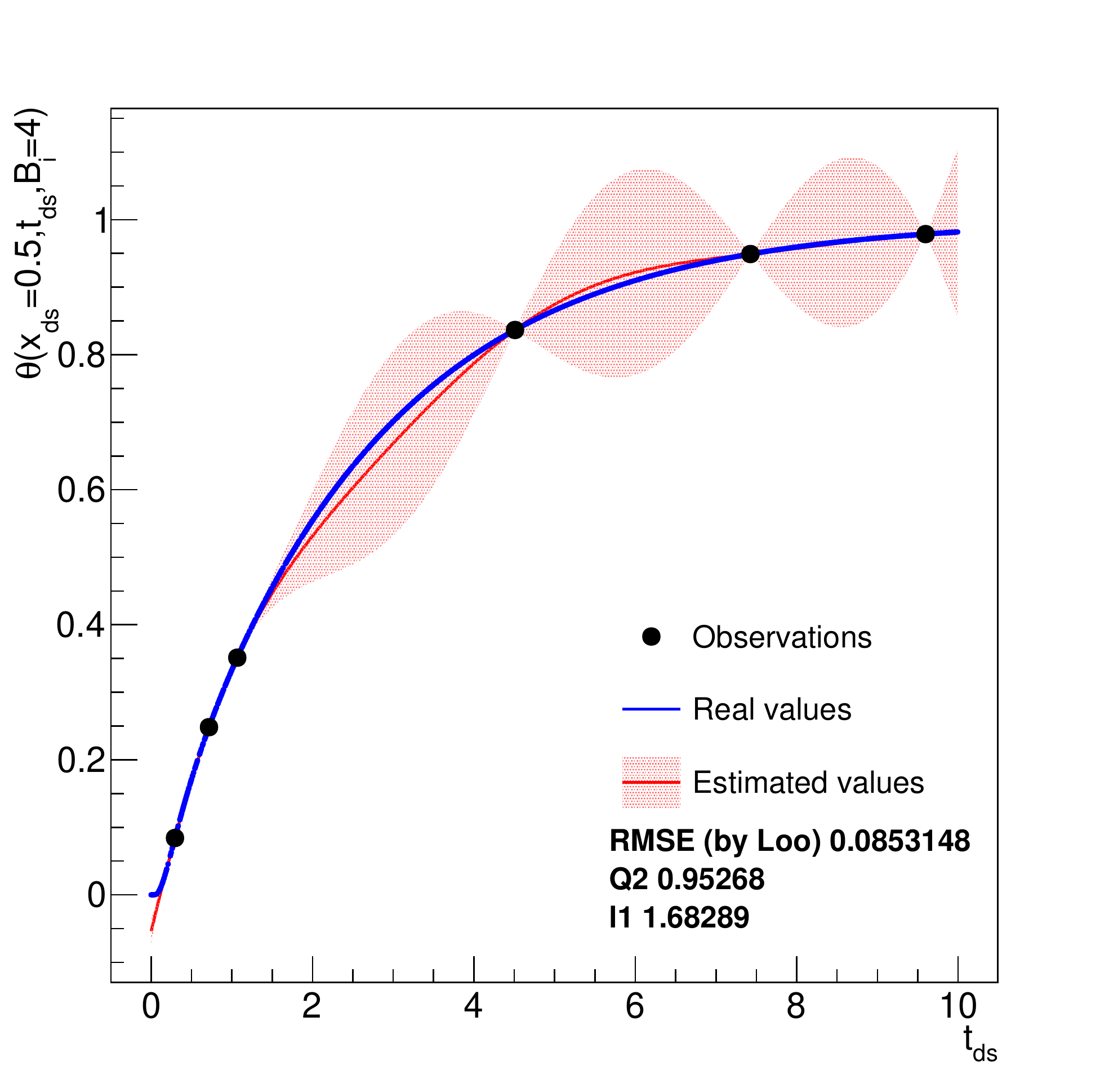}
      \label{fig_PGresults1D}
    }
    
    \subfloat[$\hat\theta$ \emph{vs.} $\theta$]{
      \includegraphics[page=1,width=.8\linewidth,keepaspectratio=true]{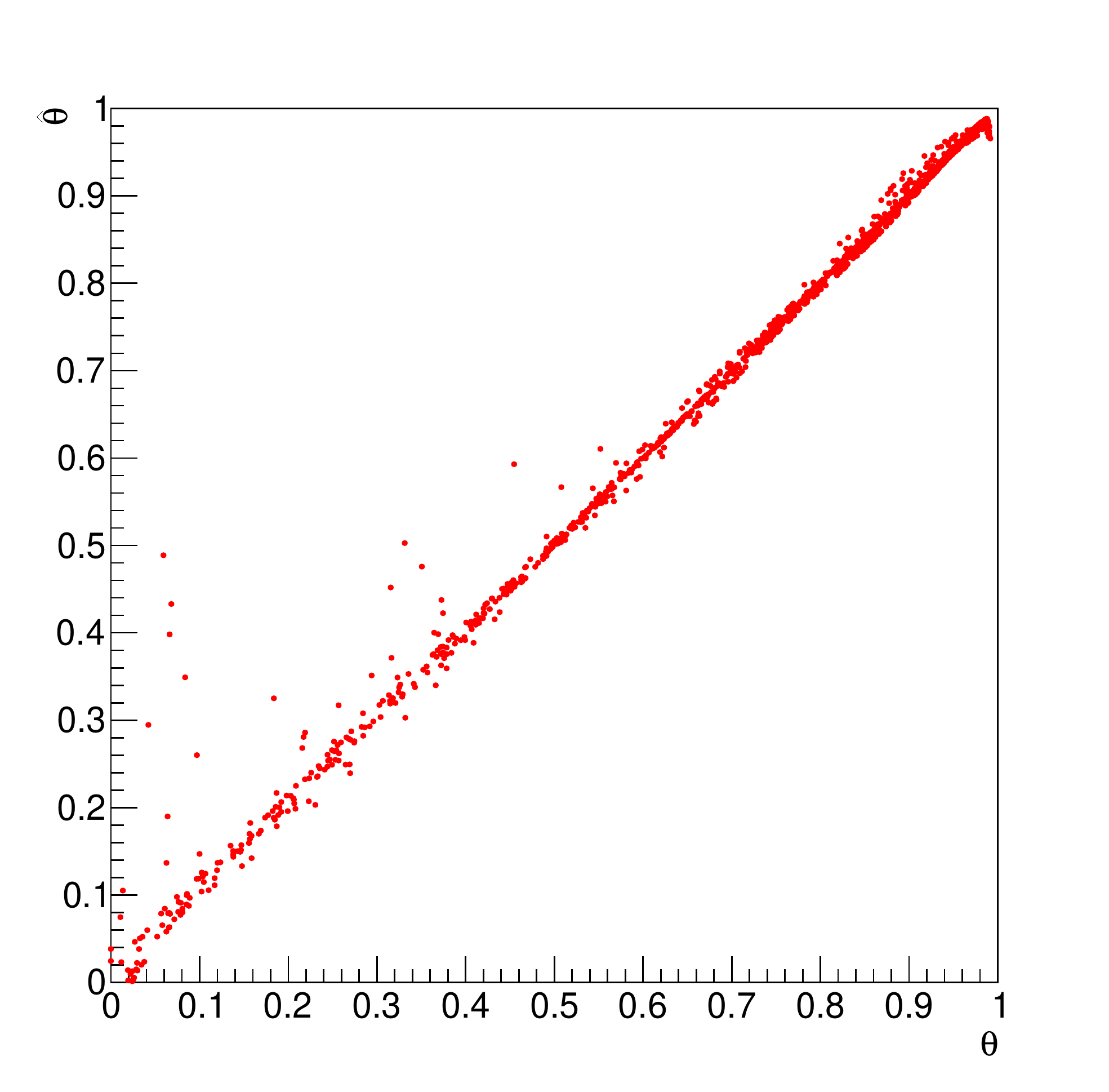}
      \label{fig_PGresults2D}
    }
  \end{center}
  \caption{Distribution of the thermal gauge values, in a test database, computed with the code (in blue) and estimated by a
    kriging model (in red) whose training database is shown as black points (a).  Distribution of the thermal gauge values
    estimated by the surrogate model ($\hat\theta$) as a function of the ones computed by the complete model ($\theta$) in a
    test database, not used for the training (b).  \label{fig_PGresults}}
\end{figure}

The kriging technique has been applied twice to illustrate its principle and the results are gathered in \Fig{PGresults}. In
the first case, it is used on a mono-dimensional thermal gauge evolution as a function of the dimensionless time, see
\Fig{PGresults1D}. On this figure, the black points represent the training database while the blue and red ones are
respectively the real output values and their estimated counterpart from the kriging model using the testing database. A good
agreement is found and confirmed by the MSE and $Q^2$ criteria. The red band represents the uncertainty on the
estimation. The kriging approximation has also been applied to $\mathcal{L}$, as for the ANN and PC, and a nice agreement is
found on the overall range, as shown in \Fig{PGresults2D}.
 
In practice, the main steps used to get the kriging model gathered in the following block:
\begin{lstcolored}[firstnumber=1,]
//Create the GPBuilder object (given a dataserver, inputs, output, correlation function and a trend)
TGPBuilder *gpb = new TGPBuilder(tds,"xad:tad","theta","matern5/2","linear");
//Find best hyper-parameters (with Subplexe algorithm with ML criterion
gpb->findOptimalParameters("ML", 1000, "Subplexe", 1000);
//Construction of the Kriging object
TKriging *kg = gpb->buildGP();
\end{lstcolored}

\subsubsection{To go further }

There are several points not discussed in this section but which can be of interest for users:
\begin{itemize}
\item other optimisation criteria. Thanks to the linear nature of the kriging model, the \Loo{} error has an analytic
  formulation~\cite{gpLib};
\item on top of the deterministic trend, an \textit{a priori} knowledge on the mean and variance of the trend parameters can
  be used to perform a Bayesian study;
\item one can take into account measurement errors when looking for the optimal hyper-parameters;
\end{itemize}

On top of the already introduced \surmod{s}, \uranie{} can provide few other solutions among which:
\begin{itemize}
\item the regression method;
\item the k-nearest neighbour method;
\item the kernel method.
\end{itemize}

%% file: UncertPropag.tex
As already stated in \Sect{intro_metho}, many analysis will start in the same way, by defining the problem investigated in
terms of number of input variables and their characteristics, setting their possible correlations\ldots From there, unless
one has an already computed set of experiments (as it was the case in \Sect{model}), it is common to generate a \doe{} as
being a set of input locations to be assessed by the code/function and that should be the most representative of the input
phase space with respect to aim of the study.

This section introduces the various mechanisms available in \uranie{} for sampling \doe, which could lead to the uncertainty
propagation from the input parameters to the quantity of interest, as shown in \Fig{UncertAna}.

\subsection{Random variable definition}
\label{variable}

\subsubsection{Defining a variable}

\uranie{} implements more than fifteen parametric distributions (continuous ones) to describe the behaviour of a given random
variable. The list of available continuous laws is given in \Tab{availLaws}, along with their corresponding adjustable
parameters. For a complete description of these laws and a set of variations of all these parameters, see \cite{metho}. The
classes, implementing these laws, give access to the main mathematical properties (theoretical ones) and they have been made
to be an interface with the sampling methods discussed in \Sect{doe}, to get a dedicated \doe.

\begin{table}[htbp]
  \begin{center}
    \begin{tabular}{lll}
      \hline
      \bfseries\textbf{Law}&\bfseries\textbf{Adjustable parameters}\\
      \hline
      Uniform&Min, Max\\
      Log-uniform&Min, Max\\
      Triangular&Min, Max, Mode\\
      Log-triangular&Min, Max, Mode\\
      Normal (gaussian)&Mean, Sigma\\
      Log-normal&Mean, Sigma\\
      Trapezium&Min, Max, Low, Up\\
      Uniform by parts&Min, Max, Median\\
      Exponential&Rate, Min\\
      Cauchy&Scale, Median\\
      GumbelMax&Mode, Scale \\
      Weibull&Scale, Shape, Min\\
      Beta& alpha,  beta, Min, Max\\
      GenPareto&Location, Scale, Shape\\
      Gamma& Shape,  Scale,  Location\\
      Inverse gamma& Shape,  Scale,  Location\\
      \hline
    \end{tabular}
  \end{center}
  \caption{List of continuous available statistical laws in \uranie, along with their corresponding adjustable
    parameters. \label{tab_availLaws}}
\end{table}

\begin{figure*}
  \begin{center}
    \subfloat[PDF]{
      \includegraphics[page=1,width=.33\textwidth]{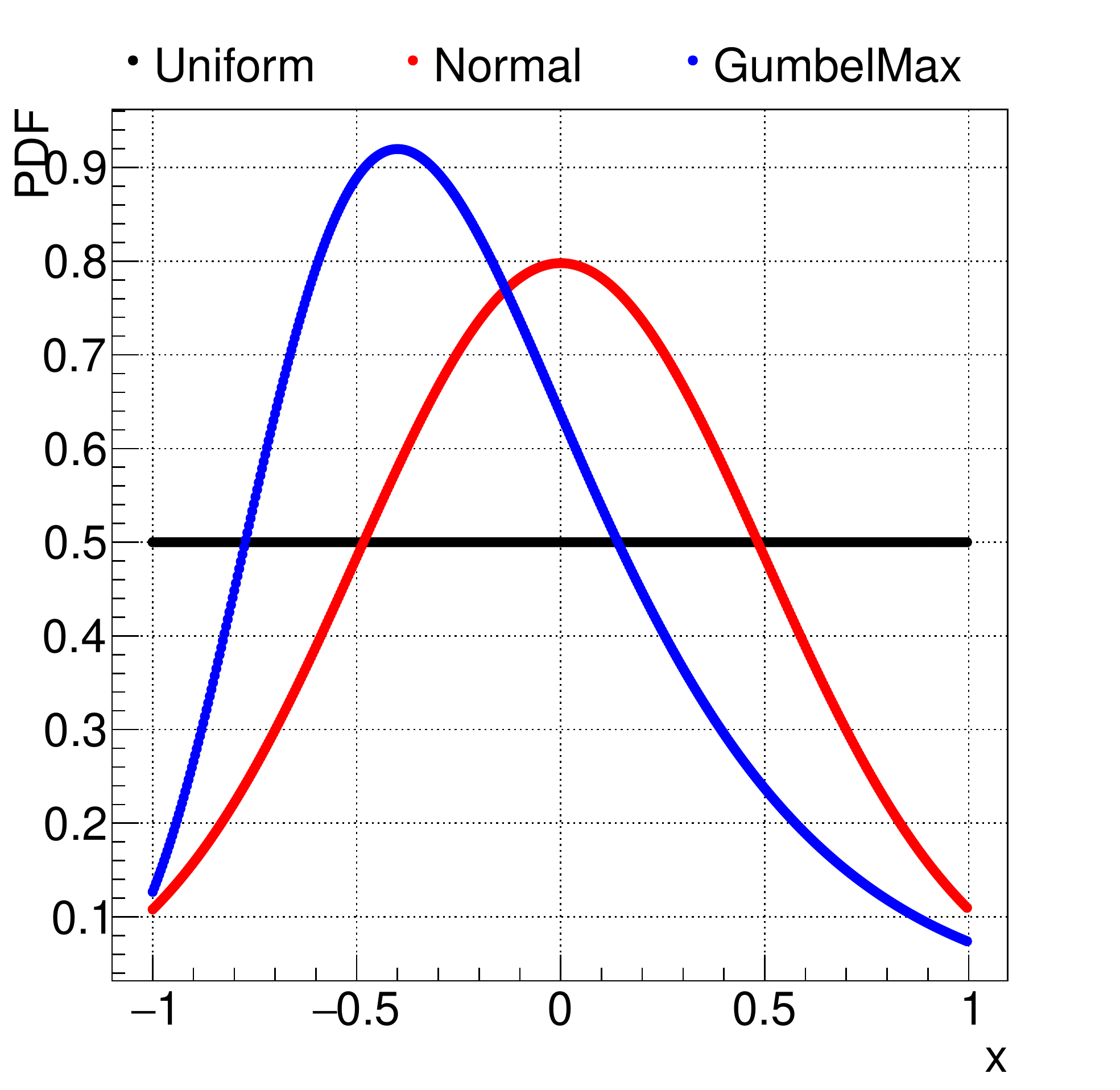}
      \label{fig_PDFExPDF}
    }
    \subfloat[CDF]{
      \includegraphics[page=2,width=.33\textwidth]{PdfCdfInverseCdf.pdf}
      \label{fig_PDFExCDF}
    }
    \subfloat[inverse-CDF]{
      \includegraphics[page=3,width=.33\textwidth]{PdfCdfInverseCdf.pdf}
      \label{fig_PDFExInvCDF}
    }
  \end{center}
  \caption{Example of PDF (a), CDF (b) and inverse-CDF (c) for an uniform law (defined between -1 and 1, in black), a normal
    law (with $\mu=0$ and $\sigma=0.5$, in red) and a gumbelmax one (with $\mu=-0.4$ and $\beta=0.4$ in
    blue).\label{fig_PDFEx}}
\end{figure*}

These classes offer also methods to compute the probability density function (PDF), the cumulative distribution function
(CDF) and its inverse-CDF. \Fig{PDFEx} shows example of PDF distributions in \Fig{PDFExPDF}, CDF distributions in
\Fig{PDFExCDF} and inverse-CDF distribution in \Fig{PDFExInvCDF}, using an uniform (black), a normal (red) and a gumbelmax
(blue) law.

On top of these definitions, it is also possible to create a new variable through a combination of already existing ones, for
instance with simple mathematical expression. This can be done independently of the origin of the original variables: either
read from a set-of-experiments without any knowledge of the underlying law, or generated from well-defined stochastic law.

\subsubsection{Correlating the laws}
\label{Correl}
Once the laws have been defined, one can introduce correlation between them. This, in \uranie, can be done with different
methods. Starting from the simplest one, one can introduce a correlation coefficient between two variables or providing the
complete correlation matrix. 

Instead of using correlation matrix to get intricated variables, one can use methods relying on copula, in order to describe
the dependencies. The idea of a copula is to define the interaction of variables using a parametric function that can allow a
broader range of entanglement than only using a correlation matrix (various shapes can be done). The copulas provided in the
\uranie{} platform are archimedian ones, with 4 pre-defined parametrisation: Ali-Mikhail-Haq, Clayton, Frank and
Plackett.

Both methods will be illustrated in the next section.

\subsection{\Doe{} definition}
\label{doe}

\subsubsection{Stochastic methods}
\label{stochaDoe}

\begin{figure*}[p]
  \begin{center}
    \subfloat[LHS drawing]{
      \includegraphics[page=3,width=.42\textwidth]{SrsAndLhsComparison.pdf}
      \label{fig_doeLHS}
    }
    \subfloat[SRS drawing]{
      \includegraphics[page=1,width=.42\textwidth]{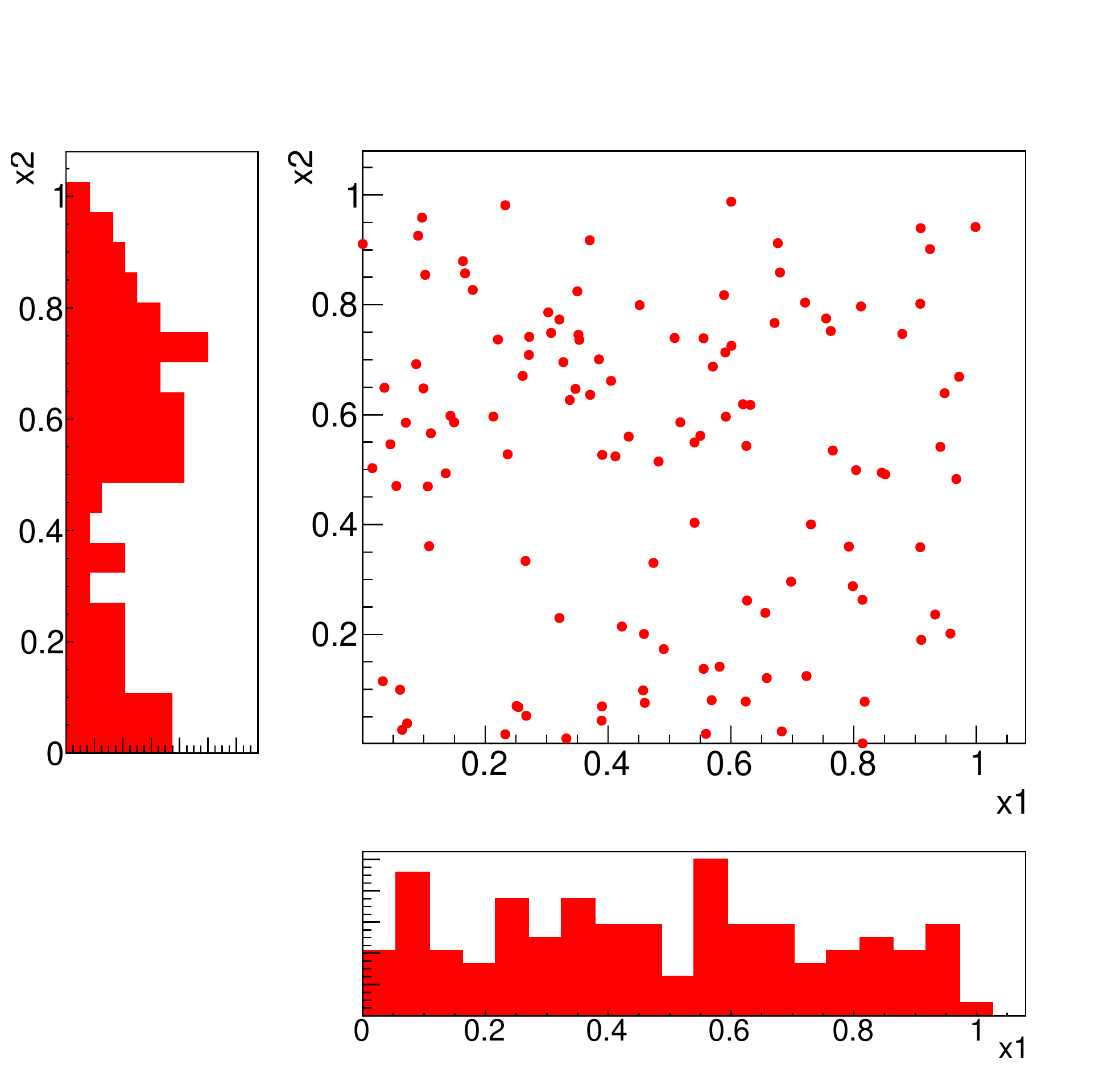}
      \label{fig_doeSRS}
    }

    \subfloat[maximin LHS drawing]{
      \includegraphics[page=4,width=.42\textwidth]{SrsAndLhsComparison.pdf}
      \label{fig_doeLHD}
    }
    \subfloat[Halton sequence]{
      \includegraphics[page=5,width=.42\textwidth]{SrsAndLhsComparison.pdf}
      \label{fig_doeHalton}
    }

    \subfloat[Sobol sequence]{
      \includegraphics[page=6,width=.42\textwidth]{SrsAndLhsComparison.pdf}
      \label{fig_doeSobol}
    }
    \subfloat[Petras algorithm]{
      \includegraphics[page=7,width=.42\textwidth]{SrsAndLhsComparison.pdf}
      \label{fig_doePetras}
    }
  \end{center}
  \caption{Drawing of the \doe{} for two uniformly-distributed variable $x1$ and $x2$, with a LHS sampling (a), a SRS one (b)
    and a maximin LHS one (c). Deterministic sampling are shown also with the Halton sequence (d), the Sobol one (e) and a
    Petras sparse grid (f). \label{fig_doe}}
\end{figure*}

In \uranie, different kind of random-based algorithms can be used to generate \doe. Here is a brief introduction of the three
main types which are illustrated in \Fig{doe} where two independent uniformly-distributed variables are used. This kind of
plot (called Tufte one) is an example of \uranie-implemented visualisation tool. The main pad, in the centre of the canvas,
shows the dependence of the two variables under consideration, while the two other pads show projection along one of the
dimension, as a mono-dimensional histogram.

\begin{description}
\item[Simple Random Sampling (SRS):] This method consists in independently generating the samples for each parameter
  following its own probability density function.  An example of this sampling when having two independent
  uniformly-distributed variables is shown in \Fig{doeSRS}. The random drawing is performed using an uniform law between 0
  and 1 and getting the corresponding value by calling the inverse CDF function corresponding to the law under study.
\item[Latin Hypercube Sampling (LHS):] This method~\cite{McKay2000} consists in partitioning the variation interval of each
  variable to obtain equiprobable segments and then get, for each segment, a representative value.  An example of this
  sampling when having two independent uniformly-distributed variables is shown in \Fig{doeLHS}. The random drawing is
  performed using an uniform law between 0 and 1, split into the requested number of points for the \doe. Thanks to this, a
  grid is prepared, assuring equi-probability in every axis-projection. Finally, a random drawing is performed in every
  sub-range. The obtained value is computed by means of the inverse CDF function corresponding to the law under study.
\item[maximin LHS:] Considering the definition of a LHS sampling, it is clear that permuting a coordinate of two different
  points creates a different \doe{} that is still a LHS one. In \uranie, a new kind of LHS sampling, called maximin LHS, has
  been recently introduced with the purpose of maximising the minimal distance calculated between every pair of two design
  locations~\cite{Morris95}. The criterion under consideration is the mindist criterion: let $D=[\mathbf{x}_1,
    \cdots,\mathbf{x}_{n_S}] \subset [0,1]^{n_X}$ be a \doe, made out of $n_S$ points. It is written as
  \begin{equation}
    \label{eq_minDist}
    \min_{i,j}{||\mathbf{x}_i-\mathbf{x}_j||_{2}}
  \end{equation}
  where $||.||_2$ is the euclidean norm. The designs which maximise the mindist criterion are referred to as maximin LHS. It
  has been observed that the best designs in terms of maximising \Eqn{minDist} can be constructed by minimising its $L^{p}$
  regularisation instead, $\phi_p$, which can be written:
  \begin{equation}
    \label{eq_LpNorm}
    \phi_p := \Big[ \sum_{i<j} || \mathbf{x}_i-\mathbf{x}_j ||_{2}^{p} \Big]^{\frac{1}{p}}.
  \end{equation}  
 The permutations done to go from a first LHS \doe{} to its maximin version are made through a simulated annealing method. An
 example is displayed, starting from the \doe{} in \Fig{doeLHS} and resulting in the one in \Fig{doeLHD}.  Both have uniform
 projections along each axis but the locations are clearly more space-filling in \Fig{doeLHD}.
\end{description}

The SRS method is a pure-random method which populates the region following the inverse-CDF of the considered probability
law. In other words, if the objective is to obtain quantiles for extreme probability values, the size of the sample should be
large for this method to be used. However, one should keep in mind that it is rather trivial to double the size of an
existing SRS sampling, as no extra caution has to be taken apart from the random seed. On the other hand, the LHS method is
built in a way that ensure that the domain of variation of each variable is totally covered in a homogeneous way. The
drawback of this construction is that it is absolutely not possible to remove or add points to a LHS sampling without having
to regenerate it completely.

From a theoretical perspective, using a maximin LHS to build a GP emulator can reduce the predictive variance when the
distribution of the GP is exactly known. However, it is not often the case in real applications where both the variance and
the range parameters of the GP are actually estimated from a set of learning simulations run over the maximin
LHS. Unfortunately, the locations of maximin LHS are far from each other, which is not a good feature to estimate these
parameters with precision. That is why maximin LHS should be used with care. Relevant discussions dealing with this issue can
be found in \cite{Pronz12}.

\begin{figure}[h!]
  \begin{center}
    \subfloat[Spearman coefficient]{
      \includegraphics[page=1,width=.85\linewidth]{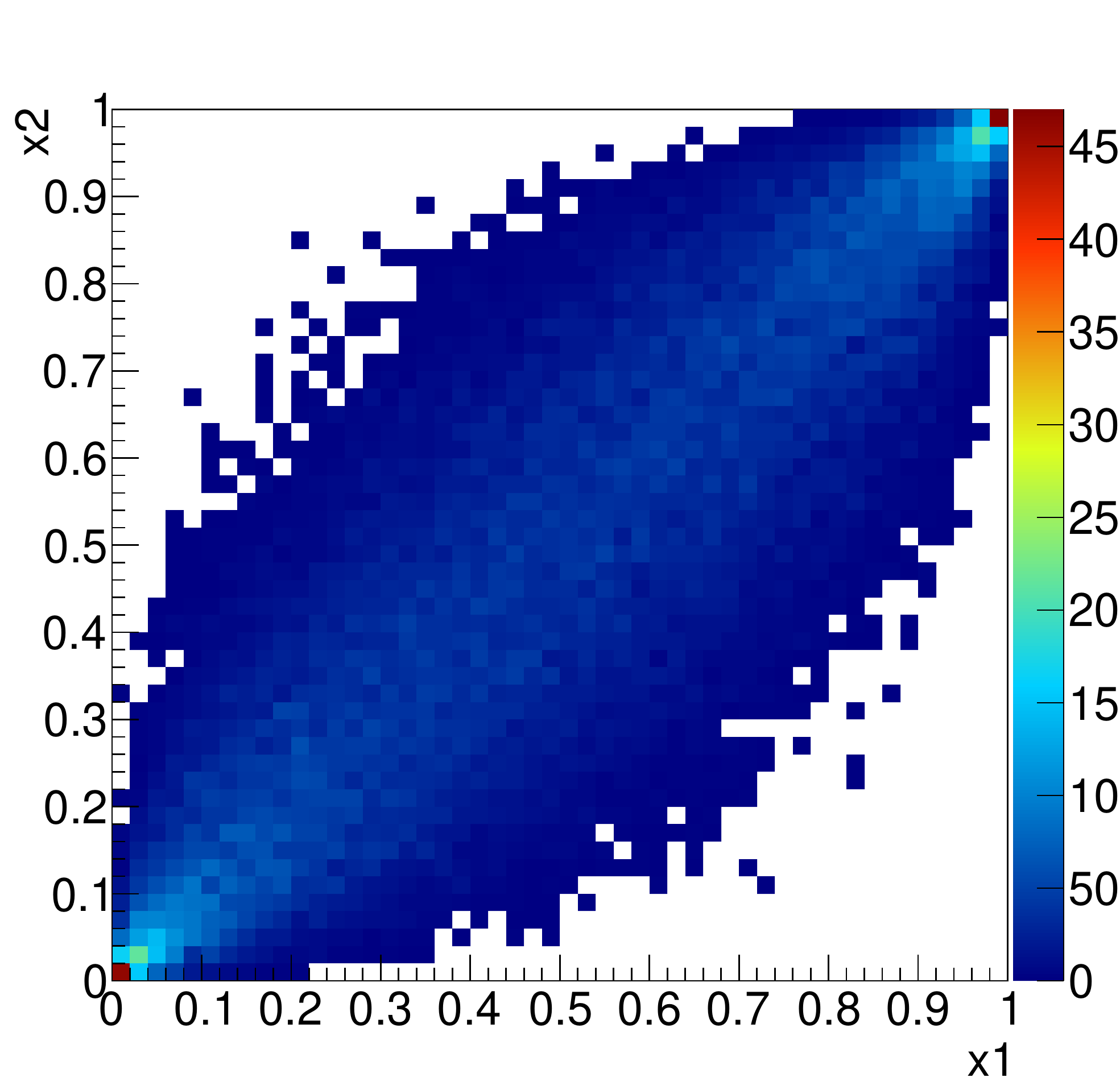}
      \label{fig_CorrelExCoe}
    }
    
    \subfloat[Ali-Mikhail-Haq Copula]{
      \includegraphics[page=2,width=.85\linewidth]{CorrelationComparison.pdf}
      \label{fig_CorrelExCop}
    }
  \end{center}
  \caption{Example of correlation introduced between two uniform distributions, using either the Spearman coefficient (a) or
    a Ali-Mikhail-Haq copula (b).\label{fig_CorrelEx}}
\end{figure}

Finally, as introduced in \Sect{Correl}, an example of correlation is provided in \Fig{CorrelEx}, both using correlation
coefficient and copula. In the first case, instead of relying on the ``Bravais-Pearson'' correlation coefficient definition,
that exclusively reflects both the degree and sign of linearity between two variables $x_i$ and $x_j$ , the method used in
\uranie{}~\cite{Iman82} takes into account the correlation on ranks, \ie{} the ``Spearman'' definition:
\begin{equation}
  \label{eq_Spearman}
  \rho^{S}(x_i,x_j)  = \rho(R_{x_i},R_{x_j}) = \frac{ {\rm Cov}(R_{x_i},R_{x_j}) } { \sqrt{ {\rm Var}(R_{x_i}) {\rm Var}(R_{x_j}) }}
\end{equation}
In this expression, $\rho^{S}$ is the Spearman coefficient, $\rho$ is the usual Bravais-Pearson definition but applied here
on $R$ which is the rank of the information under consideration.  This method can be applied only if the correlation matrix
provided by the user is positive definite. \Fig{CorrelExCoe} shows an example of correlation (set to a value of 0.9) between
two uniform distributions.

The copula, introduced in \Sect{Correl} depend only on the input variables and a parameter $\xi$. An example using two
uniform distributions is given in \Fig{CorrelExCop} for the Ali-Mikhail-Haq copula.

\subsubsection{Quasi Monte-Carlo  methods}

The deterministic samplings can produce \doe{} with specific properties, that can be very useful in cases such as:
\begin{itemize}
\item cover at best the space of the input variables
\item explore the extreme cases
\item study combined or non-linearity effects
\end{itemize}
  
There are two kinds of quasi Monte-Carlo sampling methods implemented in \uranie: the regular ones and the sparse grid
ones. The former can be generated using two different sequences:
\begin{enumerate}
\item Sequences of Halton~\cite{HaltonSeq64}
\item Sequences of Sobol~\cite{SOBOL196786}
\end{enumerate}
\Figs{doeHalton}{doeSobol} show the \doe{} obtained when having two independent uniformly-distributed variables and
can be compared with the stochastic ones (from \Fig{doeLHS} to \Fig{doeLHD}) already discussed in \Sect{stochaDoe}.  The
coverage is clearly more regular in the case of quasi Monte-Carlo sequences, but these methods can suffer from weird pattern
appearance when $n_X$ is greater than 10.  On the other hand, the sparse grid sampling can be very useful for integration
purposes and can be used in some of the meta-modelling definition, see, for instance, in \Sect{PCfurther}. In \uranie, the
Petras algorithm~\cite{Petras2001} can be used to produce these sparse grids, (shown when the level is set to 8, in
\Fig{doePetras}, that can be compared to the rest of of the \doe{} in \Fig{doe}.

In practice, the main steps used to get one of the plot shown in the \Fig{doe} are gathered in the following block:
\begin{lstcolored}[firstnumber=1,]
//Creation of the dataserver 
TDataServer *tds = new TDataServer("tds","the dataserver");
//Definition of two uniform variables, with a name, min and max value
tds->addAttribute( new TUniformDistribution("x1",0,1) ); 
tds->addAttribute( new TUniformDistribution("x2",0,1) );     
//Define the sampling (given the dataserver, the method to be used and the number of locations)
TSampling *sam = new TSampling(tds,"srs",size);
//Generation of the samples
sam->generateSample();     
//Draw the design-of-experiments (Uranie method)
tds->drawTufte("x2:x1");
\end{lstcolored}

\subsubsection{To go further}

This introduction to the \doe{} sampling is very brief with respect to the underlying complexity and possibility. It is
indeed also possible to produce with \uranie:
\begin{itemize}
\item \doe{} for integration in the conjugate Fourier space;
\item a representative set-of-points smaller than a given database to keep the main behaviour without having to run too many
  computations;
\end{itemize}

\subsection{Focusing on the PTFE case}

\begin{figure}[h!]
  \begin{center}
    \includegraphics[width=1.05\linewidth]{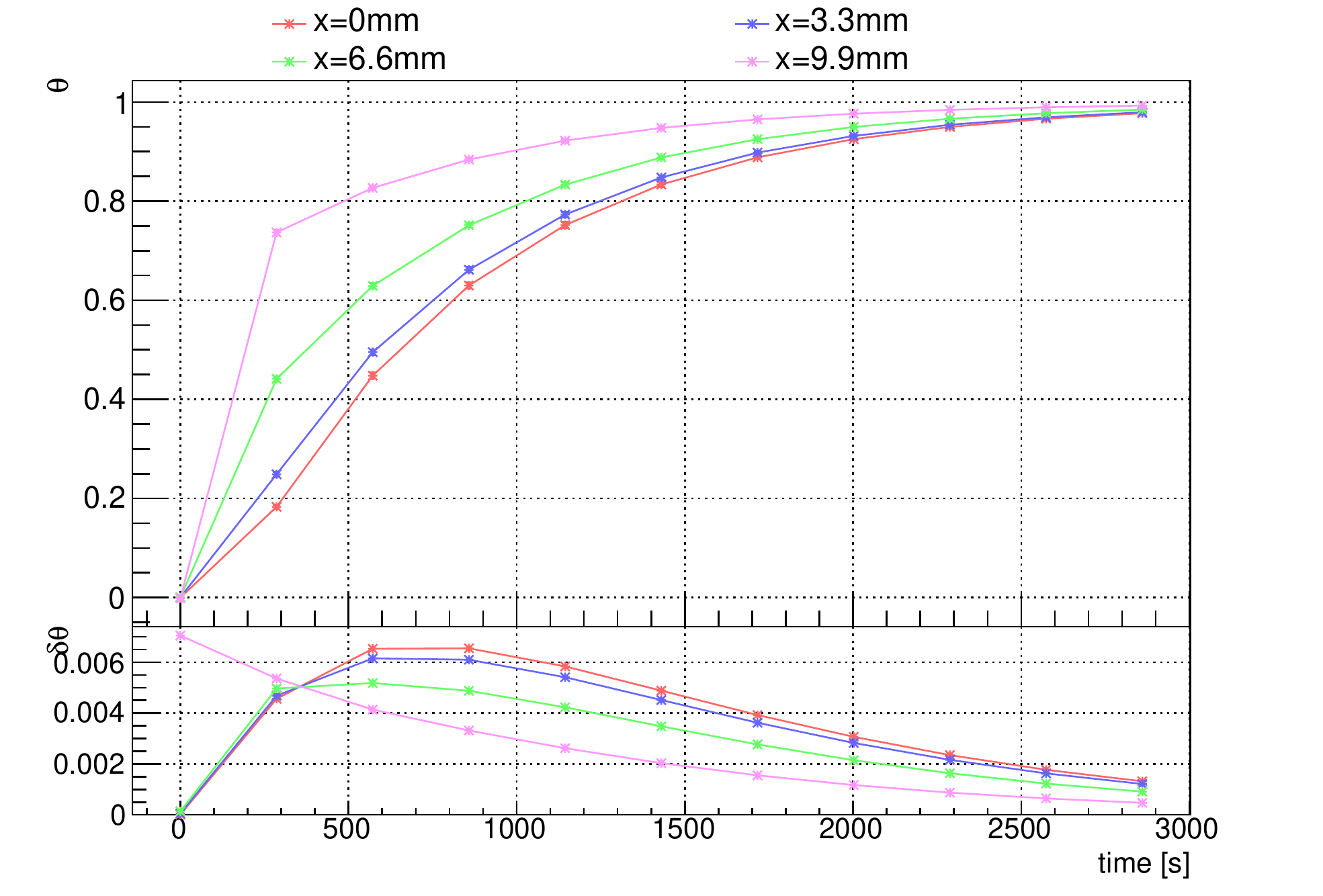}
  \end{center}
  \caption{Evolution of the thermal gauge (top pad) and its uncertainty (bottom pad), as a function of the absolute time, for
    four different values of depth within the sheet.\label{fig_UncertPropa}}
\end{figure}

In this section, the basic building blocks introduced in \Sects{variable}{doe} are put together to perform the uncertainty
propagation. The following steps are then:
\begin{enumerate}
\item create the input variables by specifying for each and every one of them a probabilistic law and their corresponding
  parameters. Here, all the input variables have been modelled using normal distributions and their nominal values and
  uncertainties have been estimated and gathered in \Tab{PTFECara}.
\item sample a LHS to be as much representative of the full input phase space as possible. No correlation between the
  parameters have been assumed. The size of this \doe{} has been set to 100 points.
\item compute 11 absolute time steps for every locations and for 4 different depths in the sheet. Every configuration (a
  configuration being a precise value of the time and depth) consists of 100 measurements where the mean and standard
  deviation have been computed. These values are then represented in \Fig{UncertPropa}.
\end{enumerate}

\begin{table}[h!]
  \begin{center}
    \begin{tabular}{lcc N}
      \cline{2-3}
      \multicolumn{1}{l}{} & Value & Uncertainty &\\[4pt]
      \hline
      Thickness: \color{blue}{e} & 10$\times$10$^{-3}$ & 5$\times$10$^{-5}$ &\\[4pt]
      Thermal conductivity: \color{blue}{$\lambda$} & 0.25 & 1.5$\times$10$^{-3}$ &\\[4pt]
      Massive thermal capacity: \color{blue}{$C_{\rho}$} & 1300 & 15.6 &\\[4pt]
      Volumic mass: \color{blue}{$\rho$} & 2200 & 4.4 &\\[4pt]
      \hline
    \end{tabular}      
  \end{center}
  \caption{Summary of the PTFE uncertain physical parameters and their corresponding uncertainty. The absolute value of the
    uncertainty are computed from the values in \Tab{metalCara} (where units are also provided). \label{tab_PTFECara}.}
\end{table}

Given the distribution obtained in \Fig{UncertPropa}, the user should decide what would be the next step in his analysis. The
following list of actions gives an illustration of the various possibilities (but it is not meant to be exhaustive, because
only provided for illustration purpose):
\begin{itemize}
\item Compare this to already existing measurements:
  \begin{itemize}
  \item check that the hypothesis are consistent with the model (in case of very surprising results for instance).
  \item move forward to a calibration or the determination of the uncertainty of physical model's parameters (through the
    Circe method for instance in \uranie).
  \end{itemize}
\item Move to a sensitivity analysis on the code or on a \surmod{} if this one is too resource consuming, (as discussed in
  \Sect{sensi}) to understand which input's uncertainty impacts the most the quantity of interest.
\end{itemize}

%% file: SensiAnalysis.tex
In this section, we will briefly remind different ways to measure the sensitivity of the output of a model to its inputs. A
brief recap of the concept of sensitivity analysis (SA) will be done, before focusing on the \usecase{} and investigating the
evolution of the sensitivity indexes through time, for two dimensionless positions: $x_{ds}=0.3$ and $x_{ds}=0.8$.\par

The \usecase{} application is done following a classical approach: starting with a screening analysis\footnote{Screening is a
constrained version of dimensionality reduction where a subset of the original variables is retained.} which is quick but not
very precise. Once conclusions are drawn from the previous step, a more meticulous investigation can be done using
quantitative methods, to get, for instance, the \sobolCs{} of the model under investigation. The starting point will always
be the definition of the input variables as gaussian-modelled objects, stored in the \code{TDataServer}.

\begin{lstcolored}[firstnumber=1,]
//Create the TDataServer object
TDataServer *tds = new TDataServer("tds", "The dataserver");
//Create the normally-distributed inputs (given name, mean and standard deviation)
tds->addAttribute( new TNormalDistribution("thickness",e,de));
tds->addAttribute( new TNormalDistribution("conductivity",lambda,dlambda));
tds->addAttribute( new TNormalDistribution("capacity",Crho,dCrho));
tds->addAttribute( new TNormalDistribution("mass",rho,drho));
\end{lstcolored}

At the end of this section, a list of the other available methods is given along with their possible improvements in the next
few years.

\subsection{Introduction to sensitivity analysis}

If one can consider that the inputs are independent one to another, it is possible to study how the output variance changes
when fixing $X_{i}$ to a certain value $x_{i}^{*}$. This variance denoted by ${\rm Var}(Y|X_{i}=x_{i}^{*})$ is called the
conditional variance and depends on the chosen value of $X_i$. In order to study this dependence, one should consider ${\rm
Var}(Y|X_i)$, the conditional variance over all possible $x_{i}^{*}$ value, which is a random variable and, as such, it can
have an expectation, $E({\rm Var}(Y|X_i))$. As the theorem of the total variance states that ${\rm Var}(Y) = {\rm
Var}(E(Y|X_{i})) + E({\rm Var}(Y|X_{i}))$ under the assumption of having $X_i$ and $Y$ two jointly distributed random
variables, it becomes clear that the variance of the conditional expectation can be a good estimator of the sensitivity of
the output to the specific input $X_i$. The more common and practical normalised index in order to define this
sensitivity is given by

\begin{equation}
  S_i=\frac{{\rm Var}(E(Y|X_i))}{{\rm Var}(Y)}.
  \label{eq_Sobol1}
\end{equation}

This normalised index is often called the \emph{first order sensitivity index} and quantifies the impact of the input $X_i$
on the output, but does not take into account the amount of variance explained by interactions between inputs. It can
actually be made with the crossed impact of this particular input with any other variable or combination of variables,
leading to a set of $2^{n_X} - 1$ indexes to compute. A full estimation of all these coefficients is possible and would lead
to a perfect break down of the output variance. It has been proposed by many authors in the literature and is referred to
with many names, such as functional decomposition, ANOVA method (\emph{ANalysis Of VAriance}), HDMR (\emph{High-Dimensional
Model Representation}), Sobol's decomposition, Hoeffding's decomposition... A much simpler index, which takes into account
the interaction of an input $X_i$ with all other inputs, is called the \emph{total order sensitivity index} or $S_{T_i}$
(\cite{Homma96} and can be computed as

\begin{equation}
  S_{T_i} = 1 - S_{\bar{i}} = 1 - \frac{{\rm Var}(E(Y|X_{\bar{i}}))}{{\rm Var}(Y)},
  \label{eq_SobolT}
\end{equation}

where $\bar{i}$ represents the group of indexes that does not contain the $i$ index. These two indexes (the first order and
total order) are referred to as the \sobolCs. They satisfy several properties and their values can be interpreted in several
ways:

\begin{itemize}
\item $\sum S_i \le 1$: should always be true.
\item $\sum S_i = 1 = \sum S_{T_i}$: the model is purely additive, or in other words, there are no interaction between the
  inputs and $S_i = S_{T_i}\; \forall i =1,\ldots,n_X$.
\item $1-\sum S_i$ is an indicator of the presence of interactions.
\item $S_{T_i} - S_i$ is a direct estimate of the interaction of the i-Th input with all the other factors.
\end{itemize}

\subsection{Screening method}

\subsubsection{Introduction to the Morris method}
\label{MorrisIntro}

\begin{figure}[h!]
  \begin{center}
    \includegraphics[width=0.95\linewidth,keepaspectratio=true]{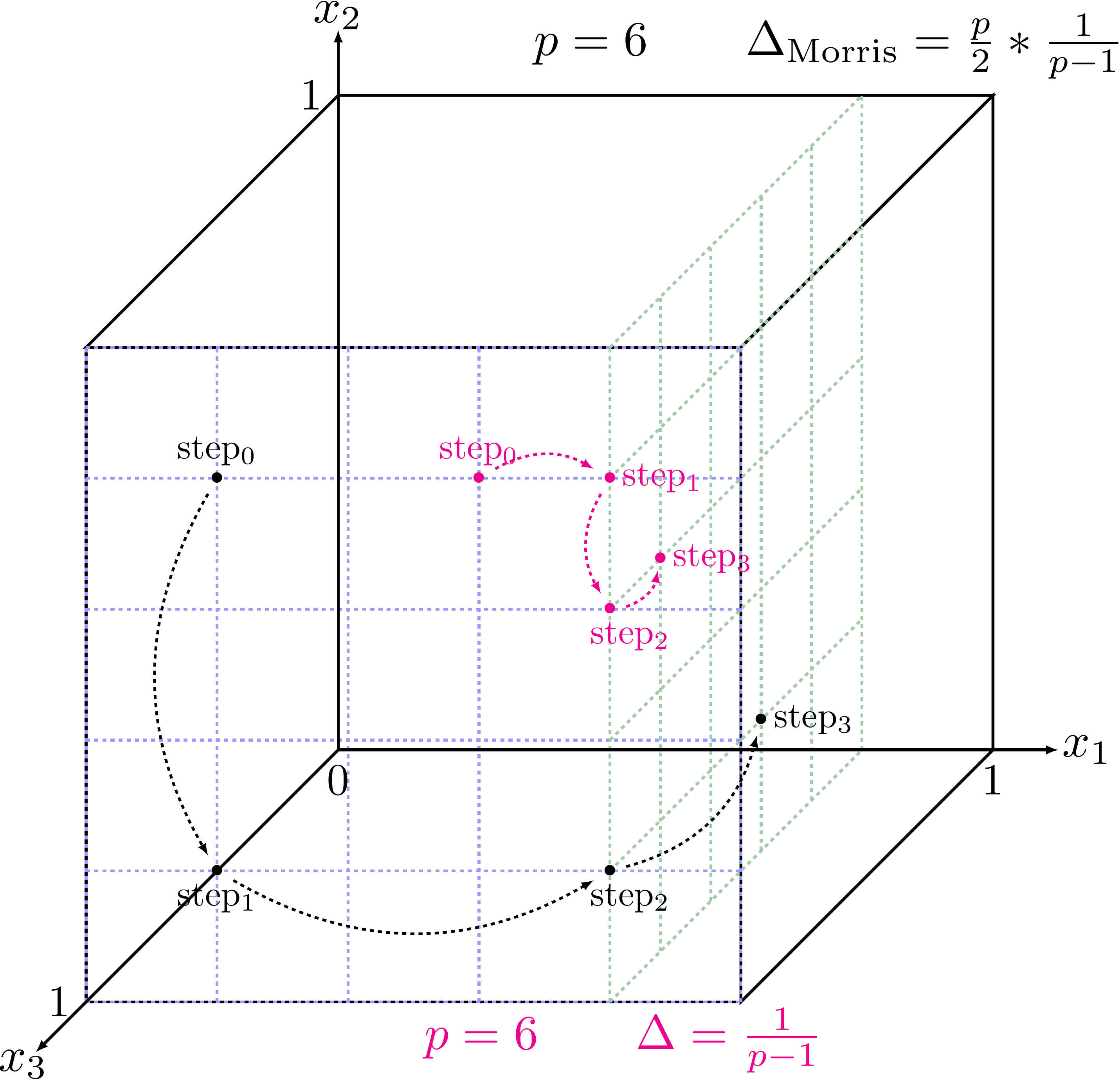}
  \end{center}
  \caption{Schematic view of two trajectories drawn randomly in the discretised hyper-volume (with p=6) for two different
    values of the elementary variation (the optimal one in black and the smallest one in pink).\label{fig_morrisprinci}}%
\end{figure}

The Morris method~\cite{Morris95} is an effective screening procedure that robustifies the
\emph{One-factor-At-a-Time} protocol (OAT). Instead of varying every input parameter only once (leading then to a
minimum of $n_X+1$ assessments of the code/function, with an OAT technique), the Morris method repeats this OAT principle $r$
times (practically, it is between 5 and 10 times), each time being called a \emph{trajectory} or a \emph{replica}. Every
trajectory begins from a randomly chosen starting point (in the input parameters space). In order to do so, it computes
Elementary effects (later on called \emph{EE}), defined as
\[ EE^t_i=EE_i(\mathbf{x}^t)  = \]\[ \frac{y(x^t_1,\ldots , x^t_i + \Delta^t_i, \ldots ,x^t_{n_X}) -  y(x^t_1,\ldots , x^t_i,\ldots ,x^t_{n_X})}{\Delta^t_i} \]
 where $\Delta^t$ is the chosen variation in the trajectory $t$. This variation can be set by the user, but the default
 (recommended, because it is said to be optimal~\cite{Salt04}) value is $\Delta = \frac{p}{2(p-1)}$, where $p$ is the
 level, that describes in how many interval, the range should be split. The resulting cost (in terms of assessments) is then
 $r(n_X+1)$.  This method is schematised in \Fig{morrisprinci} for a problem with three inputs. The hyper-volume is
 normalised and transformed into an unit hyper-cube. The resulting volume is discretised with the requested level (here,
 $p=6$) and two trajectories are drawn for different values of the elementary variation.

With the repetition of this procedure $r$ times, it is possible to compute basic statistics on the elementary effects, for
every input parameter, as
\begin{equation}
  \label{eq_MorrisVar}
  \mu_i = \frac{1}{r} \sum^r_{t=1} EE^t_i,\; \mu^*_i = \frac{1}{r} \sum^r_{t=1} |EE^t_i|
\end{equation}
and
\begin{equation}
\sigma^2_i = \frac{1}{r-1} \sum^r_{t=1} (EE^t_i - \mu_i)^2.
\end{equation}
The variable $\mu_i$ and $\sigma_i$ represents respectively the mean and standard deviation of the elementary effects of the
i-Th input parameters. In the case where the model is not monotonic some $EE_i^t$ may cancel each other out, resulting in a
low $\mu_i$ value even for an important factor. For that reason, a revised version called $\mu_i^*$ has been created and
defined as the mean of the absolute values of the $EE_i^t$~\cite{Salt08primer}. The results are usually visualised in the
($\mu^*$,$\sigma$) plane.

Even though the numerical results are not easily interpretable, their values can be used to rank the effect of one or several
inputs with respect to others, the point being to spot a certain number of inputs that can safely be thrown away, given the
underlying uncertainty model assessed.

\subsubsection[Implementation in Uranie and application to the \usecase]{Implementation in \uranie{}  and application to the \usecase}
\label{MorrisImple}

The method has been applied to the thermal exchange model introduced in \Sect{TherExchMod} which has been slightly changed
here for illustration purpose: a new input variable has been added, with the explicit name ``useless''. The idea is to shown
that it is possible to spot an input whose impact on the output can be considered so small that it can be discarded through
the rest of the analysis. \par

\begin{figure*}
  \begin{center}
    \subfloat[$x_{ds}=0.3$]{
      \includegraphics[page=1,width=0.4\textwidth]{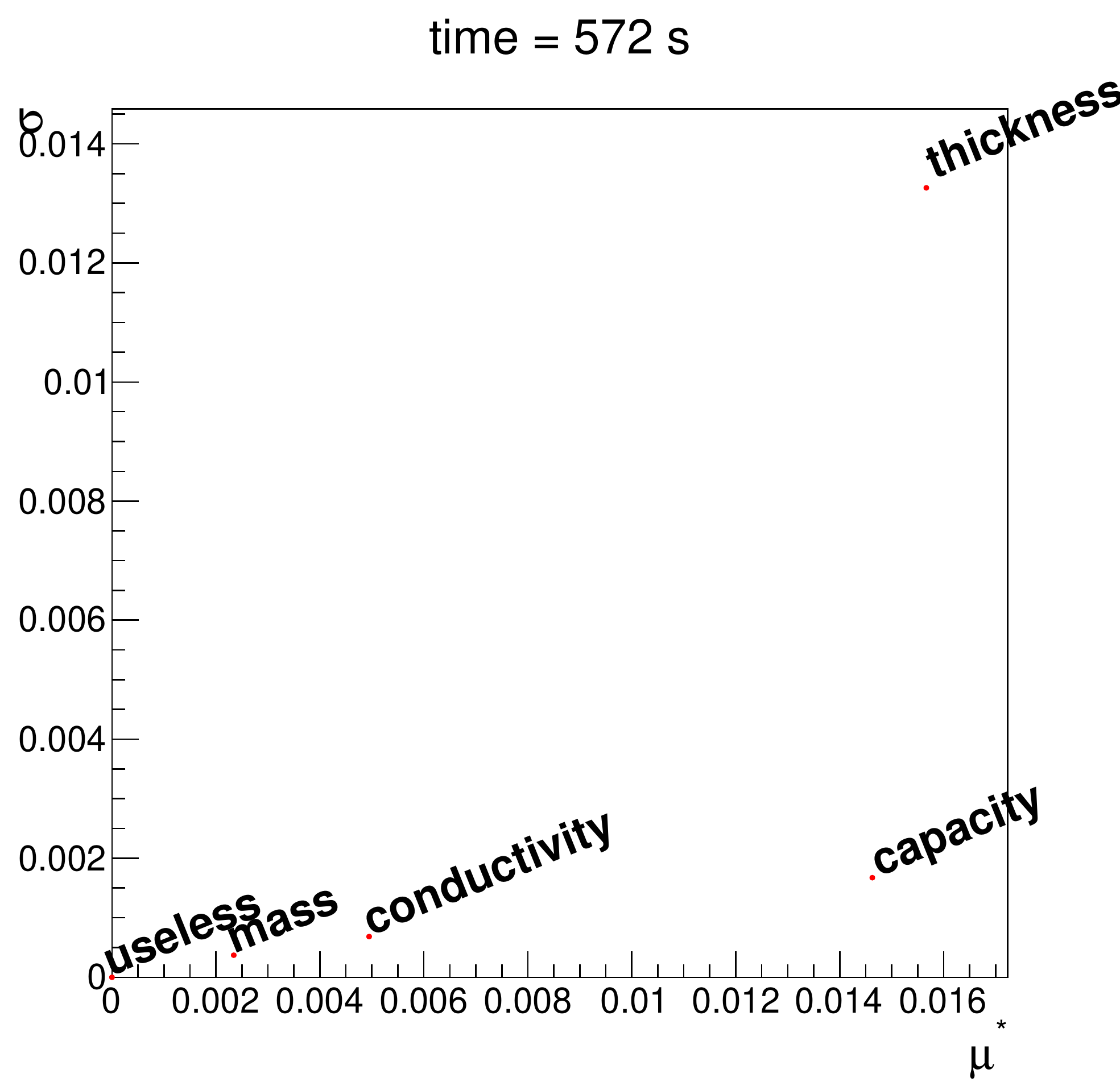}
      \label{fig_Morris3Ex}
    }
    \subfloat[$x_{ds}=0.3$]{
      \includegraphics[page=2,width=0.4\textwidth]{{PTFECaseMorrisTest_xr_0.3_0.3_s_0.1_tr_0_10_s_1}.pdf}
      \label{fig_Morris3Time}
    }
    
    \subfloat[$x_{ds}=0.8$]{
      \includegraphics[page=1,width=0.4\textwidth]{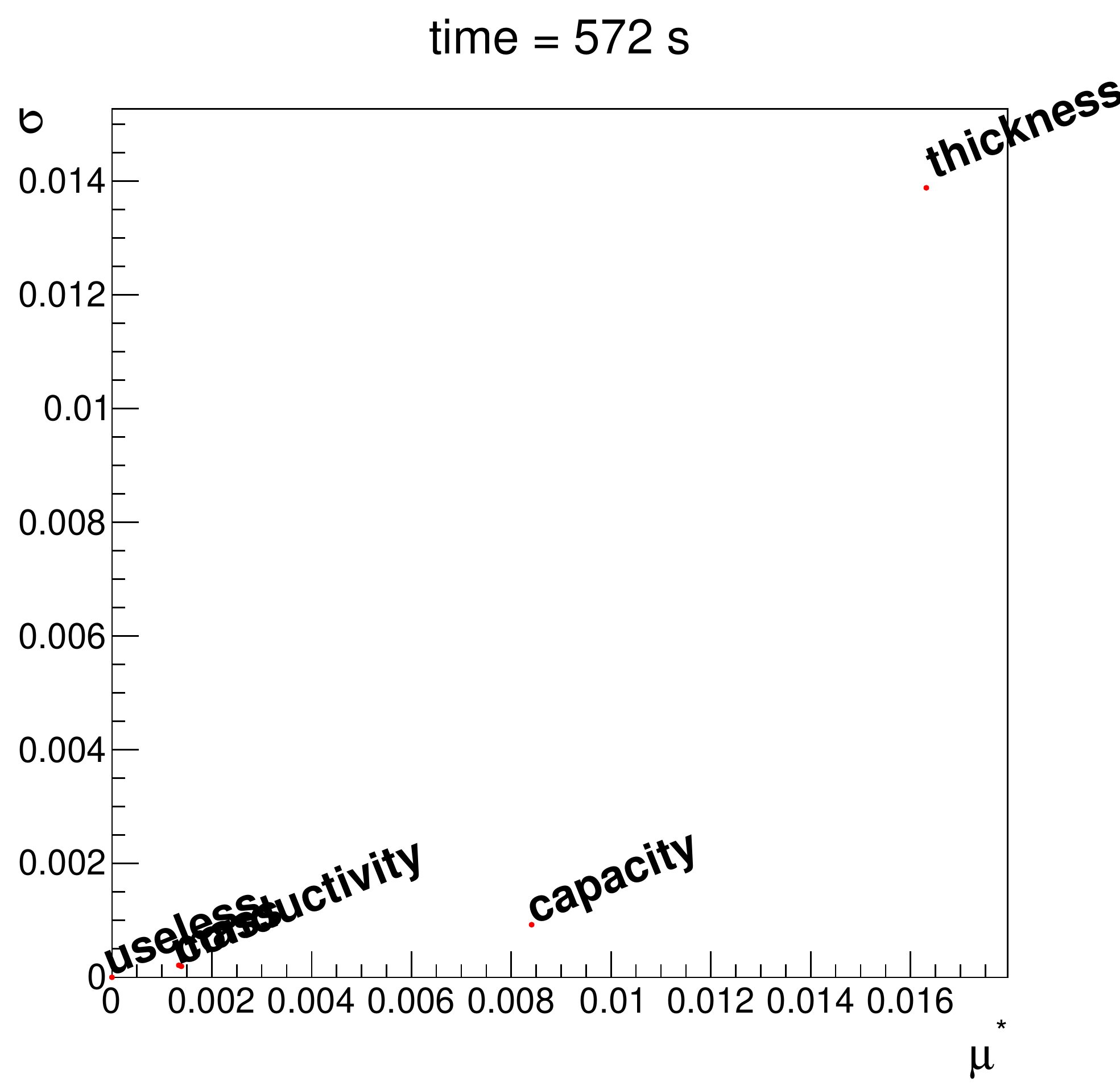}
      \label{fig_Morris8Ex}
    }
    \subfloat[$x_{ds}=0.8$]{
      \includegraphics[page=2,width=0.4\textwidth]{{PTFECaseMorrisTest_xr_0.8_0.8_s_0.1_tr_0_10_s_1}.pdf}
      \label{fig_Morris8Time}
    }    
  \end{center}
  \caption{Measurement of the Morris $\mu^*$ and $\sigma$ for $x_{ds}=0.3$ (top) and $x_{ds}=0.8$ (bottom). The a and c parts
    represent this measurement for a single value of the time, while the b and d parts show the evolution of $\mu^*$ (top
    pad) and $\sigma$ (bottom pad) as a function of the time.\label{fig_MorrisEx}}
\end{figure*}

\Figs{Morris3Ex}{Morris8Ex} represent the ($\mu^*$,$\sigma$) plane introduced in \Sect{MorrisIntro}, respectively for
$x_{ds}=0.3$ and $x_{ds}=0.8$, measured when the time is set to 572 seconds (about 2 thermal diffusion time). In both cases,
it is possible to split the plot in three parts:
\begin{itemize}
\item factors that have negligible effect on the output: both $\mu^*$ and $\sigma$ are very small. The ``useless'' input
  enters this category.
\item factors that have linear effects, without interaction with other inputs: $\mu^*$ is larger (all variations have an
  impact) but $\sigma$ is small (the impact is the same independently of the starting point). The massive thermal capacity is
  a very good illustration of this (as the thermal conductivity or the volumic mass at a smaller scale).
\item factors that have non-linear effects and/or interactions with other inputs: both $\mu^*$ and $\sigma$ are large. The
  thickness of the sheet is a perfect illustration of this.
\end{itemize}

\Figs{Morris3Time}{Morris8Time}, on the other hand, show the evolution of both the $\mu^*$ and $\sigma$ as a function of the
time for the different inputs. Here also, the ``useless'' inputs can clearly be spotted as negligible through time. Comparing
all the other curves, one has to decide the number of other inputs that can be kept into consideration, given the time and
memory consumption of a single calculation, but also the physics underlying this behaviour. For the thermal exchange example,
considering that the code is fast and the number of inputs is small, the only variable dropped thanks to this method is the
``useless'' one.

In practice, the main steps used to obtain these results are gathered in the following block:
\begin{lstcolored}[firstnumber=1,]
//Create the Morris object (given dataserver, the function, the number of trajectories and the variation (level))
TMorris * scmo = new TMorris(tds, "theta", ntrajectories, nlevel);
//Compute the indexes 
scmo->computeIndexes();
\end{lstcolored}

\subsection{The Fast method}

\subsubsection{Introduction}
\label{FastIntro}

The \emph{Fourier Amplitude Sensitivity Test}, known as FAST~\cite{MCRAE198215,SALTELLI1998445} provides an efficient way to
estimate the first order sensitivity indexes. Its main advantage is that the evaluation of sensitivity can be carried out
independently for each input factor, using just a dedicated set of runs, because all the terms in a Fourier expansion are
mutually orthogonal. To do so, it transforms the $n_X$-dimensional integration into a single-dimension one, by using the
transformation \[X_i = G_i(\sin(\omega_i \times s)),\] where ideally, $\lbrace \omega_i \rbrace$ is a set of angular
frequencies said to be incommensurate (meaning that no frequency can be obtained by linear combination of the other ones when
using integer coefficients) and $G_i$ is a transformation function chosen in order to ensure that the variable is sampled
accordingly with the probability density function of $X_i$. The parametric variable $s$ evolves in $[-\infty,\infty]$ and the
vector $(X_1(s),\hdots,X_{n_X}(s))$ traces out a curve that fills the entire $n_X$-dimensional research volume. When both
$G_i$ and $\omega_i$ are properly chosen, one can approximate the following relations:
\begin{align}
  E(Y) &= \frac{1}{2\pi} \int_{-\pi}^{\pi} f(s)ds \\
  {\rm Var}(Y)&=\frac{1}{2\pi}\int_{-\pi}^{\pi} f^2(s)ds - E^2(Y) \nonumber\\
  &\approx 2\sum_{k=1}^{\infty} (A_k^2+B_k^2),
\label{eq_FastExpAndVar}
\end{align}

where $f(s) = f(G_1(\sin(\omega_1 s)), \hdots, G_{n_X}(\sin(\omega_{n_X} s))$ and $A_k$ and $B_k$ are the Fourier
coefficients:

\begin{eqnarray}
  A_k&=&\frac{1}{2\pi}\int_{-\pi}^{\pi} f(s)\cos(k s)ds\label{eq_Akdef}\\
  B_k&=&\frac{1}{2\pi}\int_{-\pi}^{\pi} f(s)\sin(k s)ds.
\label{eq_Bkdef}
\end{eqnarray}
  
\subsubsection[Implementation in Uranie and application to the \usecase]{Implementation in \uranie{}  and application to the \usecase}
\label{FastImple}

This method is applied to the thermal exchange model. The first order coefficient is obtained by estimating the variance for
a fundamental $\omega_i$ and its harmonics, which can be done by using the second half of \Eqn{FastExpAndVar} running over
$p$ instead of $k$ and replacing the index by $p. \omega_i$. A cut-off $M$ has to be chosen for the sum and is called the
interference factor. Knowing that, the contribution to the output variance of a certain frequency, \emph{i.e.} the first
order sensitivity index, can be expressed from \Eqns{Akdef}{Bkdef} as
\begin{equation}
  \label{eq_FastSi}
  S_i= \frac{ \sum_{p=1}^{M}(A_{p. \omega_i}^2 + B_{p. \omega_i}^2) }{ \sum_{i=1}^{n_X} \sum_{p_i=1}^{M} (A_{p_i \omega_i}^2+B_{p_i \omega_i}^2)}
\end{equation}

\begin{figure}[h!]
  \begin{center}
    \subfloat[$x_{ds}=0.3$]{
      \includegraphics[width=0.8\linewidth]{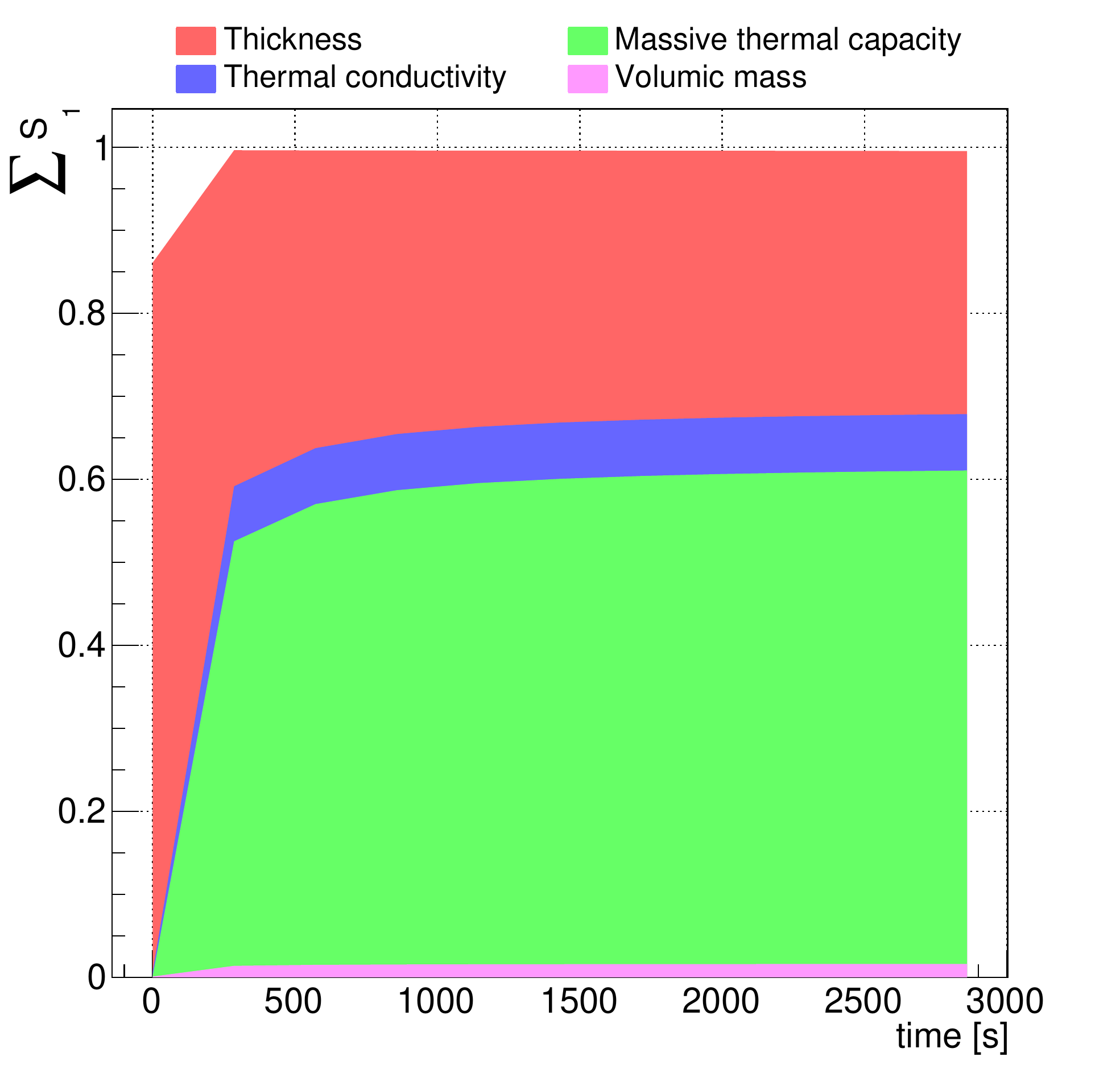}
      \label{fig_Fast3Ex}
    }
    
    \subfloat[$x_{ds}=0.8$]{
      \includegraphics[width=0.8\linewidth]{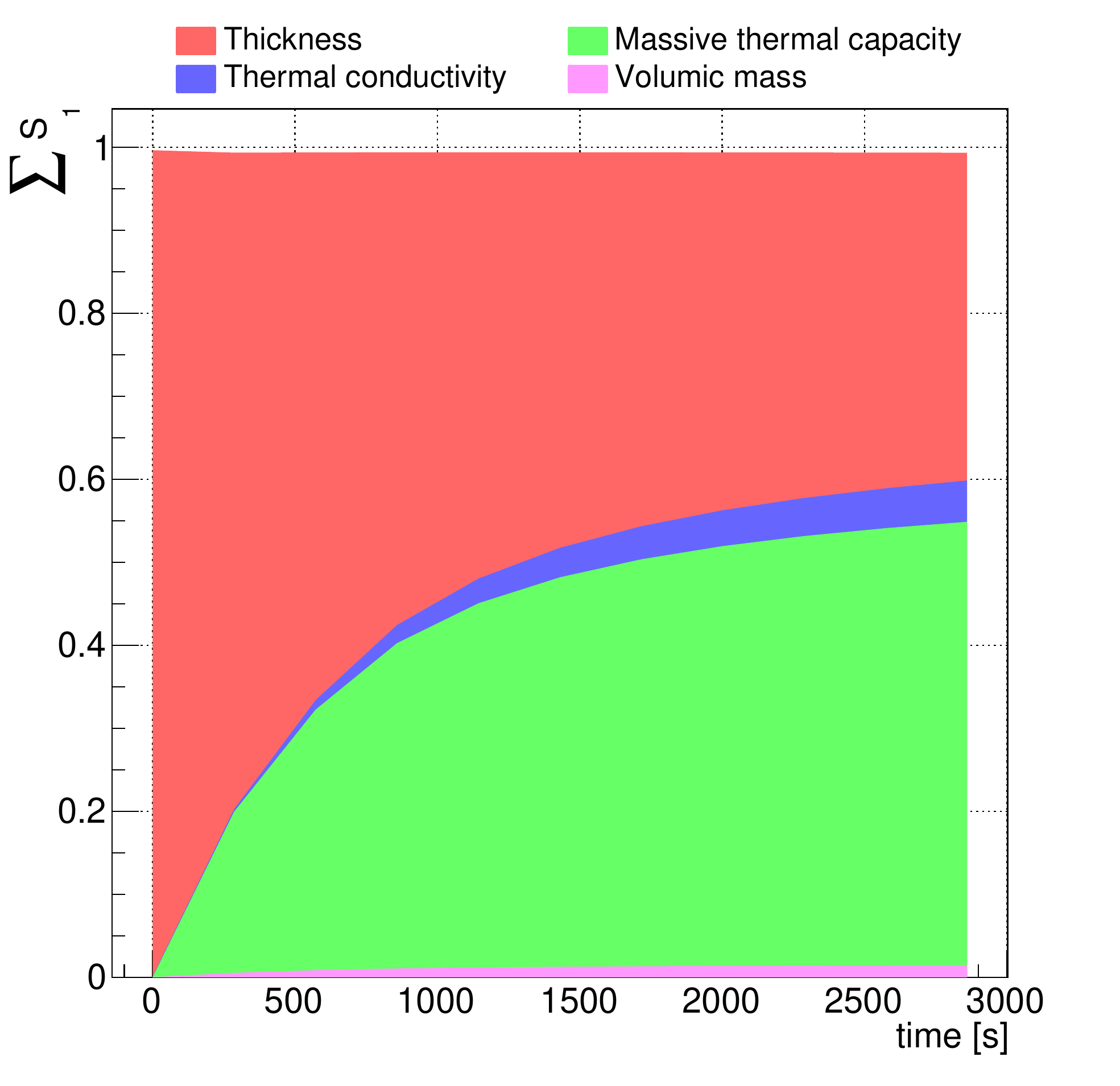}
      \label{fig_Fast8Ex}
    }
  \end{center}
  \caption{Measurement of the first order coefficients, with the FAST method, for $x_{ds}=0.3$ (a) and $x_{ds}=0.8$
    (b), as a function of time.\label{fig_FastEx}}
\end{figure}

The results are gathered in \Fig{FastEx} which shows the evolution of the first order coefficients, as a function of the
time, for the four input variables of the model. The histograms are stacked, which means that the contribution of every
inputs can be seen as the area represented by the corresponding colour, while the upper limit of the histograms is the sum of
all the contributions. \Figs{Fast3Ex}{Fast8Ex} show the evolution as a function of time respectively for $x_{ds}=0.3$ and
$x_{ds}=0.8$. The conclusions drawn here  are in agreement with the ones from the Morris method in \Sect{MorrisImple}: 
\begin{itemize}
\item the impact of the volumic mass uncertainty is negligible;
\item the two most important contributions are coming from the massive thermal capacity and thickness uncertainties;
\item the relative importance of the impact of the massive thermal capacity uncertainty with respect to the thickness one
  seems to increase once we are getting closer to the centre of the sheet.
\end{itemize}

On the other hand, by investigating the results in \Fig{MorrisEx}, the only possible statement about the impact of the
thickness uncertainty was that this factor had either a non-linear effect and/or interaction with other inputs. Here, as the
sum of the first order coefficients is equal to 1 for both dimensionless position, it seems reasonable to state that the
model has no strong interaction but that the impact of the thickness uncertainty might be a non-linear effect.

In practice, the main steps used to obtain these results are gathered in the following block:
\begin{lstcolored}[firstnumber=1,]
//Create the Fast object (givan dataserver, the function and the number of locations)
TFast * tfast = new TFast(tds, "theta", ns);
//Choose the interference level
tfast->setM(_M)
//Compute the indexes 
tfast->computeIndexes(); 
\end{lstcolored}

\subsection{The Sobol method}

\subsubsection{Introduction}

\begin{figure*}
  \begin{center}
    \includegraphics[width=0.8\textwidth]{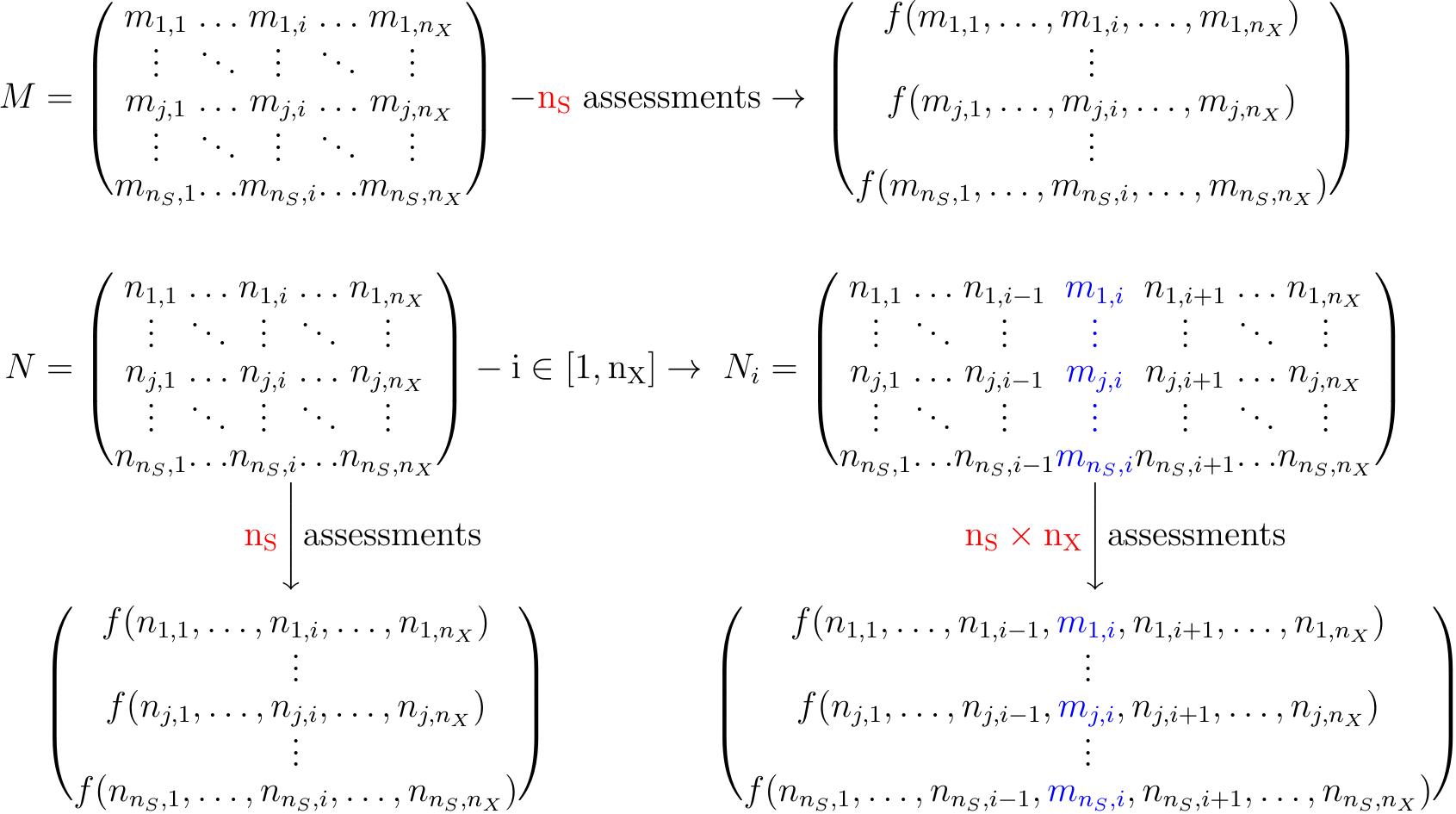}
  \end{center}
  \caption{Description of the method used to compute the \sobolCs{} from two matrices.\label{fig_SobolMethod}}
\end{figure*}

The Sobol method is a Monte-Carlo-based estimation that provides the first and total order sensitivity indexes (respectively
introduced in \Eqns{Sobol1}{SobolT}) at the cost of requiring a total of $n_S(n_X+2)$ code assessments. Instead of generating
a single \doe, the idea is to produce two of them, called M and N whom size is set to $n_S \times n_X$ (both matrices are
different and independent random samplings). A schematic view of this method, called the \emph{pick-and-freeze} method, can
be found in \Fig{SobolMethod}.
   
The first step is to compute the first order sensitivity index, based on the measurement of the numerator, ${\rm
  Var}(E(Y|X_i))$, which can be written $E(E(Y|X_i)^2) - E(E(Y|X_i))^2$. Since the second part of previous formula is
equivalent to the output expectation, the calculation of the first order indexes requires es- timates of $E(E(Y|X_i)^2)$,
${\rm Var}(Y)$ and $E(Y)$. The matrix $M$ is passed to the code and $n_S$ assessments are done to get a vector of outputs
(shown as the first line of \Fig{SobolMethod}). The i$^{th}$ column of $N$ is then replaced by the $M$'s one (\emph{pick}),
creating a new $N_i$ matrix which is provided to the code, for an additional cost of $n_S\times n_X$ assessments. This steps
is represented by the second line and the right-part of the third line in \Fig{SobolMethod} and the total cost for the first
indexes estimation is $n_S(n_X+11)$ code assessments.

Finally the total order indexes are computed starting from the right-hand side of \Eqn{SobolT}, which looks very much alike
\Eqn{Sobol1} used to compute the first order but instead of a condition on having $i$ known (\emph{frozen}), it is the exact
opposite: the condition is to freeze all the columns but $i$. It is doable as this is the only difference between the $N$ and
$N_i$ matrices. The total order indexes are thus obtained by passing the $N$ matrix to the code, leading to $n_S$ additional
code assessments, as shown by the left part of the third line in \Fig{SobolMethod}.

\subsubsection[Implementation in Uranie and application to the \usecase]{Implementation in \uranie{}  and application to the \usecase}
\label{SobolImple}
Different implementations of the \emph{pick-and-freeze} method have been proposed throughout the literature. In \uranie, a
single dedicated method gathers the results from several of them (\cite{Saltelli02,Monod06}\ldots). One of them in
particular is providing the coefficient values along with an estimation of their 95\% confidence level
(\cite{Martinez11}). By re-writing a \sobolC{} as a correlation coefficient, one can get, under certain hypothesis a
confidence level using the Fisher's transformation rule that applies on empirical correlation coefficients determination.

\begin{figure}[h!]
  \begin{center}
    \subfloat[$x_{ds}=0.3$]{
      \includegraphics[page=1,width=0.8\linewidth]{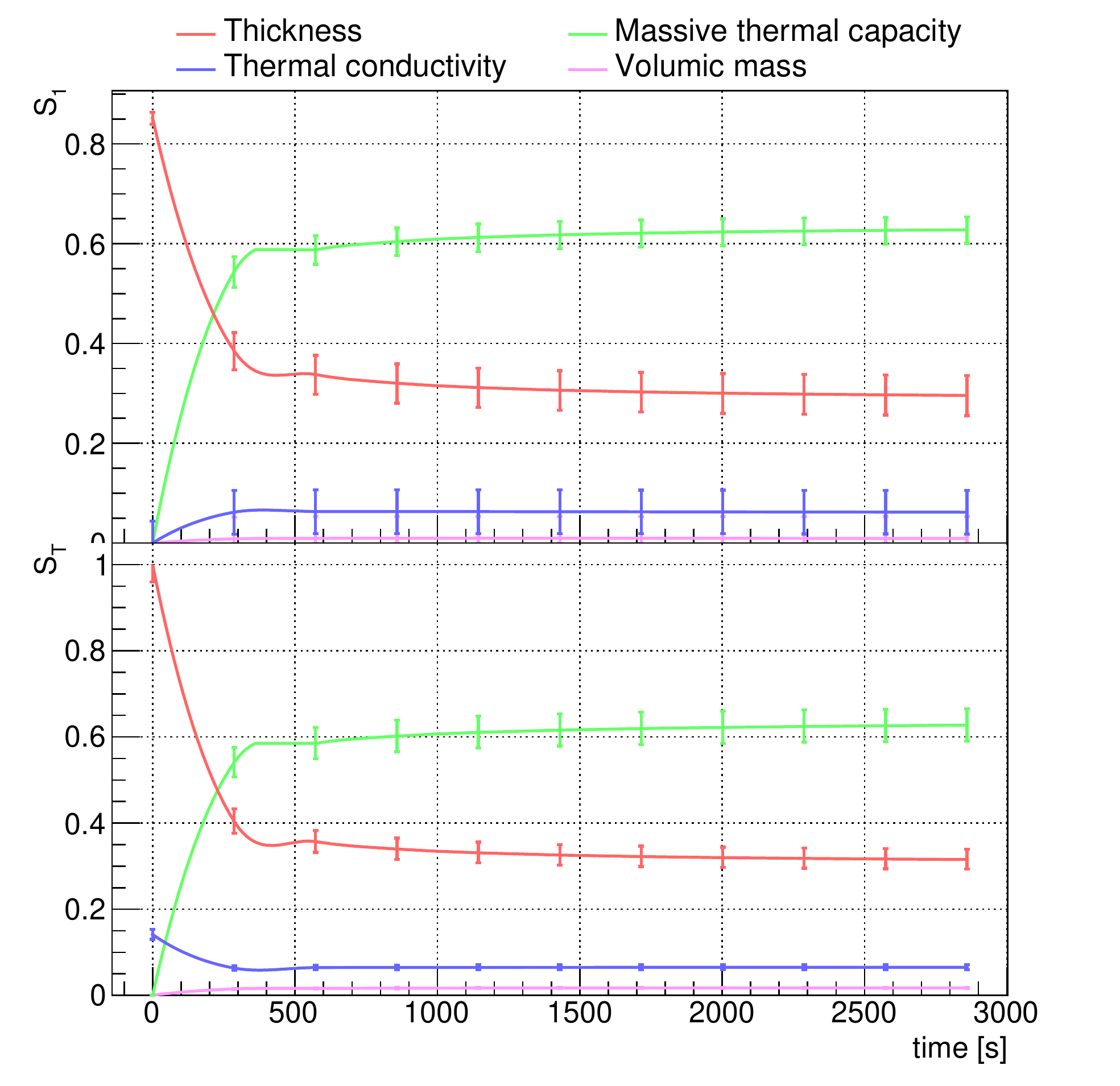}
      \label{fig_Sobol3Time}
    }
    
    \subfloat[$x_{ds}=0.8$]{
      \includegraphics[page=1,width=0.8\linewidth]{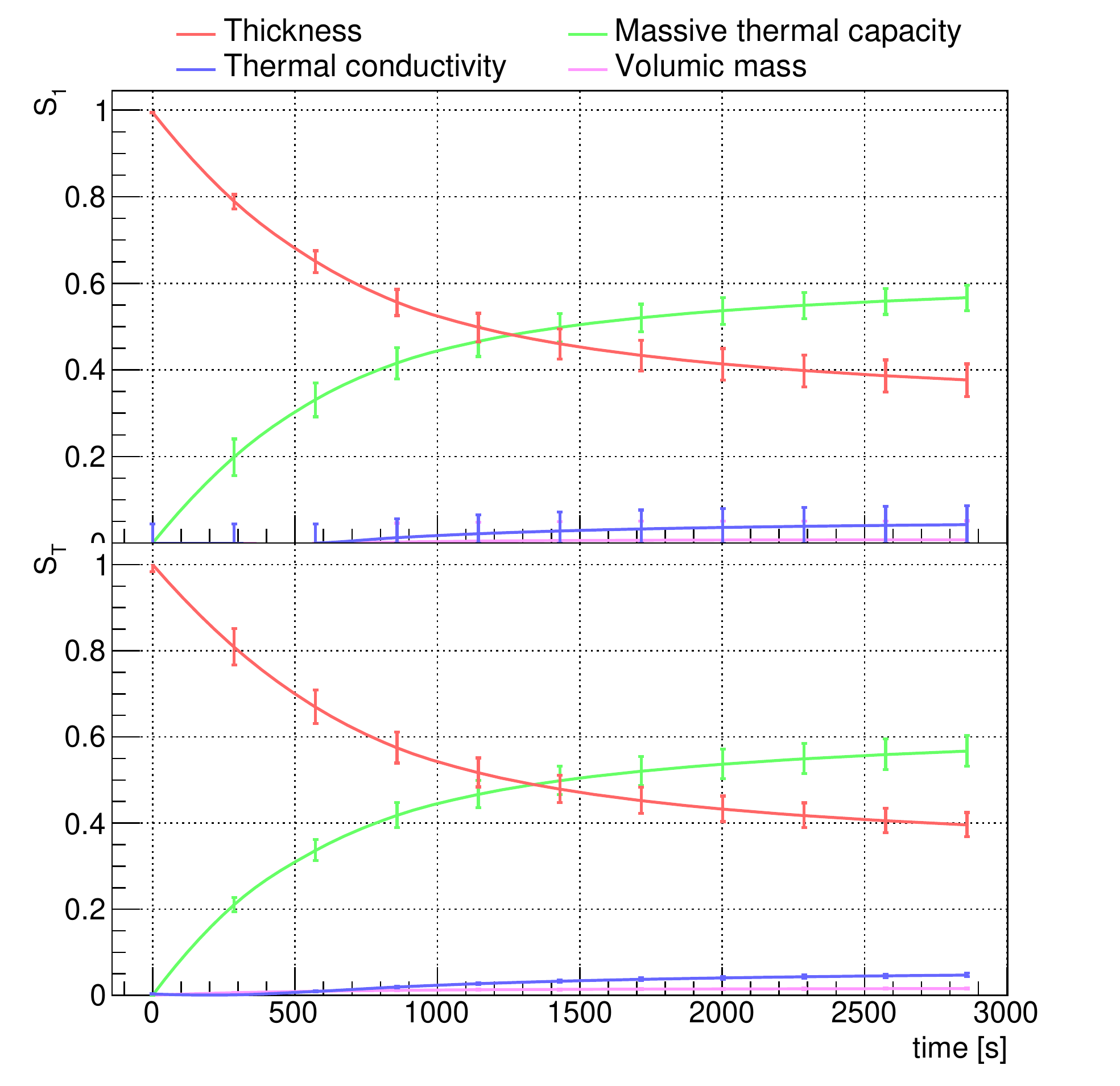}
      \label{fig_Sobol8Time}
    }
  \end{center}
  \caption{Evolution of both the first (top pad) and total order (bottom pad) coefficients, as a function of time, with the
    Sobol method, for $x_{ds}=0.3$ (a) and $x_{ds}=0.8$ (b), along with their 95\% confidence
    interval. \label{fig_SobolTime}}
\end{figure}

As for all the methods detailed in this paper, this one has been applied to our thermal exchange model to compute both the
first and total order coefficients. The results are gathered in \Fig{SobolTime} which shows the evolution of both the first
and total order coefficients, as a function of time, for the four input variables of the model, along with their 95\%
confidence interval. In \Figs{Sobol3Time}{Sobol8Time}, the upper part (the first order coefficients) and the lower one (the
total order coefficients) are displayed and a reasonable agreement between both order can be found. It leads, once more, to
the conclusion that the model has no interaction, as already stated in \Sect{FastImple} (for both $x_{ds}=0.3$ and
$x_{ds}=0.8$).

\begin{figure}[h!]
  \begin{center}
    \subfloat[$x_{ds}=0.3$]{
      \includegraphics[page=3,width=0.8\linewidth]{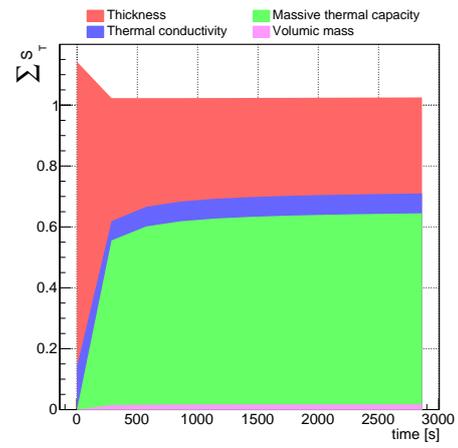}
      \label{fig_Sobol3Ex}
    }
    
    \subfloat[$x_{ds}=0.8$]{
      \includegraphics[page=3,width=0.8\linewidth]{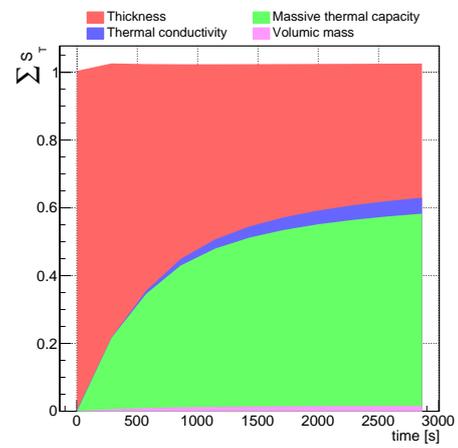}
      \label{fig_Sobol8Ex}
    }
  \end{center}
  \caption{Measurement of the first order coefficients, with the Sobol method, for $x_{ds}=0.3$ (a) and $x_{ds}=0.8$ (b), as
    a function of time.\label{fig_SobolEx}}
\end{figure}

In order to compare these results and the ones presented in \Fig{FastEx}, the first order coefficients estimated with the
Sobol method are represented as stacked histograms in \Fig{SobolEx}. Here again, the contribution of every inputs can be seen
as the area represented by the corresponding colour, while the upper limit of the histograms is the sum of all the
contributions. \Figs{Sobol3Ex}{Sobol8Ex} show the evolution as a function of time respectively for $x_{ds}=0.3$ and
$x_{ds}=0.8$.

In practice, the main steps used to obtain these results are gathered in the following block:
\begin{lstcolored}[firstnumber=1,]
//Create the Sobol object (given the dataserver, the function and the number of locations)
TSobol * tsobol = new TSobol(tds, "theta", ns);
//Compute the indexes 
tsobol->computeIndexes(); 
\end{lstcolored}

\subsubsection{To go further}
\label{FurtherSensi}
These methods to estimate either a ranking or more quantitative indicators, such as the \sobolCs, have dedicated options to
change the way the computations are done. On top of this, there are few other ways to get sensitivity indexes, such as:
\begin{itemize}
\item the regression either on values, to get \emph{standard regression coefficient} (SRC) and \emph{partial correlation
  coefficient} (PCC), or on ranks, to get \emph{standard regression rank coefficient} (SRRC) and \emph{partial correlation
  rank coefficient} (PRCC). All indexes can be estimated at once thanks to the algorithm implemented in \uranie{}~\cite{Iman85};
\item another Fourier-based algorithm, relying on a different paradigm, called \emph{Random Balance Design}
  (RBD)~\cite{Tarantola2006717};
\end{itemize}

%% file: Optimization.tex
Each optimisation study has its own peculiarities and it often requires to grope one's way forward, before finding an
interesting solution. Most commonly, when dealing with optimisation, there are:
\begin{itemize}
\item one or more objectives that one wants to minimise (or maximise).
\item decision variables that have a clear influence on the objectives.
\item some constraints either on the decision variables, on combination of some of them, or on objectives (defining
  the search domain)
\end{itemize}

For every problem, it is compulsory to choose an optimisation algorithm, which is a crucial part of the optimisation
procedure. It is possible to divide these algorithms into two different categories:
\begin{itemize}
\item local ones: they allow mono-criterion optimisation, with or without constraints. They are generally computationally
  efficient, but can not be used in parallel and tend to be trapped in local optima.
\item global ones: they allow multi-objective optimisation, with or without constraints. They are suitable for problems with
  many local optima, but are computationally expensive. However, they are easily parallelisable.
\end{itemize}

\uranie{} offers several possibilities, either by interfacing external library, as already stated in \Sect{extdep}, or
through the use of a dedicated package, called \Vizir~\cite{vizir}, developed at CEA, whose aim is to offer evolutionary
algorithms to solve multi-objective problems.

\subsection{Single-objective optimisation problem}

\subsubsection{Introduction}

In the case of a single criterion problem, the optimisation procedure is equivalent to the minimisation of a function
$f(\mathbf{x})$ which is called the \emph{cost function} or the \emph{objective function}. The optimisation leads to the
determination of a minimum (that can be called \emph{optimum}) that can either be global (there is no $\mathbf{x'}$ in the
research volume such as $f(\mathbf{x'}) < f(\mathbf{x_{min}})$) or local (same relation as before, but only in the vicinity
of $\mathbf{x_{min}}$). In the case where a maximum should be determined, all the techniques remained, but the objective is
changed (inverted) to get back to a minimum search.

In order to do so, \uranie{} offers many solutions thanks to its external dependencies:
\begin{description}
\item[\Minuit] It is \ROOT's package to perform single-objective optimisation problem, without constraint. It provides two
  algorithms
  \begin{itemize}
  \item Simplex: it does not use the first derivatives, it is insensitive to local optima, but without guarantee of
    convergence.
  \item Migrad: a fairly sophisticated gradient descent one that is able to escape from some local optima.
  \end{itemize}
\item[\NLopt] It is a library for nonlinear optimisation providing algorithms for single-objective optimisation problem, with
  or without constraint. The list of algorithms implemented in \uranie{} can be found in \cite{metho} along with a small
  description of their principle, taken from \NLopt~\cite{nlopt}.
\end{description}

\subsubsection{Application to the \usecase}
\label{optim_calib}

In this section, a calibration of some of the parameters of our thermal exchange model is performed. Indeed, performing the
calibration of a code comes down to finding the optimal set of parameters of the code which minimises the distance between
reference values and computations from the code. In \uranie, two distances are currently implemented:
\begin{itemize}
\item the root mean square deviation;
\item the weighted root mean square deviation.
\end{itemize}

The starting point is the following: one has done a set of thirty computations or measurements on a PTFE sheet without
keeping notes of the experimental conditions. Given that the sheet is made out of PTFE, several intrinsic properties are
known, such as the thermal conductivity ($\lambda$), the massive thermal capacity ($C_{\rho}$) and the volumic mass
($\rho$). On the other hand, there are two remaining unknown parameters: the thickness of the sheet ($e$) and the thermal
exchange coefficient value ($h$).

\begin{figure}[h!]
  \begin{center}
    \subfloat[Evolution of the objective]{
      \includegraphics[page=1,width=0.8\linewidth]{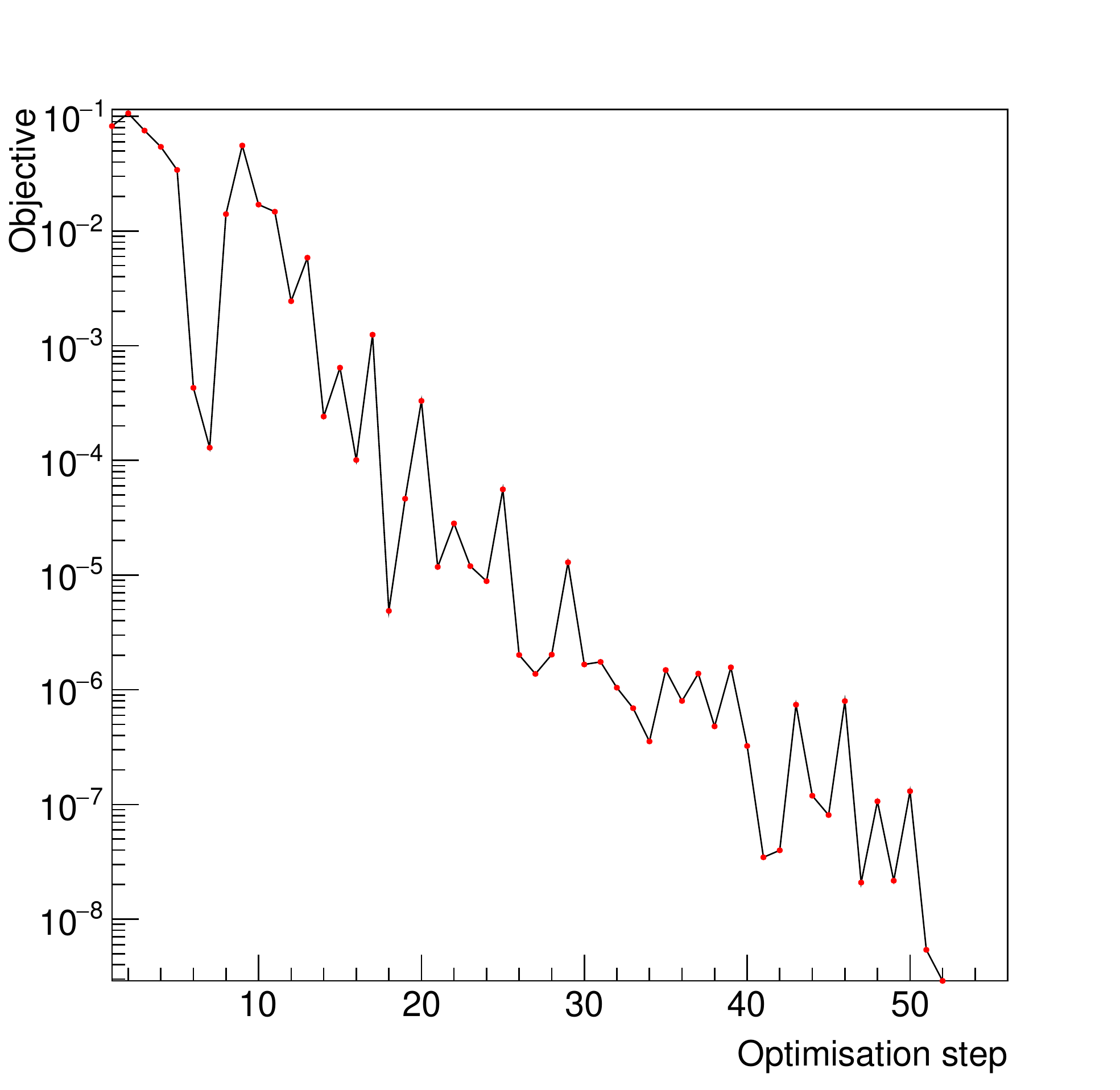}
      \label{fig_CalibObj}
    }
    
    \subfloat[$e$ and $h$ value]{
      \includegraphics[page=2,width=0.8\linewidth]{PTFECaseCalibTest.pdf}
      \label{fig_CalibPar}
      }
  \end{center}
  \caption{Evolution of the cost function as a function of the considered optimisation step (a). Evolution of the parameters
    from the initial point (red marker) to the optimal found one (blue marker) in the objective ($e$,$h$)-plane
    (b). \label{fig_Calib}}
\end{figure}

The Simplex algorithm (from \Minuit{} optimisation package) is used to minimise the root mean square deviation between the
reference thermal gauge values and the ones from every optimisation steps once the parameters under study have been
changed. Since this is a local algorithm, the starting point in the ($e$, $h$) plane has to be chosen beforehand (it is
represented with a red marker in \Fig{CalibPar}). A default step value is set for both parameters and the tolerance threshold
is chosen, along with a maximum number of calculation, both being the optimisation stopping criteria. The optimisation is run
leading to the results presented in \Fig{Calib}.

\Fig{CalibObj} shows the evolution of the objective function with respect to the iteration of the optimisation
algorithms. This evolution can be investigated along with the parameter variations shown in \Fig{CalibPar}: from the chosen
starting point in red, every optimisation steps is represented with a black marker and linked to the rest of the already done
estimation through a black line. The optimisation has stopped after 52 steps, heading to best estimated value for our
parameters of $e_{best}=.01$ and $h_{best}=100.076$ (the blue point in \Fig{CalibPar}). These values are in agreement with
the reference ones which have produced the original set of points (these values are shown in \Tab{metalCara}).

\subsubsection{Possible limitations}
\label{CalibLim}
This solution is very efficient, mainly because the code to be run is quick. In the case of a very time-memory and/or cpu
consuming code, this might have been difficult: as the Simplex algorithm is sequential, no parallelism is possible.  There
are more refined techniques to perform optimisation with less code assessments (using \surmod{} for instance), as introduced
in the following sections.

\subsection{Multi-objectives optimisation}

\subsubsection{Introduction}

The optimisation problem, in the multi-objective case, can then be expressed as the minimisation of the function
$F(\mathbf{x}) = (f_1(\mathbf{x});\,f_2(\mathbf{x});\, \ldots;\, f_n(\mathbf{x}))$ where $n$ is the number of objectives
imposed and $F$ is the complete cost function. In some cases, the objectives can be combined, for instance by doing a
weighted (or not) sum, resulting in a new objective over which the optimisation is performed. This is what is done in the
example above where the difference between the thirty output values in the reference set and the newly computed ones, for a
given set of parameters, are combined into a single objective. Unlike this case, the multi-objective hypothesis is that no
overall optimum can be determined when it is not be possible to quantify a relation between the objectives. In this case,
when two solutions $x_1$ and $x_2$ are possible, $x_1$ dominates $x_2$ if it does as good as the latter for all the
objectives and strictly better for at least one. The optimisation goal is then to get a group of solutions that are said to
be \emph{not dominated}: no solution out of this group dominates them, and in the group either. There is no best point,
unless an external constraint or preference is imposed, usually with hindsight.

\begin{figure}[h!]
  \begin{center}
    \subfloat[variable space]{
      \includegraphics[page=1,width=0.8\linewidth,keepaspectratio=true]{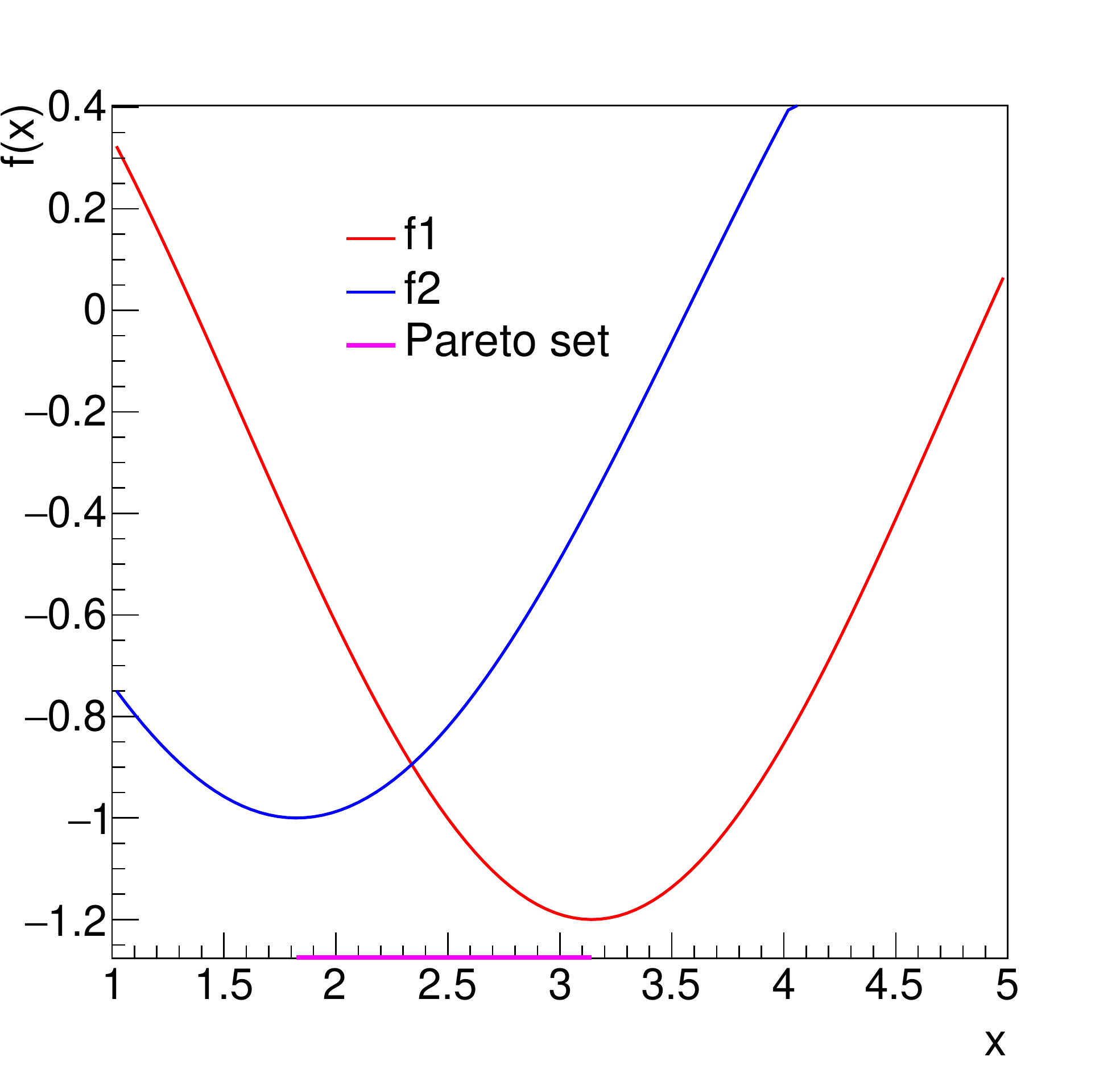}
      \label{fig_paretoVar}
    }
    
    \subfloat[objective space]{
      \includegraphics[page=2,width=0.8\linewidth,keepaspectratio=true]{Pareto_dummy_example.pdf}
      \label{fig_paretoObj}
      }
  \end{center}
  \caption{ Naive example of an imaginary optimisation case relying on two objectives that only depend on a single input
    variable. \label{fig_pareto}}
\end{figure}

The group of not-dominated solutions is called the \emph{Pareto set} and its representation in the objective space is called
the \emph{Pareto front}\footnote{Because of the discretisation, the obtained group is usually an approximation of the
  \emph{Pareto set}.}. \Fig{paretoVar} shows an academic example of a pure analytic model with two objectives depending only
on one variable. In this simple case, the Pareto set is shown in pink, as the area in between both criterion's minimum. Now
looking in the objective space in \Fig{paretoObj}, all the solutions are shown in black and the corresponding Pareto front
is, once more, depicted in pink.
 
\subsubsection[The Vizir package]{The \Vizir{} package}

\begin{figure*}
  \begin{center}
    \includegraphics[width=0.75\textwidth]{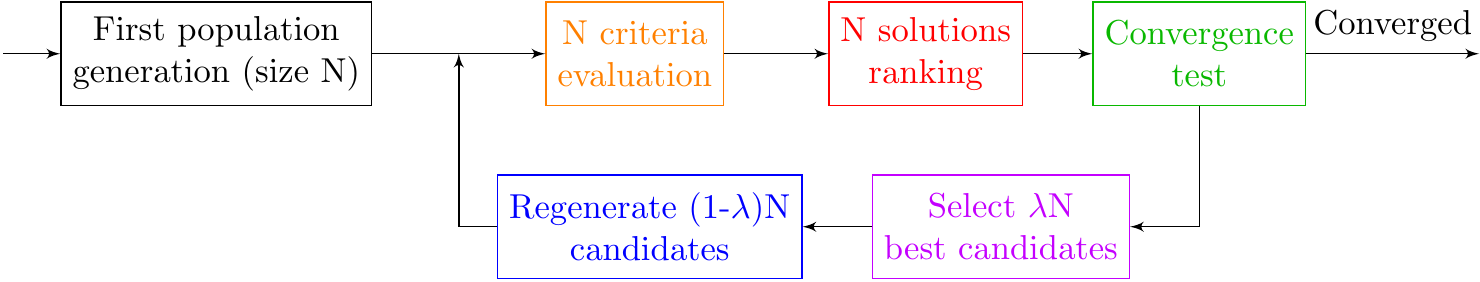}
  \end{center}
  \caption{Schematic description of the needed steps of an optimisation procedure, when this one is performed with
    \Vizir.\label{fig_VizirScheme}}
\end{figure*}

In \uranie{} multi-objectives optimisation issues are dealt with the \Vizir{} package, which gathers several solutions, all
developed at CEA, regarding the considered evolutionary algorithms and the way to make them evolve (genetic or swarm
algorithm, single or island evolution\ldots). In any case, the aim is to get a certain number $N$ of solutions to describe
correctly both the Pareto set and front, and the analysis can be described in few key steps (shown in \Fig{VizirScheme}) and
detailed below.
\begin{enumerate}
\item Initialisation. Create randomly, only using the research space definition, a population of the requested size
  ($N$). The first evaluation is performed for all candidates, meaning that the criteria and constrains will be tested and
  the results will be stored in a vector for all candidates.  This step is represented as a black box in \Fig{VizirScheme},
  followed by the evaluation shown as the orange box.
\item Ranking. The rank affected to a candidate under study corresponds to the number of other candidates that dominate
  it. The best candidates have then a rank 0 (they are not-dominated), following by those with rank 1, rank 2\ldots
\item Convergence test. This test (green box in \Fig{VizirScheme}) can reach three possible states:
  \begin{itemize}
  \item all the tested candidates are not-dominated. The algorithm has converged and the loop is stopped;
  \item not all candidates are not-dominated but the maximum number of evaluation has been reached. The algorithm has stopped
    without having converged. The optimisation should be restarted (maybe changing the configuration);
  \item not all candidates are not-dominated and the maximum number of evaluation is not reached.
  \end{itemize}
\item Re-generation. In the latter case of the convergence test, a fraction of the lowest-ranked candidates ($\lambda$) is
  kept (purple box in \Fig{VizirScheme}) and used to produce a new generation, the crossing procedure depending on the chosen
  algorithm (blue box in \Fig{VizirScheme}). This resulting population, made out of the selected ($\lambda N$) and
  re-generated candidates ($1-\lambda N$), is re-evaluated.
 \end{enumerate}

These steps are more thoroughly explained in \cite{metho}. Even though this library can be used on its own for
multi-objective optimisation, the example provided below will embedded it in the context of efficient global optimisation
(EGO).

\subsection{Efficient global optimisation}

This section layouts another optimisation possibility to look for a minimum using a global technique. The efficient global
optimisation, known as EGO~\cite{jones98EGO}, is first introduced and then applied to a simple mono-dimensional example
that will fully illustrate the principle. Finally the calibration problem discussed in \Sect{optim_calib} will be
investigated with this technique, to help appreciate the pros and cons of this method.

\subsubsection{Introduction}
\label{EGOIntro}

The EGO technique is the perfect illustration of method combination, introduced in \Sect{intro_metho}, where all the building
blocks are already implemented in \uranie. It might be useful in the case where the code is high time/cpu/memory consuming
(so one wants to limit as much as possible the computation) given that there might be several local minimum in which one
doesn't want to fall (preventing from using local optimisation such as gradient ones for instance). Instead of having to
choose a starting point (as done in \Sect{optim_calib}), the idea is to provide a training database which is supposed to be
representative of the problem, so whose size cannot be too small with respect to the dimension of the ongoing analysis. Once
done, a kriging model is constructed.

Let $f_{min} = min(y_1, \hdots, y_{n_t})$ be the current best function value ($n_t$ being the size of the kriging training
database). It is true, that before computing the output of the code for a given input vector $\mathbf{x}$, we are uncertain
about the value $y(\mathbf{x})$. On the other hand, $y(\mathbf{x})$ is not completely unknown as we can assimilate it to
$\hat{y}(\mathbf{x})$ its realisation through the kriging \surmod, which is provided along with its standard deviation
$s(\mathbf{x})$. With this hypothesis, it is possible to compute the probability of the real value to be below the actual
minimum $f_{min}$.  Different distances below the line $y = f_{min}$ , are associated with different density values. If we
weight all these possible improvements by the associated density value, we get what we call the ``expected improvement''
(EI).  The improvement for a given point $\mathbf{x}$ is $I = max(f_{min} - \hat{y}(\mathbf{x}), 0)$ which is a random
variable as $\hat y(\mathbf{x})$ models our uncertainty about the code's value at $\mathbf{x}$. To get the expected
improvement, from here, we simply take the expected value of this random variable:
\begin{align}  
  E[I(\mathbf{x})] &=& E[max(f_{min} - \hat{y}(\mathbf{x}), 0)] \nonumber\\
  &=& (f_{min} - \hat{y}(\mathbf{x})) \Phi\left(\frac{f_{min} - \hat{y}(\mathbf{x})}{s(\mathbf{x})}\right) \nonumber\\
  &+& s(\mathbf{x})\phi\left(\frac{f_{min} - \hat{y}(\mathbf{x})}{s(\mathbf{x})}\right) \label{eq_EI}
\end{align}
In this equation, $\phi(.)$ and $\Phi(.)$ are respectively the standard normal density and its cumulative distribution.
These two contributions are bringing a trade-off between exploring within a promising area and exploring where the
uncertainty of the \surmod{} is large. The latter contribution bringq back the global aspect requested.

Once done, the next step consists in searching the maximum of the expected improvement which is a positive definite
function. This is done by using the genetic algorithm, as this search is actually an optimisation: the aim is to minimise the
the opposite EI criteria, providing the best candidate. The code is then run on this specific location, which is then
included in the training database. The kriging model is re-build, using the updated training database and a new location is
determined, following the exact same recipe.

\subsubsection{Application to the thermal exchange model}
\label{EGOImple}

\begin{figure}[h!]
  \begin{center}
    \includegraphics[width=0.95\linewidth]{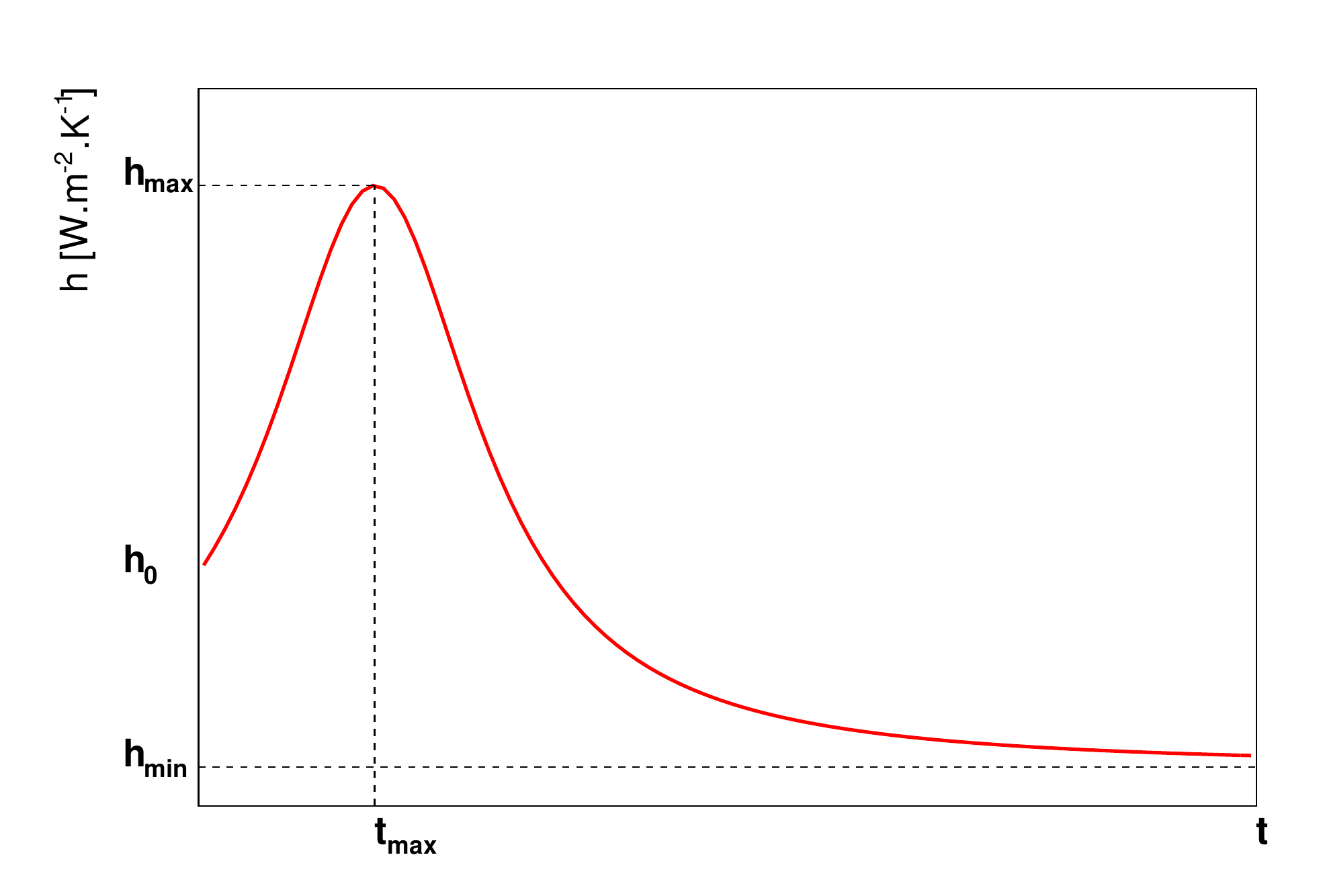}
  \end{center}
  \caption{Modelisation of the evolution of the thermal exchange coefficient as a function of the time.\label{fig_Hevol}}
\end{figure}

The idea here is to apply the EGO method on a mono-dimensional problem to get plots that would perfectly illustrate the way
this procedure works. This will be done by working on a simple function which is an extension of our thermal exchange
model. In \Sect{AnalyticModel} one of the first hypothesis to get an analytic solution was to fix the thermal exchange
coefficient $h$ to a constant value. This is known to be a rough approximation and a more rigorous way to describe the
evolution of this coefficient through time is actually given by the following equation:

\begin{equation}
  \label{eq_hevol}
  h(t) = \frac{h_{max}-h_{min}}{1+\beta(t-t_{max})^{2}}, \; \beta=\frac{h_{max}-h_{0}}{t_{max}^{2}(h_{0}-h_{min})}
\end{equation}

whose behaviour is represented in \Fig{Hevol}.

\begin{figure*}
  \begin{center}
    \subfloat[0 new points]{
      \includegraphics[page=1,width=0.49\textwidth]{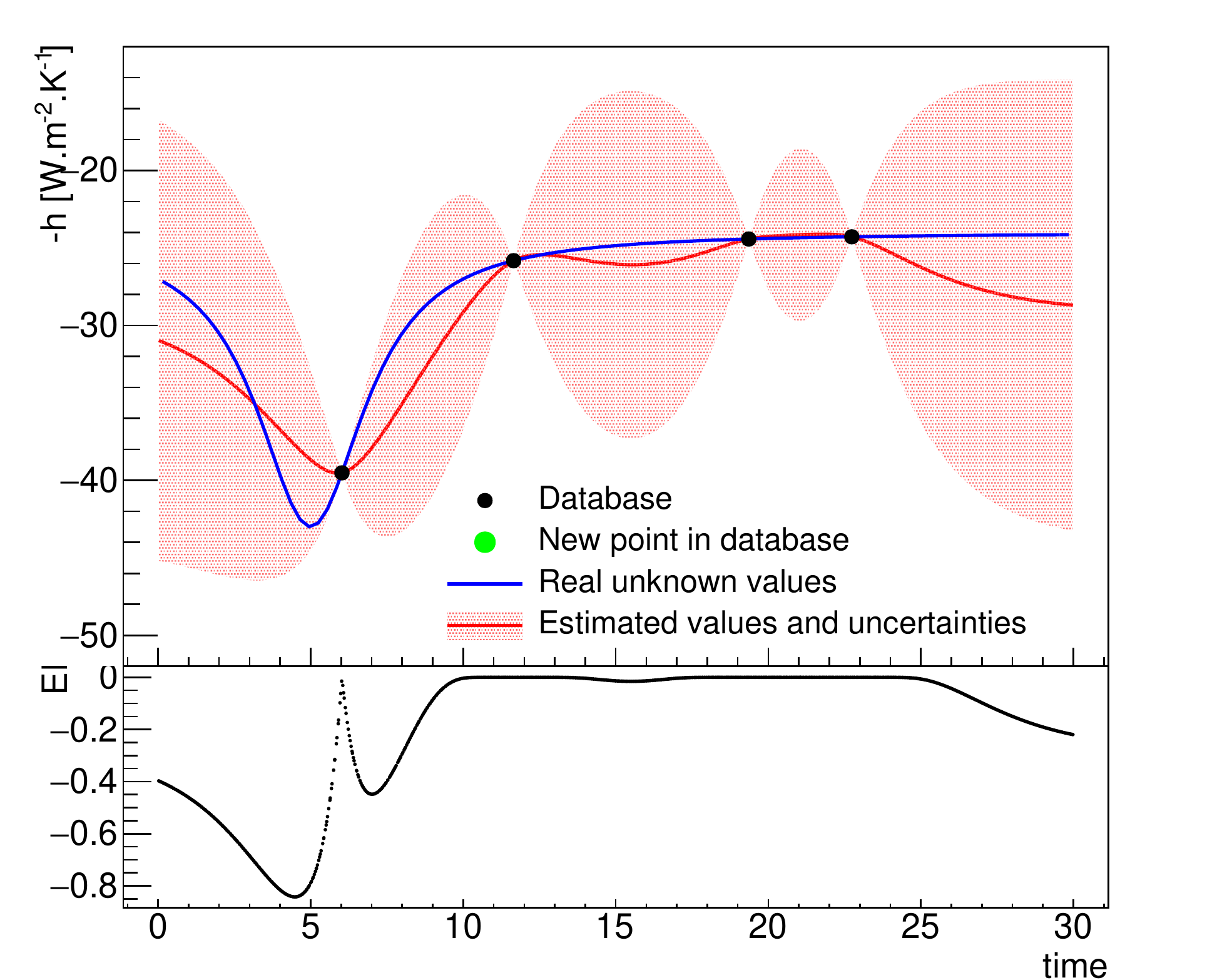}
      \label{fig_EGO0pt}
      }
    \subfloat[1 new points]{
      \includegraphics[page=2,width=0.49\textwidth]{EGOCase_NK_4_NC_10_NT_1000_EIBased_BottomPart.pdf}
      \label{fig_EGO1pt}
      }
    
    \subfloat[2 new points]{
      \includegraphics[page=3,width=0.49\textwidth]{EGOCase_NK_4_NC_10_NT_1000_EIBased_BottomPart.pdf}
      \label{fig_EGO2pt}
    }
    \subfloat[4 new points]{
      \includegraphics[page=5,width=0.49\textwidth]{EGOCase_NK_4_NC_10_NT_1000_EIBased_BottomPart.pdf}
      \label{fig_EGO4pt}
    }
    
    \subfloat[5 new points]{
      \includegraphics[page=6,width=0.49\textwidth]{EGOCase_NK_4_NC_10_NT_1000_EIBased_BottomPart.pdf}
      \label{fig_EGO5pt}
    }
    \subfloat[9 new points]{
      \includegraphics[page=10,width=0.49\textwidth]{EGOCase_NK_4_NC_10_NT_1000_EIBased_BottomPart.pdf}
      \label{fig_EGO9pt}
    }
  \end{center}
  \caption{Evolution of the kriging \surmod{} in red, compared to the real (supposed unknown) function in blue, as a function
    of the time for different number of locations in the training database. These ones are represented as black dots, expect for
    the latest-introduced one, spotted in green. The bottom part of every plot represents the evolution of the expected
    improvement.\label{fig_EGO}}
\end{figure*}

This equation depends on 4 parameters: the initial and asymptotic value of the thermal exchange coefficient (which can be
measured respectively at the very beginning and after a very long time on a given experiment). The other parameters are the
coordinate of the maximum, whose measurement can be turned into an optimisation by looking for the minimum of the opposite
function $-h(t)$. This is exactly what has been done and this analysis can be summarised in \Fig{EGO} where, in the upper
part of every pad, the blue line is the real "unknown" function used, the black points are the training database, the red
line is the approximation given by the kriging model along with its uncertainty and the green point is the latest point
included in the training database from the previous iteration. The bottom pad of every plot shows the inverted expected
improvement ($-EI$) which is minimised using the evolutionary algorithm to determine the next location to be included in the
training database.

Going though \Fig{EGO}, one can find back the different steps described in \Sect{EGOIntro}
\begin{description}
\item[\Fig{EGO0pt}] Starting point, the training database is made out of 4 locations, and the uncertainty on the model at
  several places are large. One of the location is close to the real minimum and is the current $f_{min}$. From the EI, the
  next location to be computed will be on the other side of the real minimum (where the estimated values from the kriging
  model are small and the uncertainties are large).
\item[\Fig{EGO1pt}] One more location has been computed and added to the training database (it is the current new
  $f_{min}$). The updated kriging model has changed tremendously and the lowest boundary is the next location to be
  computed. This is the perfect illustration of the global aspect of this method: a gradient would have been down to check
  for a smaller minimum, disregarding the fact that other part of the phase space might be really mis-modelled..
\item[\Fig{EGO2pt}] One more location has been computed and added to the training database upon which the kriging model has
  been updated. The next location to be computed (from the EI curve) is very close to the real minimum.
\item[\Fig{EGO4pt}] Two more locations have been added to the training database (among which a new $f_{min}$ in agreement
  with the real minimum). As the model uncertainty band for the locations around $f_{min}$ is small, the next location to be
  computed is the highest boundary, which, once more, shows the global behaviour of this protocol.
\item[\Fig{EGO5pt}] Once the highest boundary has been included, few more locations around the real minimum will be tested.
\item[\Fig{EGO9pt}] The method will start computing the closest location to $f_{min}$ with the largest uncertainty.
\end{description}

The steps described in \Fig{EGO} converge quickly toward an estimation of the parameters $t^{*}_{max}=5$ and
$h^{*}_{max}=42.9999$ which are in very good agreement with the injected values ($t^{Real}_{max}=5$,
$h^{Real}_{max}=43$). The only sensitive aspect is to be able to stop the method: one can not use a tolerance criteria as it
could prevent the exploration needed to conserve the global behaviour. The only remaining option is to set a maximum number
of location to be added, or to put a threshold on the \Loo{} $Q^2$ criterion of the kriging model. This criterion is not
really focusing on the minimum description, but more on agreement of the kriging model.

\begin{figure}[h!]
  \begin{center}
    \includegraphics[page=33,width=0.95\linewidth]{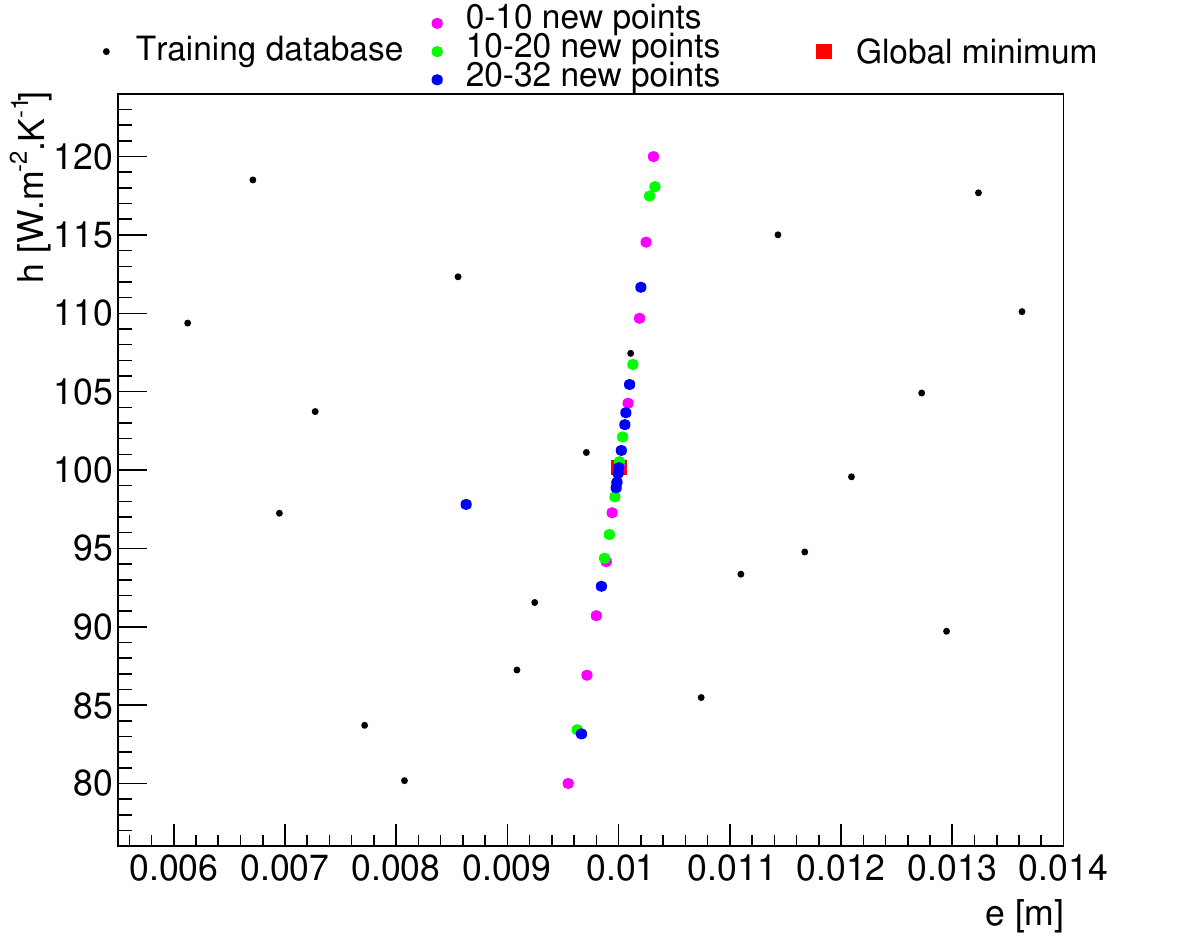}
  \end{center}
  \caption{Training database (black points) and the 32 new points computed using the EGO algorithm (purple, green and blue
    points) for the calibration problem discussed in \protect\Sect{optim_calib}. The best minimum found is represented as the
    red box.\label{fig_2DEGO}}
\end{figure}

For illustration purpose only, the same method has been applied to the calibration problem introduced in
\Sect{optim_calib}. The idea is to compare the results given in \Fig{Calib} to the one presented in \Fig{2DEGO}. A training
database of 20 locations has been generated and the algorithm has been run over 32 more computations in order to get the same
number of code assessments (1560, as already computed in \Sect{CalibLim}).

\Fig{2DEGO} shows the training database (black points) and the 32 new locations computed using the EGO algorithm (blue
points) along with the best minimum found (red box). The newly computed locations are split into three categories: the first
ten ones in purple, the following ten ones in green and the last twelve ones in blue. Thanks to this splitting, it is
possible to check that the optimisation exploration remains global as there can be green and even blue new locations computed
far away from the global minimum. However, most of the locations included by the EGO method are along a clear line, showing
the shape of the minimum valley in the ($e$,$h$)-plane (that was also visible in \Fig{CalibPar}). Here is the different level
of agreement obtained on the parameters, as a function of the number of locations computed (so the number of assessments):
\begin{description}
\item[8 new locations:] the accuracy obtained on $e$ and $h$ is respectively $\sim 0.3$\% and $\sim 1.5$\%
\item[21 new locations:] the accuracy obtained on $e$ and $h$ is respectively $\sim 0.01$\% and $\sim 0.3$\%
\item[28 new locations:] the accuracy obtained on $e$ and $h$ is respectively $\sim 0.01$\% and $\sim 0.2$\%
\end{description}

%% file: Broadening.tex
In this section, an important property, not discussed up to now but common to many \uranie's methods is briefly
discussed, along with the perspective of development considered by the team. 

\subsection{Parallelism}

The fact that the code under consideration might be very consuming in terms of time, cpu and/or memory has been raised
several times throughout this paper. This, along with the fact that some of method might need also a great deal of internal
computations, even without the external code, show the clear necessity to have parallelism strategy to benefit from the
recent hardware paradigm: the number of cpu is no-more a limiting criteria while on the other hand, cpu frequency has reached
a plateau and memory access tend to become problematic (\cite{drepper2007every}).

In order to deal with this, the \uranie{} team is working on two different aspects:
\begin{description}
\item[code assessments distribution:] this is the way an external code is called. Several strategies are implemented in
  \uranie:
  \begin{itemize}
  \item forking the code on a local node or on a list of resources listed as available at the initialisation. This is
    duplicating the code and the \launcher{} module (see \Fig{modorga}) will distribute the computations.
  \item shared-memory distribution. This technique is using the \verb?pthread? protocol to distribute the code assessments
    through the \relauncher{} module (see \Fig{modorga}). This, as all memory-shared strategy, can suffer from race conditions
    (which might only depend on whether the code used is thread-safe).
  \item split-memory distribution. This technique relies on an \pkg{mpirun} distribution of the computation through the
    \relauncher{} module (see \Fig{modorga}). This has the advantage of not suffering from race conditions, but the variables
    used can only be numerical-based (\uranie{} can also handle string variables as input/output of an external code).
  \end{itemize}
\item[internal calculation distribution:] Mainly available for the k nearest neighbour and artificial neural networks, the
  idea is to used the very high number of graphics processing unit (GPU) given in a reasonable graphical card, to perform
  internal computation (in the already-introduced case, the training of the synaptic weight for instance). This is done using
  \pkg{CUDA}, provided by the NVIDIA company (\cite{nvidia2011nvidia}).
\end{description}

The proper use of these solutions allows to benefit from the structure of the new computers and the grid upon which the
\uranie{} platform can be installed.

\subsection{Perspective}

The \uranie{} platform is in perpetual evolution to keep in touch with the latest improvement of both the academic and the
industrial world, bearing in mind the possible needs of its community as well. There are many possible fields of interests
toward which the developing team is investigating, among which:
\begin{itemize}
\item the usage of Markov Chain. Starting from the Bayesian inference needs, to be able to get the "a posteriori" parameters
  distributions, the idea would be to handle a Gibbs (\cite{casella1992explaining}) or Metropolis-Hastings
  (\cite{chib1995understanding}) implementation to move to our own Hamiltonian Markov chain (\cite{neal2011mcmc}) component
  to overcome possible slowness and convergence problems (particularly in high dimensions).
\item the development of "deep learning" capacities. The first step toward this is to change the actual perceptron to allow
  to use several outputs and to get more than one hidden layer of neurons. Other developments are considered, such as the use
  of recurrent neural network (RNN, \cite{robinson1991recurrent}) or deep belief network (DBN, \cite{hinton2007recognize}).
\item the development of many-criteria optimisation algorithms. With more than 3 criteria to optimise, the optimisation is
  said to be many-criteria, and there are several possibilities under study such as the knee point driven evolutionary
  algorithm (\cite{zhang2015knee}) or algorithm based on decomposition on a grid, as the \emph{Many Objective Evolutionary
    Algorithm based on Dominance and Decomposition} (MOEADD \cite{li2015evolutionary}). The aim is also to be able to have
  constraints on these criteria.
\end{itemize}

These leads are not exhaustive and the priority with which they might be considered can depend as well from the needs and
requests from our community.